\documentclass{article}
\usepackage[utf8]{inputenc}

\usepackage[T1]{fontenc}
\usepackage{bm}
\usepackage{graphicx}
\usepackage{amsmath}
\usepackage{url} 
\usepackage{amssymb}
\usepackage{amsfonts}
\usepackage{hyperref}
\usepackage{float}
\usepackage{amsthm}
\usepackage{xcolor}
\usepackage{algorithm2e}
\usepackage{authblk}
\usepackage{longtable}
\usepackage{enumitem}
\usepackage{appendix}
\usepackage{placeins}
\usepackage{subcaption}
\usepackage{lineno}
\graphicspath{{./FIGURES/}} 
\usepackage[scaled]{helvet}
\usepackage[superscript]{cite}
\usepackage[scale=0.8,textheight=9.5in]{geometry}
\makeatletter
\renewcommand{\maketitle}{\bgroup\setlength{\parindent}{0pt}
\begin{flushleft}
{\Large{\textbf{\@title}}}

{\small \@author}
\end{flushleft}\egroup
}
\makeatother
\renewenvironment{abstract}
{{\bfseries ABSTRACT}
\setlength{\leftmargin}{0mm}
\setlength{\rightmargin}{\leftmargin}%
\relax}
{\endlist}

\usepackage{fancyhdr}
\newcommand{\vect}[1]{\mbox{$\boldsymbol #1$}}

\title{Building surrogate temporal network data from observed backbones}
\author[1]{Charley Presigny}
\author[2]{Petter Holme}
\author[1,2,*]{Alain Barrat}

\affil[1]{Aix Marseille Univ, Universit\'e de Toulon, CNRS, CPT, Turing Center for Living Systems, Marseille, France}
\affil[2]{Tokyo Tech World Research Hub Initiative (WRHI), Tokyo Institute of Technology, Tokyo, Japan}
\affil[*]{Corresponding author: alain.barrat@cpt.univ-mrs.fr}

\date{\today}
\begin{document}
\maketitle

\begin{abstract}% max 150 words
\small
In many data sets, crucial elements co-exist with non-essential ones and noise. For data represented as networks in particular, several methods have been proposed to extract a ''network backbone'', i.e., the set of most important links. 
However, the question of how the resulting compressed views of the data can effectively be used has not been tackled. Here we address this issue by putting forward and exploring several systematic procedures to build surrogate data from various kinds of temporal network backbones. In particular, we explore how much information about the original data need to be retained alongside the backbone so that the surrogate data can be used in data-driven numerical simulations of spreading processes. We illustrate our results using empirical temporal networks with a broad variety of structures and properties.
\end{abstract}
\section{Introduction}

Many data sets coming from the world around us---transportation systems, human proximity, interactions on social media, etc.---take the form of networks~\cite{albert2002statistical,Barrat:2008}. One of network science's main objectives is to simplify complex, large-scale data sets to highlight important structures---like a map simplifies the geography of a country. There are several different perspectives one can take on how to map out a network. Perhaps the most popular direction is to detect mesoscopic structures, such as community structure~\cite{schaub2017many}. This is somewhat analogous to charting the cities of a country. In this paper, however, we start from the orthogonal problem of identifying the country's highways, i.e., the \textit{network backbone} structure. Mapping network backbones is intimately connected to another one of network science's aims---to explain and predict dynamic processes on the network~\cite{Barrat:2008}. To understand the flow of information or disease on a network, knowing the highway structure is more helpful than knowing the cities.

\begin{figure}[thb]
\includegraphics[width=.7\linewidth]{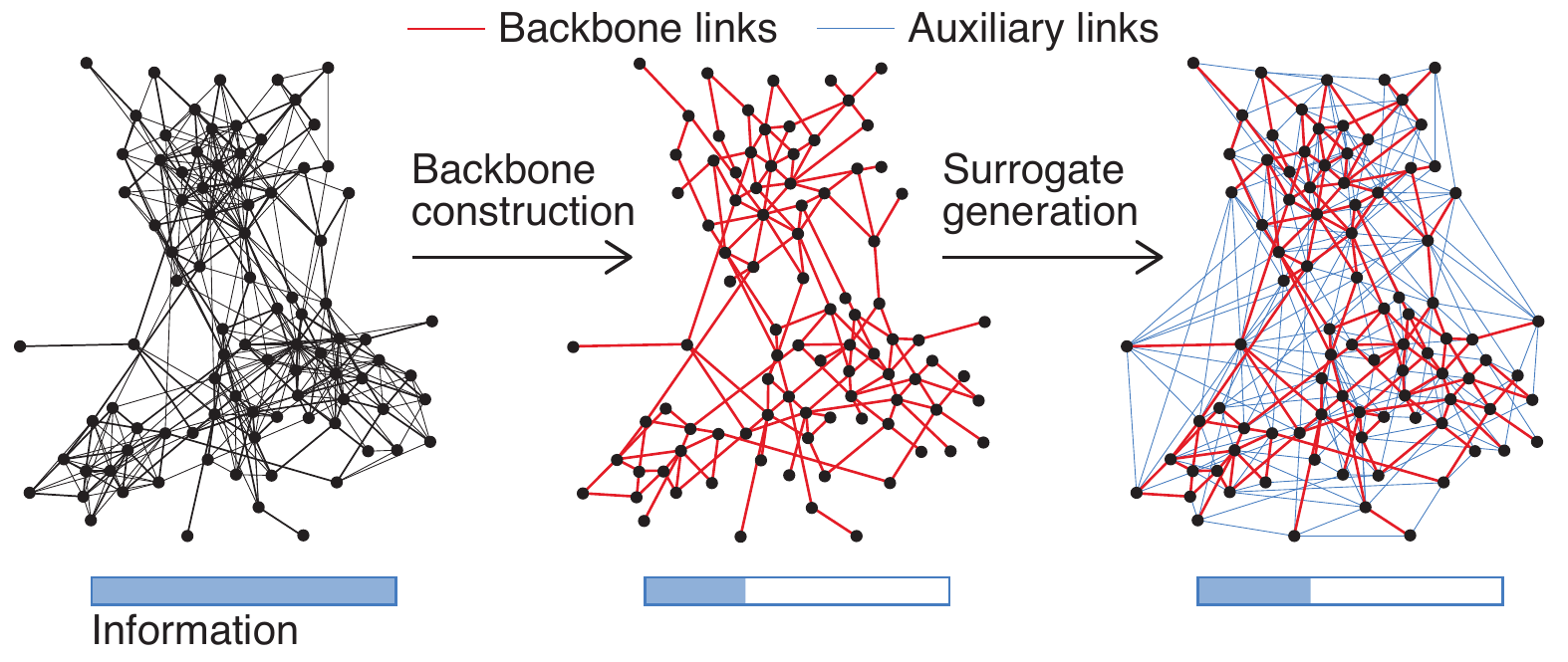}
\caption{{\bf Illustration of backbones and surrogates.} 
The backbone construction identifies the most important  links, and thereby compresses the original data. The surrogate generation models adds auxiliary links, extracted
at random using specific procedures,  
to create a network of the same size as the original. 
The more links (information) retained in the backbone construction, the more similar is the surrogate data to the original.
   \label{fig:ill}}
\end{figure}

Like statistical models, backbone extraction gives a compressed picture of a network. It can tell us many things about the original data, but it is not a model \textit{per se} without an associated decompression algorithm. In this paper, we develop and investigate such algorithms---to generate \textit{surrogate networks} from a network backbone~\cite{genois2015compensating,fournet2017estimating}. These algorithms thus model the rest of the network, apart from the backbone (See Figure~\ref{fig:ill}). In particular, we consider surrogate network construction from temporal network backbones, and investigate how to recreate the same behavior 
as the original temporal network with respect to the outcome of epidemic models unfolding on the network (note that this is a distinct problem from the one of inferring what contact network is the most likely to correspond to an observed empirical spreading pattern \cite{sah2018revealing}). Temporal networks encode not only which nodes are connected but also when the interactions between them happen~\cite{holme2012temporal,holme2016temporal,Masuda2016book}, and several studies have pointed out the necessity of including realistic topological and temporal structures to correctly describe spreading dynamics on temporal networks, including epidemic, opinion or information spreading~\cite{Barrat:2008,holme2016temporal}. 
The performance of surrogate data in correctly reproducing the outcome of spreading dynamics might thus be influenced by how such structures are taken into account in the backbone and surrogate data.

Several methods have been put forward to extract network backbones. For static weighted networks, the simplest way of filtering edges is to remove all the edges with weight below a given threshold value. More principled procedures use statistical tests based on null models to compare the weights of the edges with the ones that would be generated at random by a certain null model~\cite{serrano2009extracting,casiraghi2017relational,tumminello2011statistically,hatzopoulos2015quantifying,shekhtman2014robustness}. One then fixes a desired significance level and selects only those edges whose weight cannot be explained by the null model at the chosen significance level. These significant edges form the backbone of the network.

In the case of temporal networks, a simple approach for the extraction of backbones is to aggregate the data into a weighted static network. The weight of a link in this network is the number, or total duration, of the interactions between the involved nodes. To avoid neglecting potentially critical temporal features, it is however necessary to define an adequate temporal null model. Such a procedure makes it possible to extract a backbone of significant ties, i.e., of meaningful sequences of temporal contacts between nodes, possibly taking into account the temporal evolution of the nodes' properties
\cite{kobayashi2019structural,nadini2020detecting,nadini2020reconstructing}.

Typical backbone-method studies validate the procedures to extract backbones from static and temporal networks on synthetic benchmark tests and various empirical data sets. One explores the main properties of the resulting backbones, and compares these to known properties to understand by which network features they are influenced. However, there is most often no explicit interpretation of the ``importance'' of the links (except for the simple weight-thresholding procedure). Most importantly, it is unknown whether the information contained in the extracted backbone is enough to correctly summarize the original data and to be actionable, i.e., whether a user with access to the backbone but not to the original data can use it in data-driven applications such as simulations of dynamical processes.

In this paper, we explore this issue for backbone methods in temporal networks. Given a backbone representing only a fraction of the original data, we put forward and explore several systematic procedures to reconstruct surrogate actionable data
by adding auxiliary links to the backbone (see Figure \ref{fig:ill})
\cite{genois2015compensating,fournet2017estimating}.
These auxiliary links are extracted at random with a procedure depending on how the backbone was created.
%In each case, we investigate whether numerical simulations of spreading processes on these surrogate data give a reasonable estimation of the processes' outcome on the original data. From a general point of view, each procedure consists of starting from the backbone and adding to it random interactions in a process that is inverse to the backbone extraction. In particular, if the backbone has been obtained by comparison with a null model, one can interpret the procedure as an ``equation:'' original data = backbone + null model, or ``backbone = original data -- null model''. By adding noise corresponding to a realization of the null model to the backbone, one can thus hope to recover a network statistically similar to the original data: surrogate data $\equiv$ backbone + (random realization of the null model).
We compare several such procedures applied to backbones obtained through a simple thresholding procedure (serving as baseline) and the significant tie (ST) filter for temporal networks \cite{kobayashi2019structural}. We also propose a new version of this filter that considers the data's potential group structure. In each case, we explore how much information about the original data needs to be kept alongside the backbone (e.g., some statistical properties concerning the links that have been filtered out). To show our results' generality, we study temporal-network data with a broad range of topological and temporal structures.

\section{Results}

\subsection{Data and general methodology}

We consider data sets describing contacts between individuals
with temporal resolution, 
collected by the SocioPatterns collaboration (http://www.sociopatterns.org) in different settings:
 a workplace (office building, InVS15) \cite{genois2015data}, a high school (Thiers13) \cite{fournet2015contact}, 
 a primary school (LyonSchool) \cite{stehle2011high}
 and a scientific conference (SFHH) \cite{genois2018can}.
 These data describe close face-to-face proximity of
 individuals equipped with wearable sensors, with a temporal resolution of $20$ seconds. To limit the effect of noise, the data are moreover often aggregated over a coarser resolution of $\Delta$ minutes (e.g., in \citen{kobayashi2019structural} backbones are considered for $\Delta$ ranging from $3$ to $15$ minutes). Here we will use $\Delta = 3$ minutes, but we have obtained similar results
 for other temporal resolutions.
 Such data are conveniently represented as temporal networks in which nodes represent individuals.
 These networks are in discrete time, 
   i.e., composed by $T$ successive snapshots at times $t_0$, $t_0+\Delta, \cdots$,
 where $t_0$ is the initial time of the data set. 
 A temporal edge between two nodes $i$ and $j$ at time $t=t_0+n*\Delta$ represents the fact that the corresponding
 nodes have been in contact during the time interval $[t,t+\Delta]$. We also define a ``contact'' between $i$ and $j$ as an uninterrupted series of timestamps in which 
 there is a temporal edge between them. The duration of the contact is the length of this series.
 In each case, we also define the aggregated network as the static weighted network in which a link between two nodes denotes that these two nodes have been in contact at least once, and the weight of the link is given by the number of temporal edges between these nodes.
 Table \ref{tab:data} gives the main characteristics of each data set.
 
\begin{table}[htb] 
  \begin{tabular}{|c|c|c|c|c|c|c|c|c|c|}
    \hline
     Data set & Location & Year & $N$  & Duration & 
    $N_g$ & $E$ & $T$ & $E_T$ & 
    Ref. \\
    \hline
    InVS15 & Office building & 2015 & 217  & 2 weeks &  12 & 4,274 & 2,307 & 28,950 & 
    \citen{genois2018can}\\
    LyonSchool  & Primary school  & 2009 & 242  & 2 days  & 10 & 8,317 & 345 & 64,419 &\citen{stehle2011high}  \\
    SFHH & Conference & 2009 & 403  & 2 days  & None  & 9,565 & 421 & 73,620 & \citen{genois2018can}   \\
    Thiers13 & High school & 2013 & 326  & 1 week  & $9$ & 5,818 & 811 & 59,372 & \citen{fournet2015contact} \\
    \hline
  \end{tabular} 
  \caption{ \label{tab:data} {\bf Data sets considered.}
  $N$ is the number of participants, "Duration" the total duration of the data collection, $N_g$ the number of groups in the population, $E$ the number of ties (i.e., links in the aggregated network), $T$ the number of time stamps (once nights and week-ends, with no activity, have been removed), 
  $E_T$ the number of temporal edges. Here the temporal resolution is  $\Delta = 3\:{\rm min}$. }
\end{table}

These data sets were collected in very different contexts, so that the resulting structural and temporal properties
of the contact network differ strongly. School and high school populations are divided into classes of similar sizes, 
with a strong community structure and interactions between
classes only during the breaks (occurring with similar patterns in different days) \cite{stehle2011high,fournet2015contact}. 
In the office building, individuals are divided into departments of unequal sizes, and interactions are not limited by strict schedules \cite{genois2018can}. In the conference, a homogeneous aggregated contact network is observed \cite{stehle2011simulation}.

For each data set, we first extract their backbones
according either to a simple thresholding procedure or using the significant tie filter \cite{kobayashi2019structural} (see below and Methods). Each backbone contains only a tunable
fraction $f$ of the original ties (we will use
$f=40\%$, $10\%$ and $5\%$). In addition to the list of backbone ties (and possibly the corresponding
lists of temporal edges), we assume that some additional statistics of the original data sets are
conserved, such as the total number of temporal edges, 
the distributions of contact and
inter-contact durations (or simply the parameters of their fit
to simple functional forms such as power-laws \cite{cattuto2010dynamics,machens2013infectious,genois2015compensating}).
Whenever the data presents a group structure, 
the corresponding metadata can also be conserved
alongside the backbones.

We then consider several methods to reconstruct surrogate data from the backbones. Each method consists in adding temporal edges to the backbone in a way tailored to reproduce
several statistical features of the original data (see below and Methods).
For the resulting surrogate data, we investigate whether
they are suitable to feed numerical simulations of dynamical processes, i.e., whether 
the outcome of dynamical processes simulated on top of the surrogate data is close to the one obtained when using the original data. Specifically, we focus on the
paradigmatic susceptible-infectious-recovered (SIR) 
model of epidemic propagation. In this model, a susceptible (S) node becomes infectious (I) at rate $\beta$ when in contact with an infectious node. Infectious nodes recover spontaneously at rate $\nu$ and enter an immune recovered (R) state. We quantify the outcome of these processes, i.e., the epidemic risk, by two quantities: (i) the basic reproductive number $R_0$ (the average number of secondary infections by the source) and (ii) the average final size $\Omega$ of the spread, i.e., the fraction of nodes that have been in the infectious state at any time,
and we explore a wide range of parameter values (See Methods for details on numerical simulations and measures.)

In the following, we will show in the main text the results for the Thiers13 data set. As we indeed observe a robust phenomenology across data sets, the results for the other data sets are shown in the Supplementary Material.

\subsection{Backbones}

To extract a backbone of a given size from a temporal network data set, we consider the Significant Ties (ST) filter
\cite{kobayashi2019structural}. In this method, the actual number of
temporal edges between two nodes is compared to the one
of a temporal null model. The significant ties at significance level $\alpha$ are the ones such that their number of temporal edges cannot be explained by the null model at significance level $\alpha$. Specifically, the null model is defined
as follows: 
an ``activity level'' $a_i$ is associated to each node $i$, and two nodes $i$ and $j$ have a temporal edge at each time with probability $a_i a_j$. The activity levels of the nodes are obtained from the data by maximum likelihood estimation (see Methods and \citen{kobayashi2019structural}).
Tuning $\alpha$ makes it possible to select backbones representing a specific fraction $f$ of the ties of the original data. 

Moreover, we extend the ST filter to take into account the group structure of several data sets. The resulting GST filter
is obtained by modifying the temporal null model as follows:
the probability of a temporal edge between $i$ and $j$ is equal to $a_i a_j$ if $i$ and $j$ belong to the same group, 
and to $p a_i a_j$ if they belong to different groups.
The node activities and parameter $p$ are obtained by maximum
likelihood estimation as for the ST filter (see Methods).
Note that $p <1$ corresponds to cohesive group structures, while $p > 1$ would be obtained for disassortative structures. It would also be possible to use several values of $p$ depending on the respective groups of $i$ and $j$, but we consider here for simplicity only one parameter.

In addition, we consider as baseline
the simplest method to extract ties that can be  
interpreted as the most important in a network: we
order the ties according to their weight in the aggregated
network, as given by their number of temporal edges
(in the context of contact networks, this corresponds to the total duration of the contacts between the two nodes forming
the tie). The ``threshold'' backbone (TB) of the original data is then given by the fraction $f$ of ties with the largest weights.

We report in Table~\ref{tab:backbones}, for backbones
formed of a fraction $f= 40\%$, $10\%$ and $5\%$ of the original network, the corresponding number of 
temporal edges for each backbone extraction method.
Moreover, Figure~\ref{fig:distrib_backbones} in the Supplementary Material shows how
some statistics of the backbones compare to the ones of the original data.
As already discussed in \citen{kobayashi2019structural}, the ST backbone ties tend to have large weights, with distributions clearly shifted to large values with respect to the original data. 
However, while
this happens by definition in the threshold backbone, the distribution of weights in the ST backbone is smooth
and does not have a sharp cutoff at a minimal value.
Moreover, when the group structure is included (GST backbone), the distribution of weights becomes notably broader. This is due to inter-group ties that tend to have lower weights \cite{genois2015compensating}: these ties appear as significant only when we take into account, through the adequate null model (i.e., through the use of the parameter $p$), that pairs of individuals belonging to different groups have an a priori tendency to form less temporal edges than individuals of the same group. In fact, the ST filter tends to filter out most ties joining nodes of different groups~\cite{kobayashi2019structural}; the GST filter instead keeps ties both within and between groups.
We also note that both the clustering coefficient 
and the modularity of the partition in groups, when
measured in the backbones, can strongly deviate from the values in the original data (see Tables \ref{tab:cc}
and \ref{tab:modularity} in the Supplementary Material).
On the other hand, the 
distributions of contact and inter-contact durations are 
close to the ones observed in the original data
(see the Supplementary Material).

\subsection{Building surrogate data}

Backbones are by definition composed of a much smaller number of temporal edges and ties than the original data. 
As discussed above, their statistical properties are not identical to the ones of the data. 
It is therefore expected that numerical simulations of spreading processes on top of a backbone largely underestimate their outcome. We illustrate this in 
Figure~\ref{fig:orig_back_r0_omega_values} and in
the Supplementary Material. Note that the underestimation is not as strong as the one that would be obtained by a random sampling of the events, as the backbone ties tend to have large weights.

\begin{figure}[thb]
\includegraphics[width=.95\linewidth]{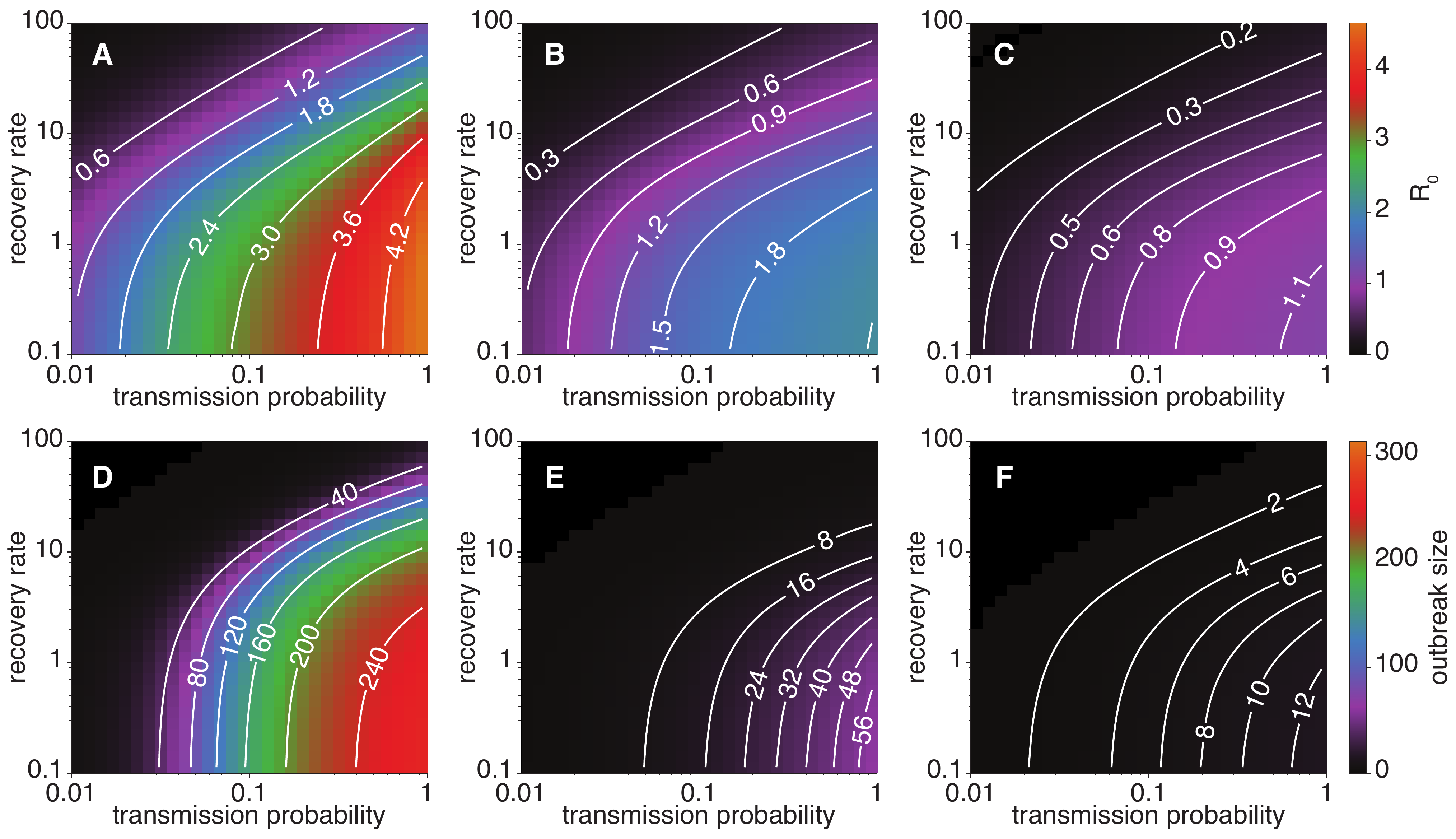}
%\centering
%\includegraphics[width=.3\linewidth]{Figures/r0/values/DATA_3min/tij_Thiers13.png}        
%\includegraphics[width=.3\linewidth]{Figures/r0/values/BACKBONES/Thiers13_tempback_groups_size_01}
%\includegraphics[width=.3\linewidth]{Figures/r0/values/BACKBONES/Thiers13_tempback_groups_size_005} 
% \includegraphics[width=.3\linewidth]{Figures/omega/values/DATA_3min/tij_Thiers13.png} 
%\includegraphics[width=.3\linewidth]{Figures/omega/values/BACKBONES/Thiers13_tempback_groups_size_01}
%\includegraphics[width=.3\linewidth]{Figures/omega/values/BACKBONES/Thiers13_tempback_groups_size_005}  
\caption{{\bf Original vs. backbone.} 
$R_0$ (top row) and $\Omega$ (bottom row) values obtained from the simulations on the original data and on the backbones, as a function of the SIR
parameters, for the Thiers13 data set.  Left: original data.  Middle: GST backbone with $f=10\%$. Right: GST backbone with $f=5\%$.
   \label{fig:orig_back_r0_omega_values}   }
\end{figure}

We therefore put forward several methods to construct surrogate data that are statistically more similar to the original data and, most importantly, yield more accurate estimations of processes' outcomes. Starting from a backbone composed of $E_b$ ties and $E_{bT}$ temporal edges, we want to recreate a temporal network with approximately $E$ ties and $E_T$ temporal edges. To this aim, we need to use complementary information, in addition to the list of temporal edges composing the backbone. For instance, it is quite clear that we cannot guess from the backbone itself the correct numbers of ties and temporal edges to be added. Thus, this additional information should be kept alongside the backbone to make it a usable summary of the data. Here we consider several procedures, highlighting in each case the necessary type and amount of information. Note that the resulting list of procedures does not pretend to be exhaustive but addresses a wide range of possibilities in terms of available information. Each procedure can be separated into two steps: (i) choosing ties (not included in the backbone) that interact in the surrogate data, and (ii) building timelines of interactions on the chosen ties. Procedure (ii) might also need to be performed on the backbone ties if the temporal information of the backbone ties is not available.

For step (i), we consider three distinct methods
for backbones extracted using the ST or GST method.
\begin{description}
\item[(G)ST-OA,] where ``OA'' stands for ``original activities''. We assume that the parameters of the null model used to extract the backbones are available, namely the original node activities $\{a_i, i=1,\cdots,N\}$ (and the parameter $p$ for the GST). Moreover we assume that $E_T$ is known and, for the GST, that the number of temporal edges between groups and within groups, $E_{T,{\rm inter}}$ and 
$E_{T,{\rm intra}}$, are also known, as well as the group to which each node belongs. 

In this procedure, for each pair of nodes $(i,j)$ not in the backbone, we add a temporal edge between $i$ and $j$ at each timestamp with probability $\alpha a_i a_j$, calibrating $\alpha$ so 
that the obtained total number of temporal edges is close to $E_T$ (see Methods).

For the GST, we use at each time the probabilities 
$\alpha_{\rm intra} a_i a_j$ if $i$ and $j$ are in the same group
and $\alpha_{\rm inter} p a_i a_j$ if they are not, 
calibrating $\alpha_{\rm inter}$ and $\alpha_{\rm intra}$ to get approximately the correct number of temporal edges both at the inter-group and intra-group levels.

\item[(G)ST-RA,] where ``RA'' stands for ```recomputed activities''. If the parameters of the null model (i.e., the activities $\{ a_i \}$ of the nodes) 
are not known, we use the fact that applying the MLE equations
to the backbone itself yields activity parameters correlated to the original ones (see Table \ref{tab:act_correl} in the Supplementary Material). 
We thus compute the activity $\tilde{a}_i$ of each node $i$  
(and the parameter $\tilde{p}$ if the group structure is known) in the backbone; we then add at each time a temporal edge between $i$ and $j$ with
probability $\alpha \tilde{a}_i \tilde{a}_j$, calibrating $\alpha$ to get approximately the correct number of temporal edges (we assume as previously that $E_T$ 
is known). For the GST case, we use probabilities
$\alpha_{\rm intra} \tilde{a}_i \tilde{a}_j$
and $\alpha_{\rm inter} \tilde{p} \tilde{a}_i \tilde{a}_j$
and calibrate $\alpha_{{\rm inter}/{\rm intra}}$ as in the previous method, assuming $E_{T,{\rm inter}}$ and $E_{T,{\rm intra}}$ are known.

\item[(G)ST-RT,] where ``RT'' stands for ``random ties''. 
We moreover consider a baseline in which
we add to the backbone the correct number of ties at random
(i.e., $E-E_b$), with weights drawn from the list of weights of the non-backbone ties.
Note that here we do not consider simply adding the correct number of temporal edges at random between nodes, because that would result in a very large number of ties with only one or few temporal edges, a structure very different from the original data. We thus assume that the number of ties in the original data $E$ is known ($E_{\rm inter}$ and $E_{\rm intra}$ if there are groups),
 in addition to the original number of temporal edges.
 Moreover, as the distribution of the backbone weights 
 is very different from the original data (see Figure 
 \ref{fig:distrib_backbones} and \citen{kobayashi2019structural}), we do not have a simple
 functional form for the weights of the non-backbone 
ties. We thus assume that the list of weights of the
non-significant ties has been kept.

\end{description}

Finally, for the backbones consisting of the ties with the largest weights (TB), as there is no underlying null model, we only consider the baseline reconstruction method 
which we denote by (G)TB-RT: we proceed here exactly as for the (G)ST-RT procedure.

Once the ties and the number of temporal edges on each tie have been chosen by one of these procedures, we can create surrogate timelines (step (ii) of the procedure)
in various ways. In each case, 
for each tie $(i,j)$ with number of temporal edges
$n_{ij}$, the aim is to 
choose $n_{ij}$ timestamps out of the $T$
possible ones.

\begin{description}
\item[Poisson:] if no temporal information on the original data is available, the simplest procedure consists in choosing totally at random the timestamps of the temporal edges for each tie.

\item[BTL-Poisson:] if the actual timelines of the 
backbone ties are known, one can keep these actual timelines
and choose at random the timestamps of temporal edges only for the surrogate ties.

\item[Stats:] we can instead assume that some information on the statistics of contact and inter-contact durations are known, as these properties have been shown to be extremely robust \cite{cattuto2010dynamics,genois2018can} 
(see also Figure~\ref{fig:distrib_backbones} in the Supplementary Material). They can moreover be approximately fitted to (truncated) power-law forms, meaning that the whole list of values is not needed, but only the parameters of the fit. We can then build a timeline of temporal edges for each tie using contact and inter-contact durations generated randomly from these fitted distributions.

\item[BTL-Stats:] if the actual timelines of the backbone ties are known, we keep these actual timelines, and proceed as in the Stats case for the surrogate ties only.

\end{description}

We note here that each step of the procedure is stochastic, with random choices of ties and temporal edges. 
Thus, repeating the same procedure multiple times yields an ensemble of surrogate temporal networks. In the Methods section, we provide a summary table of these procedures and the corresponding data used.

\subsection{Structural and temporal statistical properties of the surrogate data}

\begin{figure}[ht!]
\includegraphics[width=.7\linewidth]{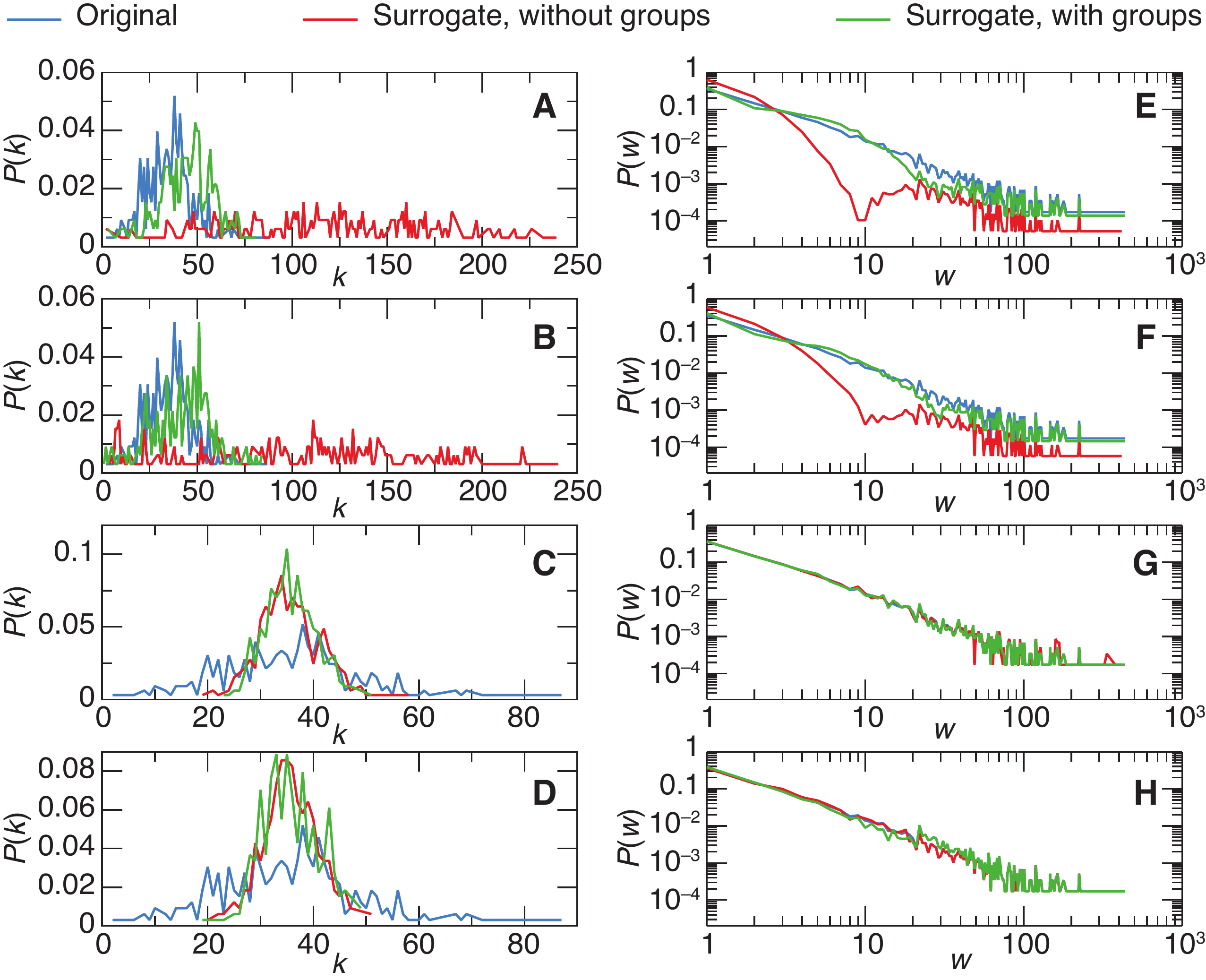}
\caption{{\bf Distributions of (aggregated) 
degrees (left columns) and weights (right column) in the surrogate data obtained by various methods.}
From top to bottom: (G)ST-OA, (G)ST-RA, (G)ST-RT, (G)TB-RT.
In each case, the blue line shows the distribution
for the original data, the red and green line for the surrogate built respectively without and with group information. Using group information yields
distributions closer to the original ones.
The surrogates were built from backbones with $f=10\%$ 
of the ties of the original data.
}
\label{fig:distrib_surrogate}
\end{figure}

Figure \ref{fig:distrib_surrogate} shows distributions of degrees and weights for the aggregate networks resulting from surrogate data created by several methods for the Thiers13 data set and $f=10\%$. Similar figures are shown in the Supplementary Material for the other data sets as well as a table with the relative values of the clustering coefficient and of the modularity of the aggregated surrogate networks.

The surrogates based on adding ties according to the ST null model tend to overestimate the node degrees, with the whole distribution shifting to larger values than in the original data and becoming broader. This effect is very strong for the ST-OA and ST-RA, but taking into account groups (GST-OA and GST-RA) leads to much weaker deviations from the data. Using group data also leads to distributions of weights close to the original ones. At the same time, ST-OA and ST-RA have a substantial depletion of the distribution at intermediate weight values (the tail of the distribution being correctly represented as most ties with large weights belong to the ST backbone). Note that these distributions emerge from the surrogate's construction, as the initial distribution is not assumed to be known here.

For surrogates created using random ties, (G)ST-RT and (G)TB-RT, the average degree is well reproduced by design, as the information about the original data number of ties is assumed to be known. On the other hand, the distribution of degrees is much narrower than the original one. The distribution of weights is almost identical to the original data since the list of the actual weights of the non-significant ties is assumed to be known.

In terms of clustering and modularity, the procedures
in which group information is known and used all lead
to values that are close to the original ones, while
ignoring group information can yield large discrepancies (see Supplementary Material).

Finally, the distributions of contact and inter-contact durations depend only on the way in which the timelines of ties are built in the surrogate data: they are exponential for the Poisson procedure, and very close to the original data distributions for the Stats procedures (not shown).

\subsection{Outcome of SIR processes on surrogate data sets}

\begin{figure}[thb]
    \centering
    
\includegraphics[width=.95\linewidth]{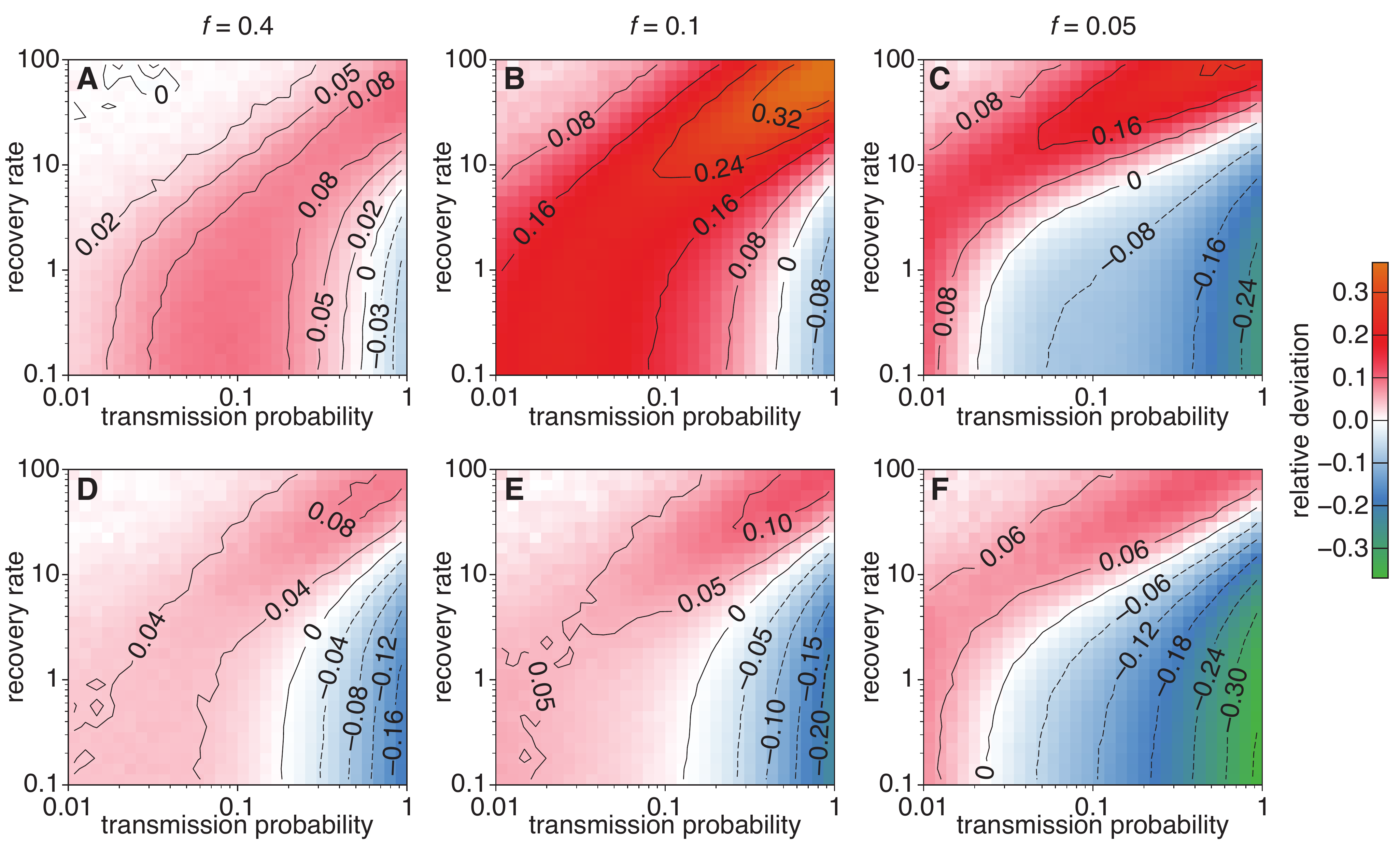}
 \caption{\textbf{Effect of taking group structure into account when building the surrogate data, for the Thiers13 data set.} Each panel shows the  relative difference in $R_0$ obtained from the simulations on surrogate data with respect to simulations on the original
 data. 
 Top: ST-RA. Bottom: GST-RA. Left $f=40\%$, middle $f=10\%$; right $f=5\%$. In each case the backbone timelines are kept, and timelines respecting the statistics of contact and inter-contact durations are built for the surrogate ties (BTL-Stats method).
\label{fig:group_reldiff_r0}}
\end{figure}

\begin{figure}[h!]
\centering
\includegraphics[width=.95\linewidth]{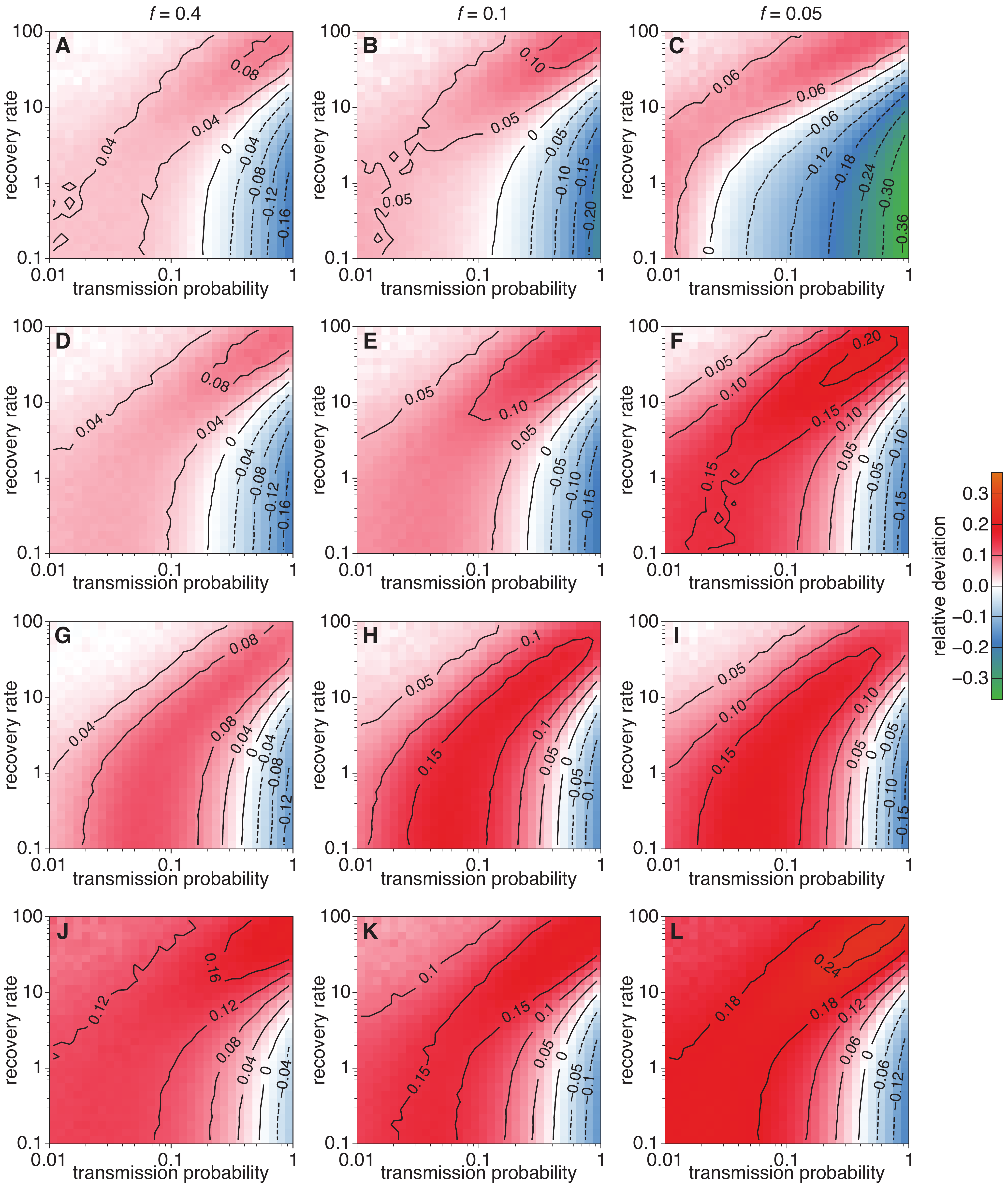}
\caption{{\bf Outcome of SIR processes on surrogate data
obtained by various reconstruction methods, for the Thiers13 data set.} Relative difference in the values of $R_0$ measured in simulations on the surrogate and on the original data. 
First row: GST-RA. Second row: GST-OA. Third row: GST-RT. Fourth row: GTB-RT. Left column $f=40\%$, middle $f=10\%$; right $f=5\%$. In each case the backbone timelines are kept, and timelines respecting the statistics of contact and inter-contact durations are built for the surrogate ties (BTL-Stats method).
\label{fig:sir_reldiff_r0}}
\end{figure}

We present our main results in Figures \ref{fig:group_reldiff_r0} -- \ref{fig:sir_reldiff_r0}: each panel of the figures shows, as a function of the
parameters $\beta$ and $\nu$, a color plot of the relative differences in the outcomes of SIR processes simulated either on surrogate data or on the original data. The outcome is measured here by the basic reproductive number $R_0$, and we show similar results for the epidemic size $\Omega$
in the Supplementary Material.

Let us first note a general pattern: $R_0$ tends to be underestimated, when using surrogate data, at large $\beta$ and small $\nu$, i.e., at very large $R_0$ and epidemic size. At large $\beta$ and $\nu$ on the other hand, the tendency is to overestimate the epidemic outcome. Finally, smaller deviations with respect to the simulations on the original data are observed in parameter regions where $R_0$ is close to $1$, i.e., close to the epidemic threshold.

Let us now consider in more details the results obtained with various types surrogate data and the effect of the choices made in the reconstruction procedure. 

Figures \ref{fig:group_reldiff_r0} and \ref{fig:group_reldiff_omega} highlight the impact of using information on the group structure of data. The surrogate data gets more ties between groups when group information is not taken into account (see also the degree distribution in
Fig.~\ref{fig:distrib_surrogate}), leading to larger values of $R_0$ and $\Omega$. As a result, the range of parameters in which $R_0$ is underestimated is slightly smaller. Still, on the other hand, both $R_0$ and $\Omega$ can be strongly overestimated in some parameter regions and in particular close to the epidemic threshold.

We furthermore examine---see the Supplementary Material (Figures \ref{fig:timelines_R0} and \ref{fig:timelines_omega})---the effect of different timelines reconstruction methods, at a fixed procedure for choosing the surrogate ties. As could be expected, better results are obtained when more statistical information about the actual data timelines is used. In particular, using Poisson timelines leads to stronger overestimations. On the other hand, using timelines with random contact and inter-contact durations reflecting the original data statistics leads to smaller deviations, and using these statistics to create surrogate timelines 
even on the backbone ties does not have a strong impact.

We thus consider all the surrogate reconstruction methods that take into account the group structure of the data and
 use the BTL-Stats method for the timelines. Figure~\ref{fig:sir_reldiff_r0} shows the results for $R_0$, while the results for $\Omega$ are shown in the Supplementary Material. The main result underlined by the panels is that all methods give rather good results. The deviations with respect to the original data naturally tend to increase as $f$ decreases, but wide ranges of parameters with small variations are still observed even at $f=5\%$. We also see that recomputing the activities leads to worse underestimations than if the original activities are known. Despite being based on the simple procedure of adding random ties and not using data on the nodes' activities, the RT methods produce results of comparable quality. However,  
their costs in terms of conserved information is much higher (as we use the list of weights of non-backbone links
in the surrogate construction method).

\section{Discussion}

In this paper, we have considered how to bridge the gap between the backbone of a network and its actual use, particularly in data-driven numerical simulations of dynamic processes. In other words, how to turn network backbone extraction into the production of surrogate network data. Several backbone extraction procedures have indeed been put forward in the literature to extract a network's most important ties, which are supposed to summarize the most important information in the network. The issue of whether this summary suffices for actual data-driven uses has not been explored. 

Here, we have tackled this issue for several types of backbones of a temporal network, by proposing systematic ways to construct surrogate data from the backbone information. We have then used these surrogate data in numerical simulations of epidemic processes and investigated how well the outcomes of simulations and the measure of epidemic risk match the simulations on the original data.

We have considered a wide variety of procedures, with different amounts and types of information on the original data kept in the summary of the original data formed by the backbone and completed by additional statistical information. The threshold backbone arbitrarily selects the links with the largest weights, while the significant ties filters are more principled and retain ties that cannot be explained by a null model. In all cases, the data summaries need to be informed by the number of temporal edges in the original data set. Still, the amount of other additional data they contain can vary significantly. In particular, these summaries might include information on the network structure and retain, or not, the values of the node activities computed on the whole data set. If these values are unknown,  we have shown that it is possible to recompute approximate values by applying the ST filter null model to the backbone itself. Alternatively, it is possible to add ties at random between nodes to reach the original number of ties contained in the data, at the cost of also keeping the list of link weights of the non-backbone links. The same procedure can be used to build surrogate data from the threshold backbone. 

Most procedures yield surrogate data that allow us to obtain a reasonable approximation of the original outcome when used to simulate epidemic spreading processes. The quality of the approximation, however, depends on the surrogate's method. In particular, the information on the data's group structure turns out to play an important role, in line with other results showing its importance in diffusion processes\cite{smieszek2013lowcost,genois2015compensating}. Using realistic activity timelines of temporal edges also yields better results \cite{genois2015compensating}. Once group information and realistic timelines are included in the construction of surrogate data, all methods give good results. The largest discrepancies between the original and surrogate data outcomes are obtained at large spreading and recovery parameters. This is not surprising as these parameters correspond to fast processes. In this case, the outcome can depend on the data's details \cite{isella2011what} and temporal structures at short timescales that are not present in the surrogate data. For instance, in school data, temporal edges between classes occur in a synchronized way during the breaks \cite{gauvin2014detecting}, creating activity patterns that would need to be put by hand in the surrogate data and thus be contained in some way in the data summary.

Our results give hints on how to summarize complex data sets best so that they remain actionable. Moreover, as the construction of surrogate data is a  stochastic process, each of the procedures discussed here yields an ensemble of surrogate data with similar statistical properties. This highlights an interesting potential application of our results. Indeed, collecting data sets is an expensive task, and several data properties depend on context, making modeling of realistic temporal networks a problematic task.  Simultaneously, the availability of data with real properties is crucial to inform data-driven models of diffusion processes such as epidemics of infectious diseases. Moreover, collected data typically have a limited duration, and merely repeating the data might create undesired biases \cite{stehle2011simulation}. The various procedures we have described here make it possible to create synthetic surrogate data with properties very similar to empirical data without modeling assumptions. By tuning the backbone size, and hence the amount of surrogate data needed to be added to it, one can moreover tune the similarity between the original data and the surrogate replicas.

Our work has some limitations that also indicate the way for future work. First, we have limited our study to data describing contact between individuals. However, these data cover a broad range of contexts, have widely different temporal properties \cite{genois2018can}, and are particularly relevant for simulations of epidemic spread.  Second, we have considered only a limited number of backbone and surrogate construction methods. We sought to keep the methods parsimonious, so one could consider other backbone extraction methods, taking, e.g., temporal variations of the activities into account \cite{nadini2020detecting}. Finally, networks could support other types of processes, such as synchronization or complex contagion,  which might also involve higher-order structures going beyond ties \cite{iacopini2019simplicial,battiston2020networks}. Correctly reproducing the outcome of such processes from a network summary might require the development of backbones of significant structures and corresponding new surrogate data construction methods.

%\clearpage
%\newpage

\section*{Data and Methods}

\subsection*{Data}

We use state-of-the-art publicly available datasets describing contacts between individuals in different settings, with high spatial and temporal resolution. 
All data were collected by the SocioPatterns collaboration, using an infrastructure based on wearable sensors that exchange radio packets, detecting close proximity ($\leq 1.5m$) of individuals wearing the devices~\cite{cattuto2010dynamics}, with 
temporal resolution of $20$ seconds. The data can be downloaded from the SocioPatterns website www.sociopatterns.org.
\begin{itemize}
\item The InVS15 data set contains the temporal network of contacts between individuals recorded in office buildings in France in 2015. The population is divided into twelve departments
of varying sizes, but individuals of different departments
can mix during the day with no time constraints \cite{genois2018can}.

\item The LyonSchool data set contains the contact events between 242 individuals (232 children and 10 teachers) in a primary school in Lyon, France, during two days in October 2009 \cite{stehle2011high}. The children are divided into 
ten classes of similar sizes (two classes per grade) 
and follow strict schedules, with mixing between classes limited to the breaks \cite{stehle2011high}.

\item The Thiers13 data set gives the interactions between 327 students of nine classes of similar sizes within a high school in Marseille, during five days in December 2013 \cite{fournet2015contact}.

\item The SFHH conference data set describes the face-to-face interactions of 405 participants to the 2009 SFHH conference in Nice, France (June 4--5, 2009) \cite{stehle2011simulation,genois2018can}. No metadata on the participants was collected and the resulting contact network does not show any group structure \cite{stehle2011simulation}.

\end{itemize}

\subsection*{Significant ties backbones} 

For completeness, we recall here the procedure   to extract the significant ties at a given significance level $\alpha$ from a temporal network \cite{kobayashi2019structural}. 

We first define a temporal fitness model in which each node $i$ has an activity level $a_i$, and  
the probability $u$ that nodes $i$ and $j$ interact 
 during any given time interval is  given by the product of their activity levels, $u(a_{i},a_{j}) = a_{i} a_{j}$. 
 
Given a data set of $N$ nodes and temporal length $T$, we estimate the node activity levels 
$\vect{a}\equiv (a_1^*,\ldots , a_N^*)$
within the temporal fitness model from the $N$ maximum likelihood equations
$$\sum_{j:j\neq i}\frac{m_{ij}^{\rm o}- T a_{i}^{*}a_{j}^{*}}{1-a_{i}^{*}a_{j}^{*}} = 0, \; \forall \: i = 1,\ldots, N,
$$
that can be solved by standard numerical algorithms.
We then compute for each pair of nodes $i$ and $j$ the probability distribution of their total number of interactions $m_{ij}$ in the null model, which is simply given by the  binomial distribution $g(m_{ij}|a_{i}^*,a_{j}^*) 
   = \begin{pmatrix} T \\ m_{ij}\end{pmatrix} u(a_{i}^*,a_{j}^*)^{m_{ij}} (1-u(a_i^*,a_j^*))^{T-m_{ij}} $.
  
Let $m_{ij}^{c}$ denote the $c$-th percentile $(0 \leq c \leq 100)$ of $g(m_{ij}|a_{i}^*,a_{j}^*)$.
If the actual empirical number of interactions 
$m_{ij}^{\rm o}$ between $i$ and $j$ is larger than $m_{ij}^{c}$, it means that this empirical number
cannot be explained by the null model at significance level 
$\alpha \equiv 1-c/100$: in other words, $i$ and $j$ are connected by a significant tie at significant level $\alpha$. 

For a given value of $\alpha$, the set of significant ties and the corresponding temporal edges form the ST backbone of the network. As $\alpha$ decreases, the number of significant ties obviously decreases, and one can tune $\alpha$ in order to obtain a backbone formed by a given fraction $f$ of ties.
Note that, as the significant ties tend to have large number of interactions, the relative sizes of backbones 
in terms of number of temporal edges are higher than in terms of number of ties (see Table \ref{tab:backbones}).

\begin{table}[h!]
 \begin{tabular}{|c|c|c|c||c|c|c||c|c|c|}
\hline
\bf Data set & \multicolumn{3}{|c||}{\bf Threshold} & 
    \multicolumn{3}{c||}{\bf ST} & 
    \multicolumn{3}{c|}{\bf GST} \\
\hline
$f$ & 40\% & 10\% & 5\% & 40\% & 10\% & 5\% & 40\% & 10\% & 5\% \\
\hline
InVS15     &  25,417  & 16,541 & 12,014 & 24,738 & 16,166 & 9,084 & 20,890 & 13,581 & 9,522\\
LyonSchool  & 56,807 & 33,205 & 22,912 & 55,773 & 32,346 & 18,746 & 41,032 & 24,732 & 16,431\\
SFHH    & 18,802 & 12,650 & 10,253 & 17,257 & 11,950 & 9,763 & -- & -- & -- \\
Thiers13 & 53,992 & 39,171 & 30,239 & 52,975 & 30,981 & 12,678 & 43,834 & 34,613 & 22,572 \\
\hline
\end{tabular}
\caption{\label{tab:backbones}{\bf Number of 
temporal edges $E_{bT}$ of the various backbones}, for various values of the fraction $f$ of ties forming the backbone.}
\end{table}

When the nodes are divided into groups, we moreover 
consider a modified null model in which the probability
of interaction at each time between $i$ and $j$ is
$u_p(a_i,a_j) \equiv a_i a_j  (\delta_{g_i,g_j} 
+p (1 -  \delta_{g_i,g_j}) )$ 
where $g_i$ indicates the group of node $i$ and $\delta$ is the Kronecker symbol.

For a given data set, we can write the maximum likelihood equations (MLE) to estimate the 
node activity levels 
$\vect{a}\equiv (a_1^*,\ldots , a_N^*)$ and the parameter $p^*$,
similarly to the procedure of \citen{kobayashi2019structural}:
given the null model, 
the number of times temporal edges are formed between nodes $i$ and $j$ over $T$ time intervals 
is a random variable $m_{ij}$ that follows a binomial distribution with parameters $T$ and $u_p(a_{i},a_{j})$. Therefore, the joint probability function leads to
\begin{align}
p(\{m_{ij}\}|\vect{a},p) = \prod_{i,j: i\neq j}\begin{pmatrix} T \\ m_{ij}\end{pmatrix} u_p(a_{i},a_{j})^{m_{ij}} (1-u_p(a_{i},a_{j}))^{T-m_{ij}},
\end{align} 
and the $N+1$ MLE equations are
$$
\sum_{j:j\neq i}
\frac{m_{ij}^{\rm o}- T a_{i}^{*}a_{j}^{*} 
 (\delta_{g_i,g_j} +p^* (1 -  \delta_{g_i,g_j}) ) }
{1-  (\delta_{g_i,g_j} 
+p^* (1 -  \delta_{g_i,g_j}) ) a_{i}^{*}a_{j}^{*}} = 0, 
\;  \forall \: i = 1,\ldots, N,
$$
and
$$
\sum _{i,j: g_i \ne g_j}
\frac{m_{ij}^{\rm o}- T p^* a_{i}^{*}a_{j}^{*} }
{1- p^* a_{i}^{*}a_{j}^{*}} = 0 .
$$

The (G)ST filter can be applied to the original data set but also to the backbone itself. In Table \ref{tab:act_correl} we give the correlation coefficients between the activities obtained by solving the MLE equations for a data set and for its extracted backbone representing a fraction $f$ of ties.

\begin{table}[h!]
\begin{tabular}{|c|c|c|c||c|c|c||c|c|c|}
\hline
\bf Data set & \multicolumn{3}{c||}{\bf ST} & 
\multicolumn{3}{c|}{\bf GST} \\
\hline
$f$ & 40\% & 10\% & 5\% & 40\% & 10\% & 5\% \\
\hline
InVS15     &   0.99 & 0.92 & 0.62 & 0.97 & 0.87 & 0.75\\
LyonSchool  & 0.99 & 0.90 & 0.60 & 0.96 & 0.80 & 0.64\\
SFHH    &  0.97 & 0.93 & 0.90 & -- & -- & -- \\
Thiers13 & 0.99 & 0.73 & 0.25 & 0.97 & 0.92 & 0.68 \\
\hline
  \end{tabular}
\caption{\label{tab:act_correl}{\bf Correlation between
original activities and activities recomputed using the backbone ties.}
}
\end{table}

\subsection*{Surrogate data}

As described in the main text, we have put forward several methods to build surrogate data starting from a backbone. These methods consist of two steps, first choosing the surrogate ties and then creating timelines of  temporal edges on each tie. 

In Table~\ref{tab:surrogate_summary}, we summarize each method's main points for each type of backbone, the data needed in addition to the backbone information, and the size of these additional data. Note that the random links methods need several inputs of the order of the number of ties in the original data, while the methods based on the null model instead use an input scaling with the number of nodes. The method needing the least extra data is the one recomputing the activities applying the ST filter methodology on the backbone data itself. 

In the methods based on the (G)ST null models, we need to calibrate the parameter $\alpha$ (or the two parameters $\alpha_{\rm intra}$ and $\alpha_{\rm inter}$). To this aim, we first try at each timestamp to add a temporal edge with probability $a_i a_j$ for each $(i,j)$ not in the backbone. This creates a total number of temporal edges $E_T'$. The actual number of surrogate temporal edges needed is actually $E_T - E_{bT}$---i.e., the difference between the number of temporal edges in the data and in the backbone. Therefore, we set $\alpha = (E_T - E_{bT})/E_T'$ and we use as probabilities of creation of  temporal edges $\alpha a_i a_j$.  When the data group structure is taken into account, the procedure is performed separately for intra- and inter-group ties. We note that the final number of temporal edges in the surrogate data is not strictly fixed by this procedure but remains a stochastic outcome. The number of ties is not fixed either but is also an outcome of the procedure, contrary to the procedures based on adding random links.

\begin{table}[h!]
\begin{tabular}{|p{0.09\textwidth}|p{0.09\textwidth}|p{0.32\textwidth}|p{0.32\textwidth}|p{0.1\textwidth}|}
\hline
Backbone type & Surrogate type & Method summary & Extra data needed & Size of extra data needed\\
\hline
ST & ST-OA &  
For each $(i,j)$ not in backbone, at each timestamp
add a temporal edge with probability 
$\alpha a_i a_j$, with $\alpha$ scaled to adjust the 
total number of temporal edges
&	List of original node activities; 
Number of temporal edges & $N+1$ \\  \cline{2-5}
& ST-RA &  
Compute the activity $\tilde{a_i}$ of each node 
 with the ST backbone
method applied on the backbone itself; add surrogate 
ties as for the ST-OA method, using the recomputed activities
& Number of temporal edges & $1$  \\ \cline{2-5}
& ST-RT&  Add ties at random in order to reach the 
number of ties of the original data, with weights 
extracted at random from the list of weights of the
non-backbone ties
&  Number of ties;
List of weights of ties not in backbone  & $E-E_b+1$  \\ \hline
TB & TB-RT & Same as for ST-RL
& Number of ties;
List of weights of ties not in backbone  & $E-E_b+1$ \\
\hline \hline
GST & GST-OA & For each $(i,j)$ not in backbone, at each timestamp add a temporal edge with probability 
$\alpha_{\rm intra} a_i a_j$ if $i$ and $j$ are in the same group, $\alpha_{\rm inter} p a_i a_j$ else, with $\alpha_{{\rm intra}/{\rm inter}}$ scaled to adjust the 
total number of temporal edges within and between groups
& 
List of original node activities and parameter p;
Group membership;
Number of temporal edges within groups
and between groups
& $2N+3$ \\  \cline{2-5}
& GST-RA&  
Compute the activity $\tilde{a_i}$ of each node and the parameter $\tilde{p}$ with the GST backbone
method applied on the backbone itself; add surrogate 
ties as for the GST-OA method, using the recomputed activities.
& Group membership;
Number of temporal edges within groups
and between groups & N+2 \\  \cline{2-5}
& GST-RT&  Add ties at random in order to reach the 
same number of ties within and between groups
as in the original data, with weights 
extracted at random from the list of weights of the
non-backbone ties & Group membership;
Number of ties between and within groups; 
List of weights of non-backbone ties, between and within groups &  $N+E-E_b+2$\\ \hline
TB & GTB-RT & 
Same as for GST-RT
& Group membership; 
Number of ties between and within groups; 
List of weights of non-backbone 
ties, between and within groups & $N+E-E_b+2$ \\\hline
  \end{tabular}
\caption{\label{tab:surrogate_summary} 
Summary of the various methods to choose 
the surrogate ties.
}
\end{table}

Finally, to construct surrogate timelines respecting the 
data statistics of event and interevent durations, 
we proceed as follows, for each tie $(i,j)$ with
number of temporal edges $n_{ij}$:
\begin{enumerate}
\item we extract a random initial time $t_0$ in $[0,T]$; all the times are then considered modulo $T$; we set $n=n_{ij}$;
\item we iterate the procedure
\begin{enumerate}
\item extract a random duration $\tau$ from the fitted distribution of event durations
\item check that $\tau \le n$, else replace $\tau$ by $n$    
\item add $\tau$ temporal edges between $i$ and $j$, namely on the interval $[t_0,t_0 + \tau -1]$
\item extract a random interevent time $\Delta t$ 
from the fitted distribution of interevent durations    
\item replace $t_0$ by $t_0+\tau+\Delta t$ and $n$ by $n-\tau$
\end{enumerate}
until $n=0$, i.e., until $n_{ij}$ temporal edges have been created.
\end{enumerate}

\subsection*{Simulations of the epidemic spread}
 
For the simulation of the SIR model on temporal networks we use the approach and code presented in Ref.~\citen{holme2020tsir}. We start the simulation with all nodes susceptible and introduce the disease at a random node at a random time (uniformly chosen between the beginning and end of the temporal network). Then if there is an event between a susceptible and an infectious, a contagion occurs with a probability $\beta$. The infected person recovers with a rate $\nu$, i.e., the time to recovery is a random variable $\delta$ extracted from the distribution $\nu \exp(-\nu \delta)$\cite{holme2020tsir}. Finally, we assume that an individual that gets infected at time $t'$ cannot infect anyone else until $t>t'$. For every pair of parameter values $(\beta,\nu)$, we run this algorithm $10^7$ times for averages.

We calculate the basic reproductive number $R_0$ directly from the simulations as the average numbers of individuals infected directly by the source. Calculating the average outbreak size $\Omega$ is a similarly straightforward average over the number of nodes in the recovered state when the outbreak is extinct. If the outbreak is not extinct when the simulation reaches the end of the data set, the outbreak size is the number of nodes in either the infectious of the recovered state at the last time stamp of the data.
 
\clearpage
\newpage

\bibliographystyle{abbrv}
\bibliography{biblio}

\section*{Acknowledgements}

A.B. was supported by the ANR project DATAREDUX (ANR-19-CE46-0008) and JSPS KAKENHI Grant Number JP 20H04288. P.H. was supported by JSPS KAKENHI Grant Number JP 18H01655.

\section*{Author contributions statement}

A.B. and P.H. conceived the study.
C.P., P.H., A.B. designed and conducted the numerical experiments and analysed the results.  
All authors reviewed the manuscript. 

\section*{Competing interests}

The authors declare no competing interests.

\clearpage
\newpage

\section{Supplementary Material}

\setcounter{section}{0}
\setcounter{table}{0}
\setcounter{equation}{0}
\setcounter{figure}{0}

\renewcommand{\thetable}{S\arabic{table}}
\renewcommand{\thefigure}{S\arabic{figure}}
\renewcommand{\thesection}{Supplementary Note \arabic{section}.}

\section{Statistics}

\begin{figure}[ht!]
\includegraphics[width=\linewidth]{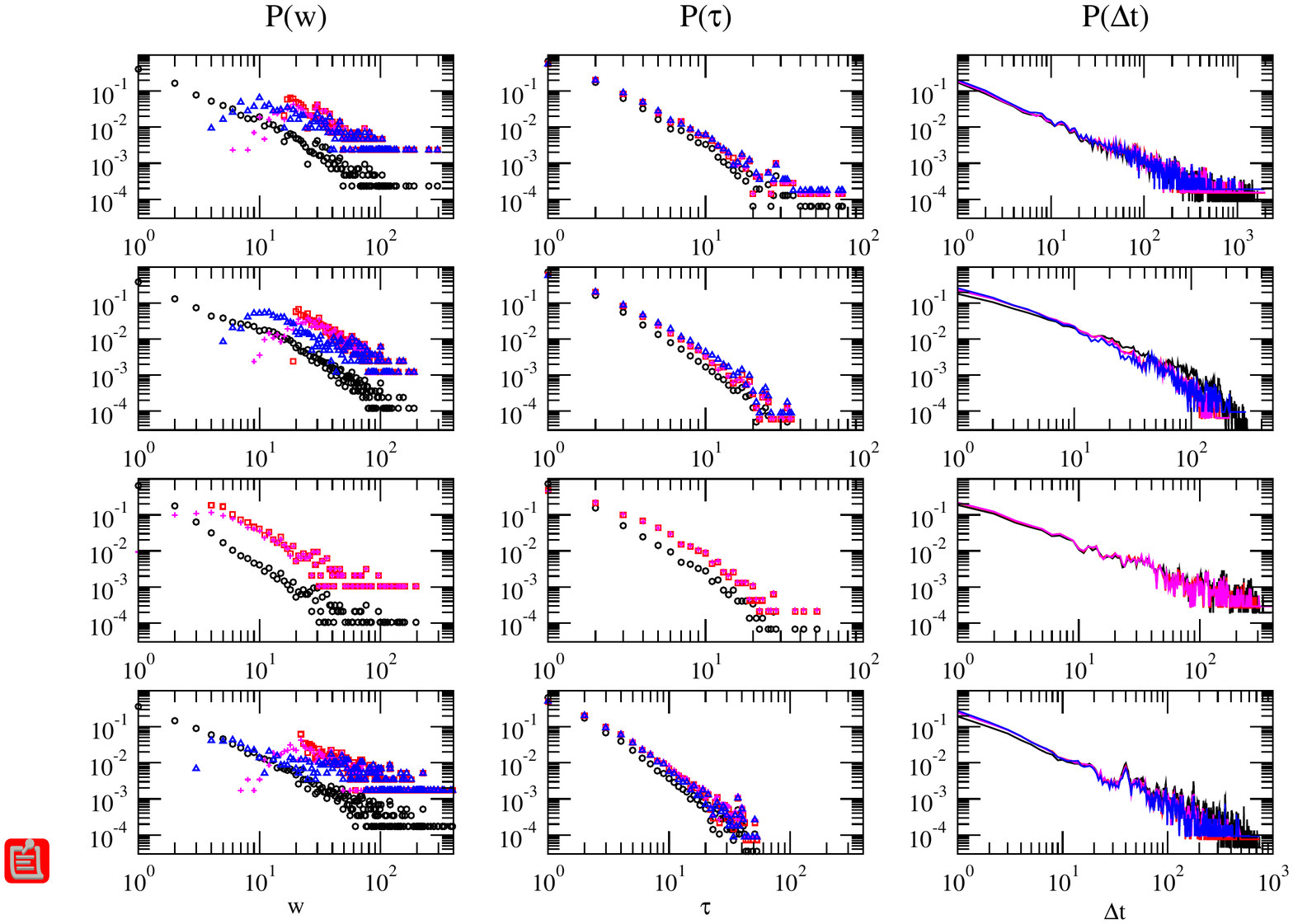}
\caption{{\bf Backbone statistics for $f=10\%$.}
Left column: distributions of weights.
Middle column: distribution of contact durations.
Right column: distribution of intercontact durations. 
The black circles (resp. lines for the right column) correspond 
to the distributions for the original data sets.
Red squares (resp. lines) are for the threshold backbone,
magenta crosses (resp. lines) for the ST filter and
blue triangles (resp. lines) for the GST filter.
From top to bottom:
InVS15, LyonSchool, SFHH, Thiers13.
}
\label{fig:distrib_backbones}
\end{figure}

\begin{table}[htb] 
\hspace{-1cm}
  \begin{tabular}{|c|c||c|c|c||c|c|c|c||c|c|c|c|}
    \hline
    Data set &  $f$ & ST & GST & TB & ST-OA & ST-RA & ST-RT & TB-RT & GST-OA & GST-RA & GST-RT & GTB-RT \\
    \hline
    InVS15  & 0.4 & 0.84 & 0.48 & 0.98 & 0.82 & 0.82 & 0.55 & 0.58 & 0.99 & 0.99 & 0.66 & 0.65 \\
    & 0.1 & 0.72 & 0.31 & 0.64 & 1.37 & 1.47 & 0.48 & 0.60 & 1.15 & 1.22 & 0.48 & 0.60 \\
    & 0.05 & 0.36 & 0.16 & 0.39 & 1.64 & 1.83 & 0.48 & 0.49 & 1.18 & 1.40 & 0.60 & 0.59 \\
    \hline
    LyonSchool & 0.4 & 1.23&   0.50 & 1.22 & 0.82 & 0.82 & 0.62 & 0.62& 0.77 & 0.77 & 0.69 & 0.65 \\
    & 0.1 & 0.71 & 0.54 & 0.79 & 1.32 & 1.34 & 0.55 & 0.54 &    0.87 & 0.92 & 0.63 & 0.62 \\
    & 0.05 & 0.37 & 0.35 & 0.41 &    1.48 & 1.51 & 0.54 & 0.55 &    0.96 & 1.05 & 0.62 & 0.62 \\ \hline
Thiers13 & 0.4 &  0.89 & 0.40 & 0.97 & 0.49 & 0.49 & 0.32 & 0.33 & 1.11 & 1.10 & 0.85 & 0.74 \\
  & 0.1 & 0.50 & 0.41 & 0.57 &  1.04 & 1.16 & 0.22 & 0.22 &  1.01 & 1.01 & 0.72 & 0.71 \\ 
 & 0.05 &  0.24 & 0.23 & 0.29 & 1.27 & 1.47 & 0.22 & 0.22 & 0.99 &  1.03 & 0.72 &  0.71 \\ \hline
 SFHH & 0.4  &  0.52 & &  0.90 & 0.76 & 0.74 & 0.46 & 0.52 & & & & \\
 & 0.1& 1.04& & 1.05&  1.05& 1.27&0.43& 0.43 & & & & \\
 & 0.05 & 1.11&  &1.04& 1.16& 1.52& 0.42& 0.43& & & & \\
\hline
  \end{tabular} 
  \caption{ \label{tab:cc} {\bf Clustering coefficients
  in backbones and surrogates, normalized by the value
  of the clustering coefficient in the original data.}
   }
\end{table}

\begin{table}[htb] 
\hspace{-1cm}
  \begin{tabular}{|c|c||c|c|c||c|c|c|c||c|c|c|c|}
    \hline
    Data set &  $f$ & ST & GST & TB & ST-OA & ST-RA & ST-RT & TB-RT & GST-OA & GST-RA & GST-RT & GTB-RT \\
    \hline
    InVS15  & 0.4 & 0.89 & 0.25 & 0.85 & 0.94 & 0.94 & 0.95 & 0.96 & 0.99 & 0.99 & 0.96 & 1.02 \\
    & 0.1 &  1.06 & 0.50 & 1.13 & 0.69 & 0.69 &0.69 & 0.70 &    0.99 & 0.99 & 0.93 & 1.05 \\
    & 0.05 & 1.03 & 0.53& 1.22 &0.38 & 0.35 & 0.41 & 0.54 & 0.99 & 0.99 & 0.91 & 1.05 \\
\hline
    LyonSchool & 0.4 &  1.03&0.22& 0.93 &    0.99 & 0.98 & 0.99 & 0.98 &     1.00 & 1.01 &  1.01 & 1.05  \\
    & 0.1 & 1.26 & 0.45 & 1.24 &    0.63 & 0.62 & 0.62 & 0.64 &     1.00 & 0.99 &  1.00 & 1.14 \\
    & 0.05 &     1.24 & 0.52 & 1.32  & 0.35 & 0.33 & 0.38 & 0.47 &      1.00 & 0.98 & 1.01& 1.12    
    \\ \hline
Thiers13 & 0.4 &  0.98& 0.32& 0.97& 0.92 &
0.92 & 0.92 & 0.94 & 1.00 & 1.00 &1.00 & 1.02\\
  & 0.1 & 1.05& 0.77& 1.06 & 0.55 & 0.54 & 0.57 & 0.69 & 1.00 & 0.99 & 1.01 & 1.03 \\ 
 & 0.05 & 1.03&  0.53&1.22 & 0.38 & 0.35 & 0.41 & 0.54 &
0.99& 0.99& 0.91 & 1.05\\
\hline
  \end{tabular} 
  \caption{ \label{tab:modularity} {\bf Modularity
  in backbones and surrogates, normalized by the value
  of the modularity in the original data.}
  Here the modularity is computed imposing as partition
  the known group structure of the data (as it represents the ground truth), rather than 
  using community detection algorithm.
   }
\end{table}

\clearpage
\newpage

\section{Original vs. backbone}

\begin{figure}[hb]
    \centering
    \includegraphics[width=.3\linewidth]{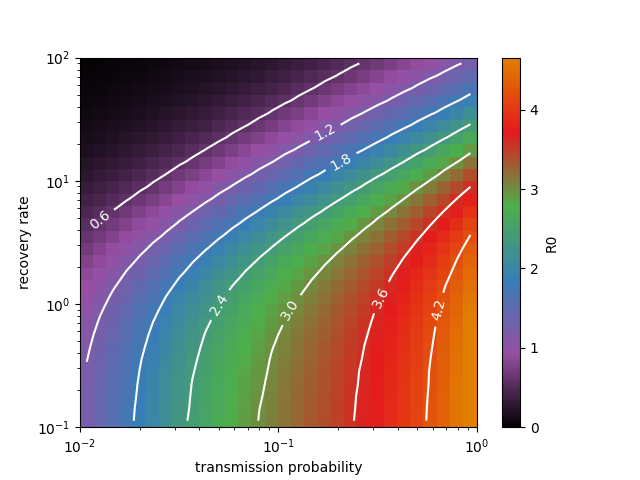}\\
        \includegraphics[width=.3\linewidth]{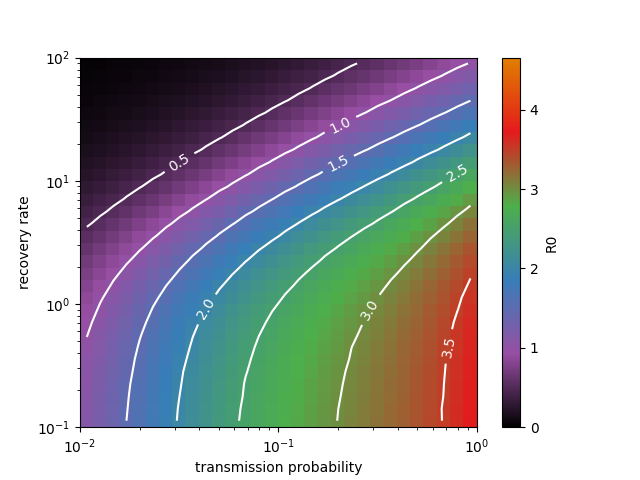}
        \includegraphics[width=.3\linewidth]{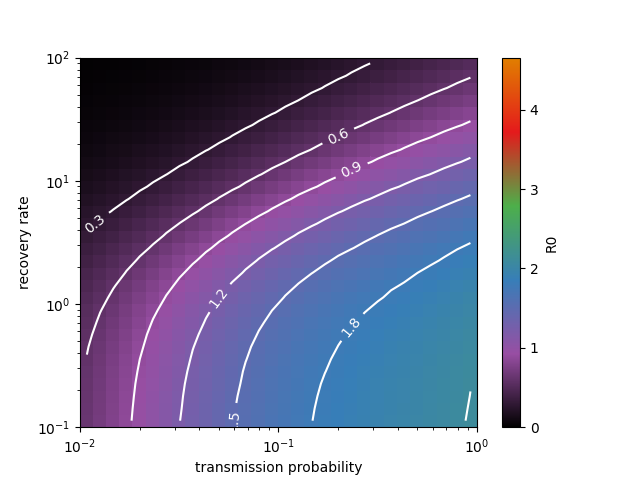}
        \includegraphics[width=.3\linewidth]{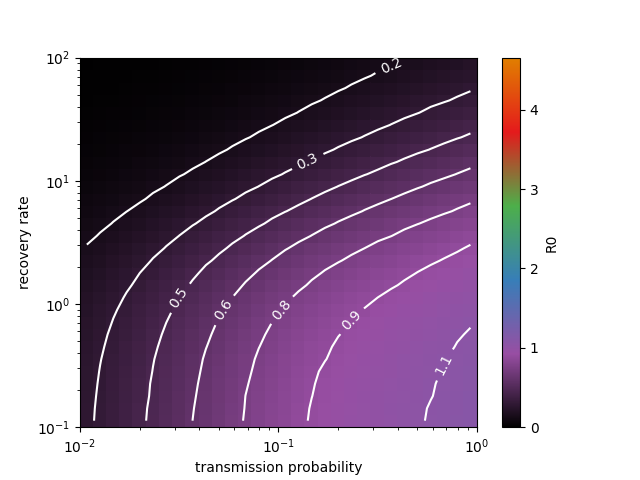}
     \includegraphics[width=.3\linewidth]{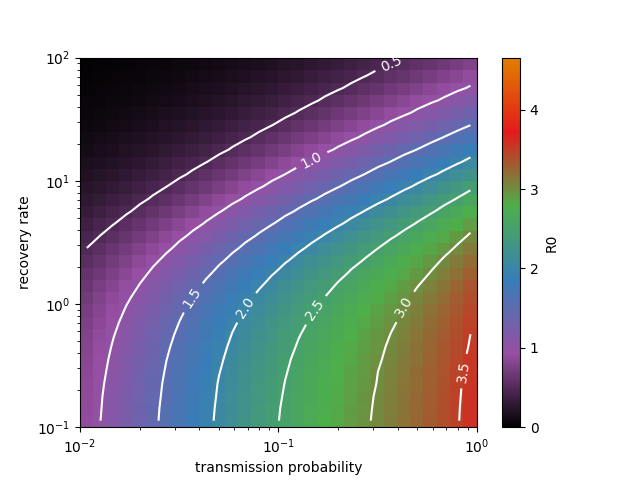}
    \includegraphics[width=.3\linewidth]{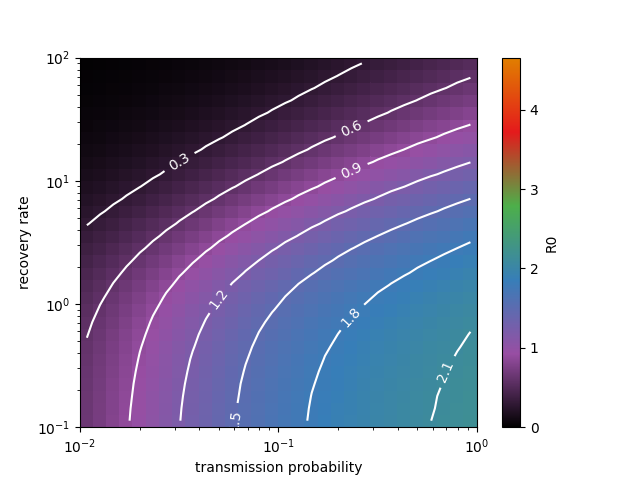}
        \includegraphics[width=.3\linewidth]{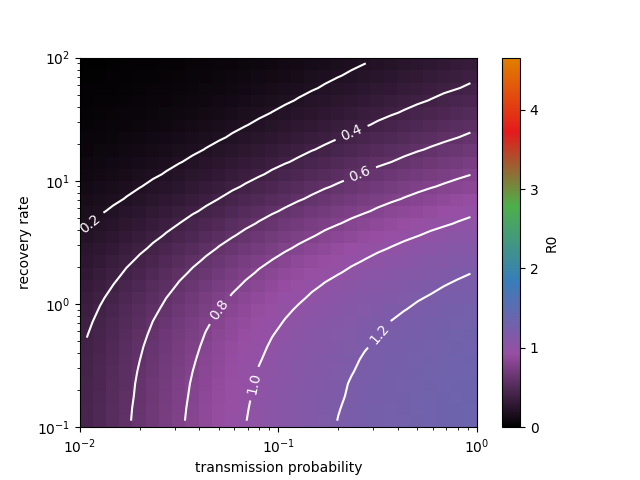} 
                \includegraphics[width=.3\linewidth]{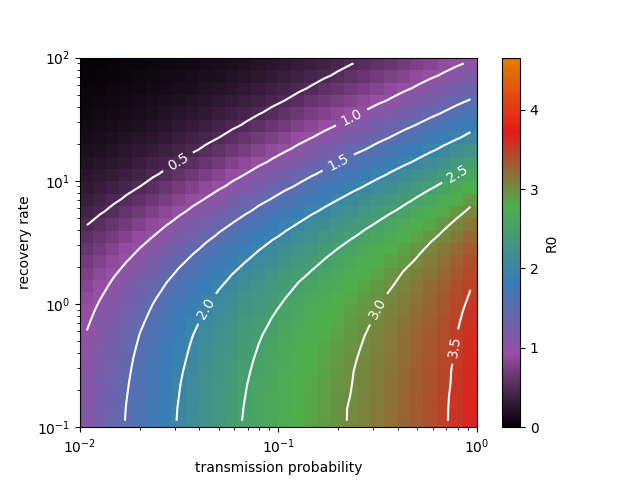}
        \includegraphics[width=.3\linewidth]{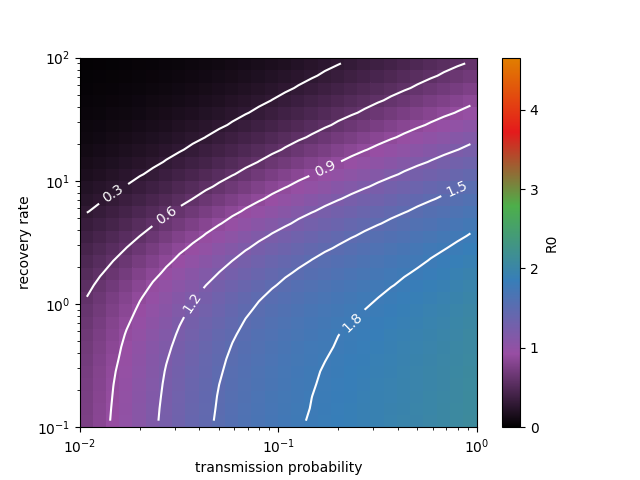}
        \includegraphics[width=.3\linewidth]{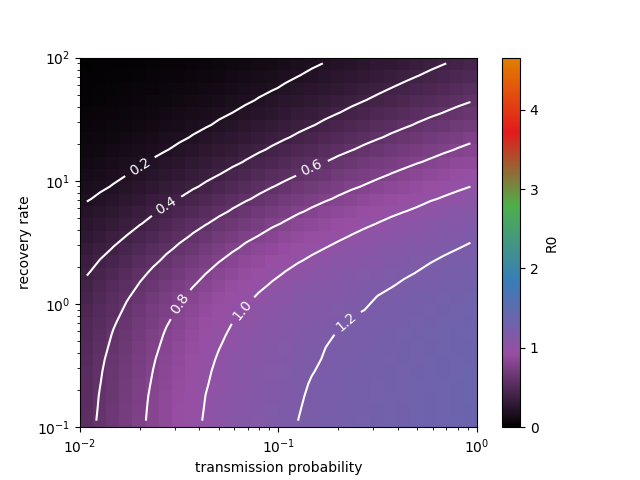}  
    \caption{{\bf Original vs. backbones.} $R_0$ values obtained from the simulations on the original data and on the backbones, for the Thiers13 data set. Top: original data.  Second row: ST backbone; third row: GST backbone; fourth row: TB.
    Left: $f= 40\%$. Middle : $f=10\%$.  right: $f=5\%$.      \label{fig:orig_back_r0_values}}
\end{figure}

\begin{figure}[thb]
    \centering
        \includegraphics[width=.3\linewidth]{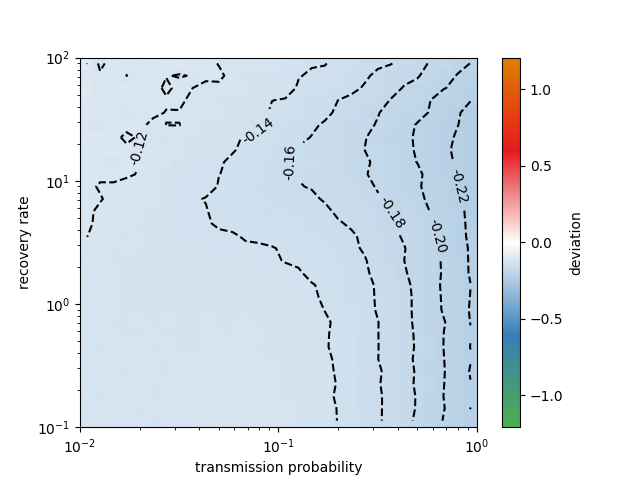}
        \includegraphics[width=.3\linewidth]{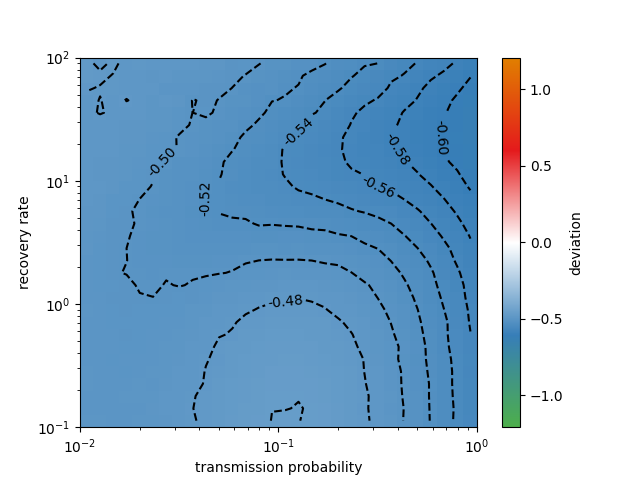}
        \includegraphics[width=.3\linewidth]{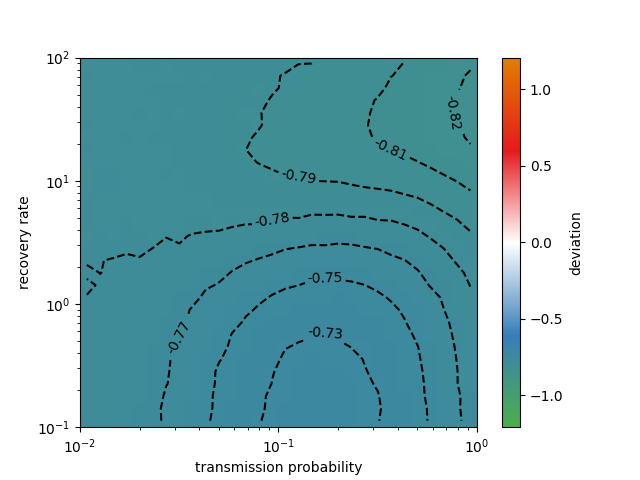}
              \includegraphics[width=.3\linewidth]{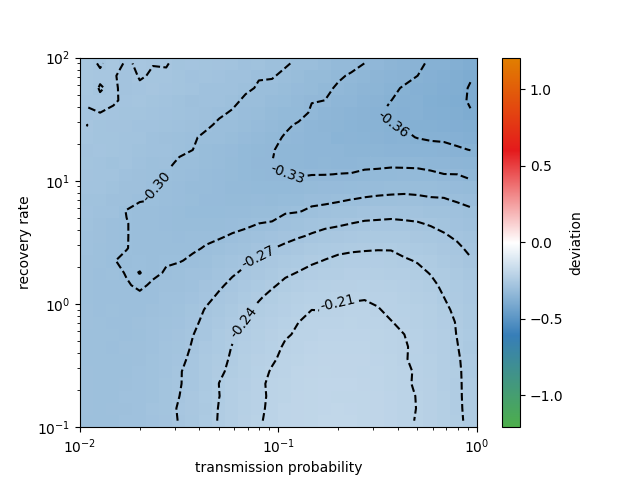}
        \includegraphics[width=.3\linewidth]{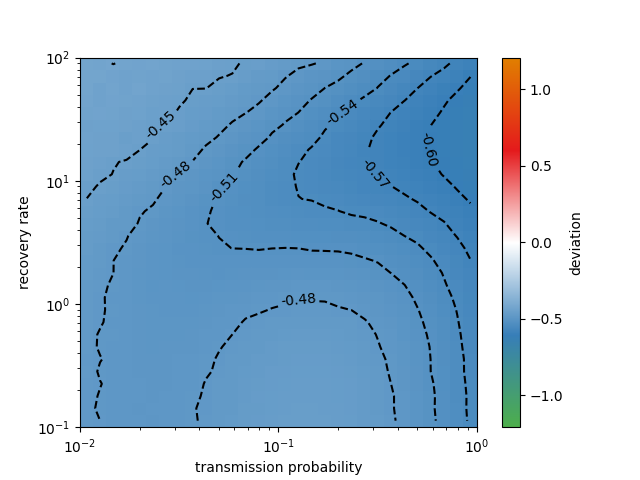}
        \includegraphics[width=.3\linewidth]{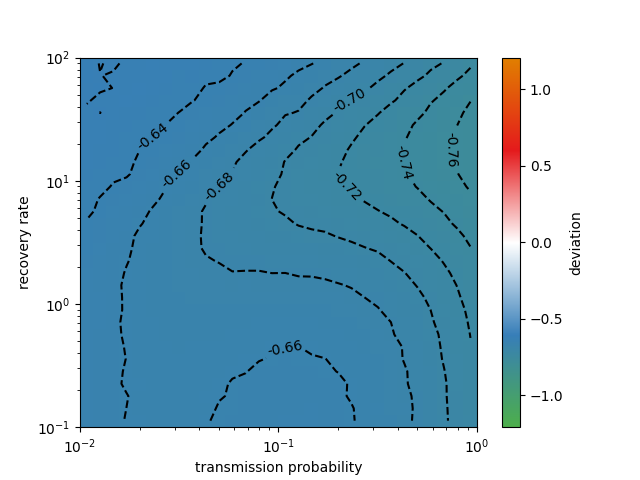} 
                \includegraphics[width=.3\linewidth]{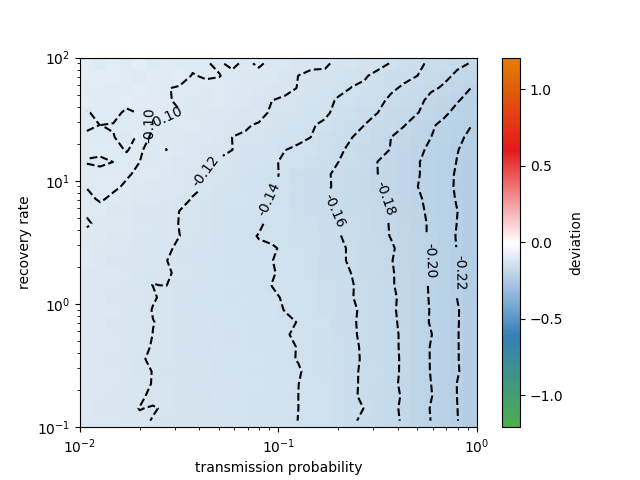}
        \includegraphics[width=.3\linewidth]{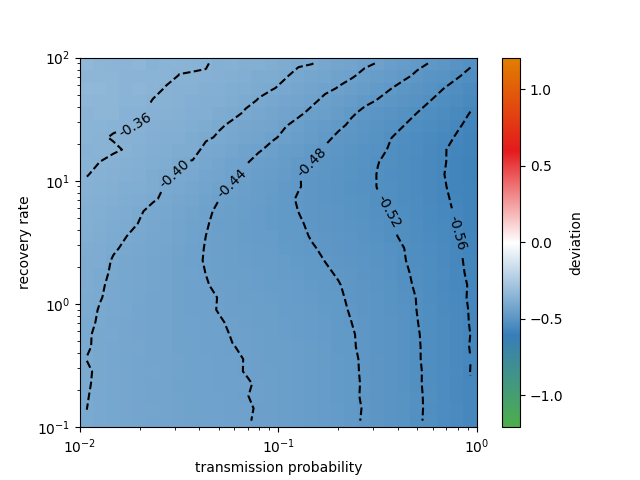}
        \includegraphics[width=.3\linewidth]{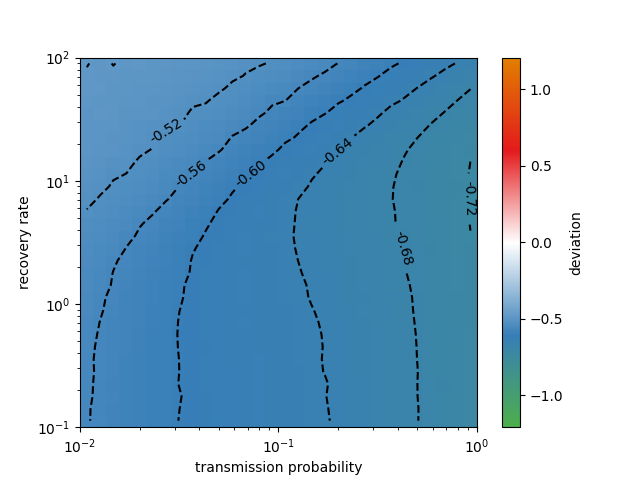}  
    \caption{{\bf Original vs. backbone.} Relative difference in $R_0$ values obtained from the simulations on the original data and on the backbones, for the Thiers13 data set.  
    First row: ST backbone; 2nd row: GST backbone; 3rd row: TB. Left: $f= 40\%$. Middle : $f=10\%$.  right: $f=5\%$.      \label{fig:orig_back_r0_reldiff}}
\end{figure}

\begin{figure}[thb]
    \centering
    \includegraphics[width=.3\linewidth]{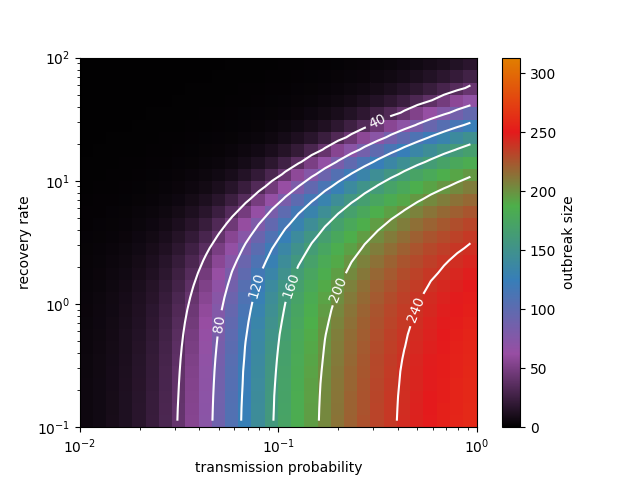}\\
        \includegraphics[width=.3\linewidth]{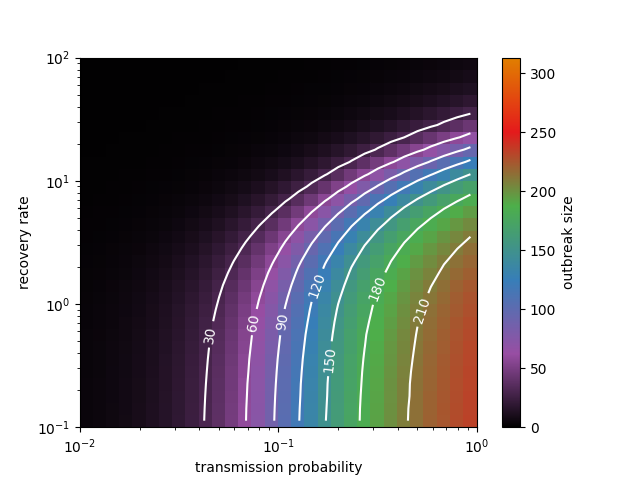}
        \includegraphics[width=.3\linewidth]{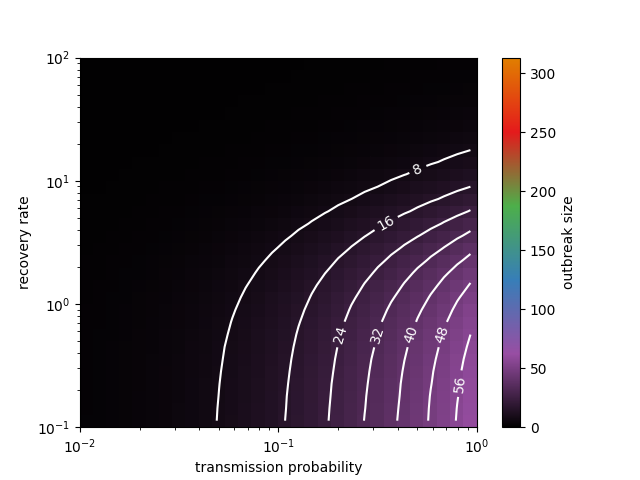}
        \includegraphics[width=.3\linewidth]{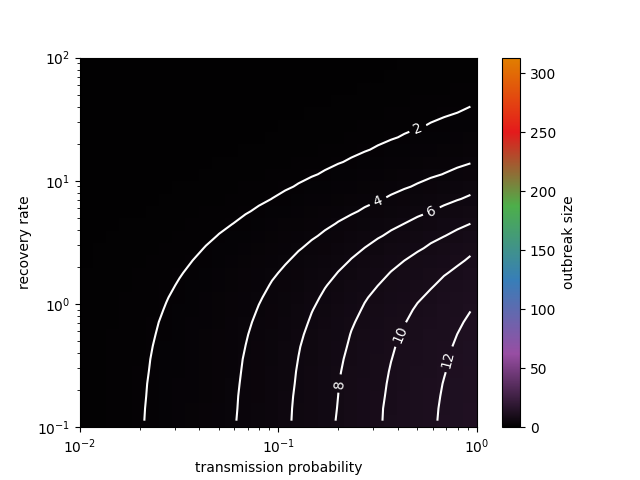}
              \includegraphics[width=.3\linewidth]{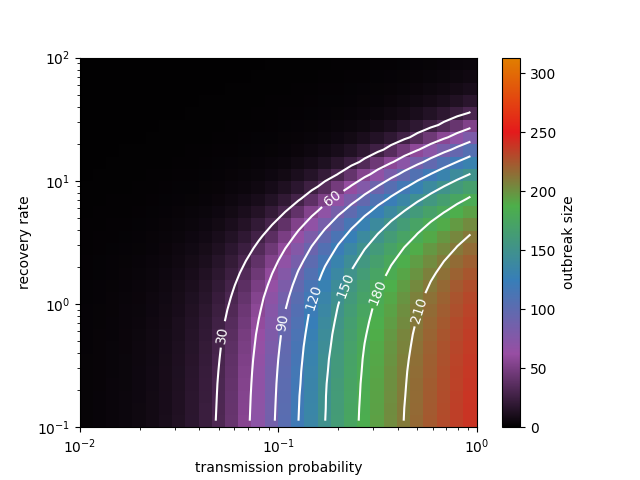}
        \includegraphics[width=.3\linewidth]{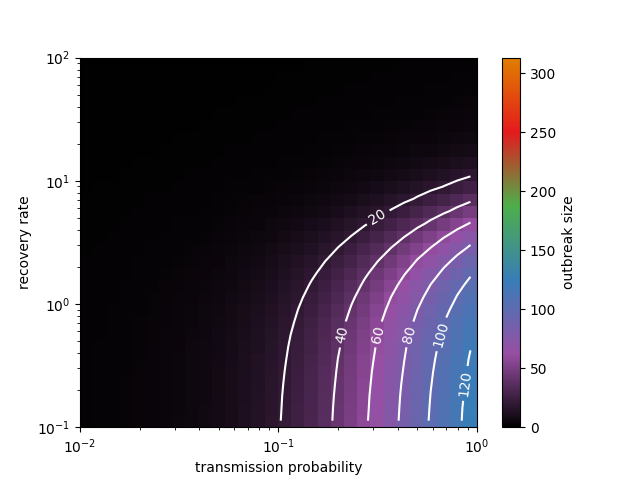}
        \includegraphics[width=.3\linewidth]{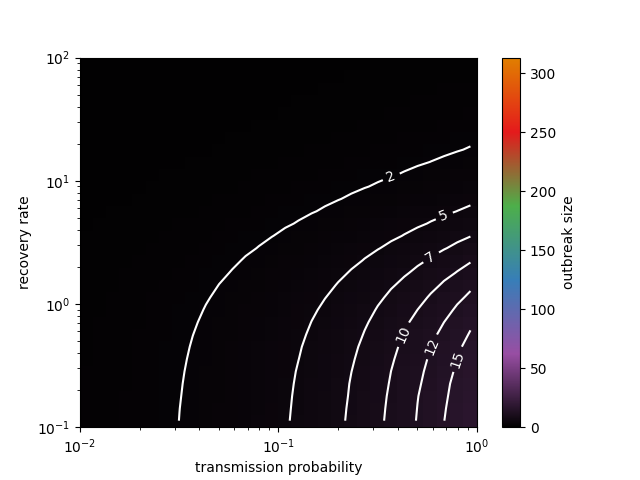}
                \includegraphics[width=.3\linewidth]{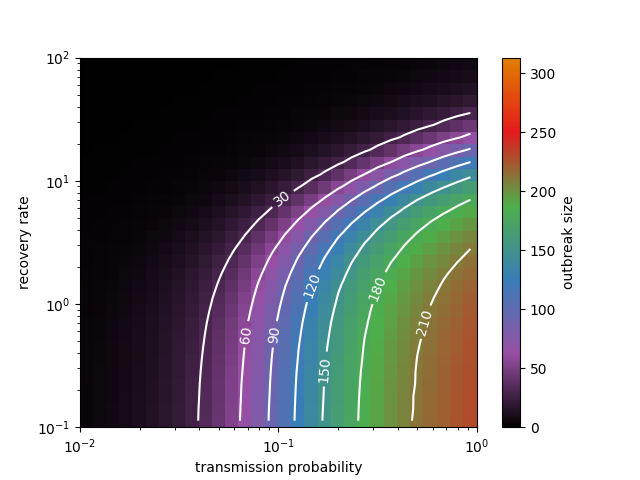}
        \includegraphics[width=.3\linewidth]{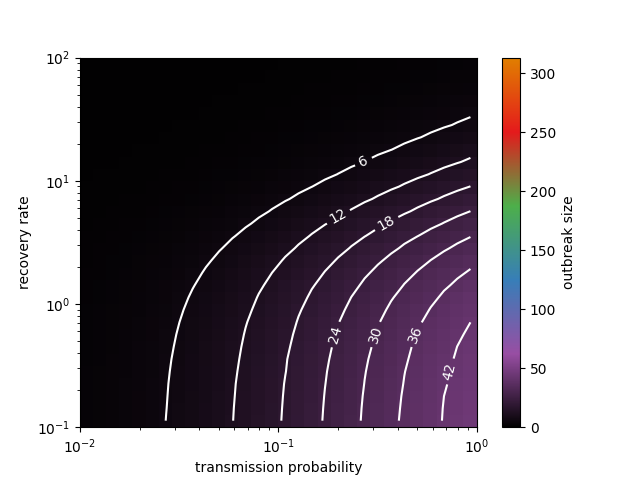}
        \includegraphics[width=.3\linewidth]{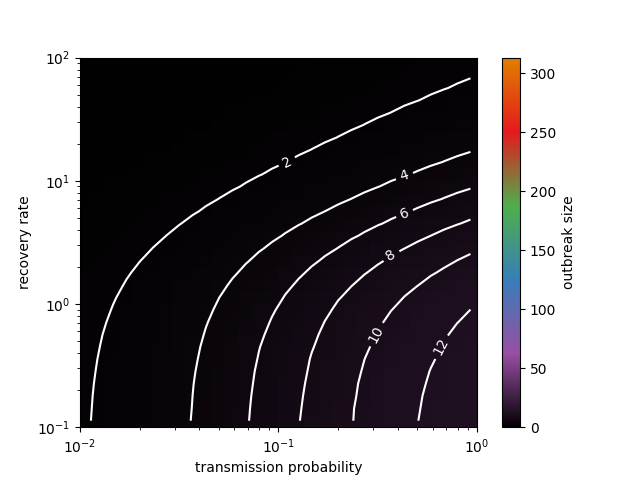}  
\caption{{\bf Original vs. backbone.} 
$\Omega$ values obtained from the simulations on the original data and on the backbones, for the Thiers13 data set.
 Top: original data.  Second row: ST backbone; third row: GST backbone; fourth row: TB. Left: $f= 40\%$. Middle : $f=10\%$.  right: $f=5\%$.      \label{fig:orig_back_omega_values}}
\end{figure}

\begin{figure}[thb]
    \centering
        \includegraphics[width=.3\linewidth]{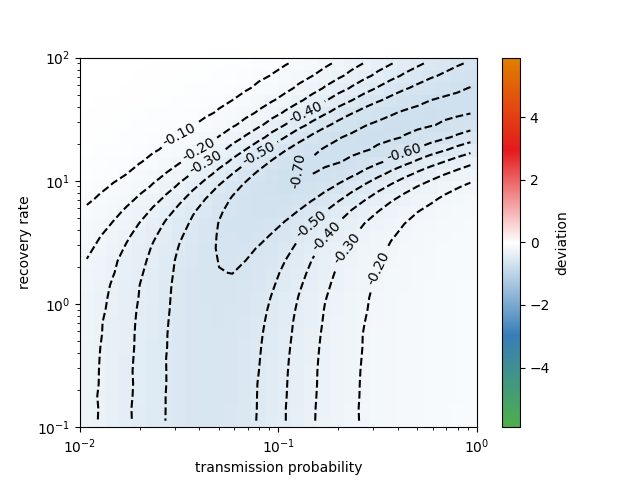}
        \includegraphics[width=.3\linewidth]{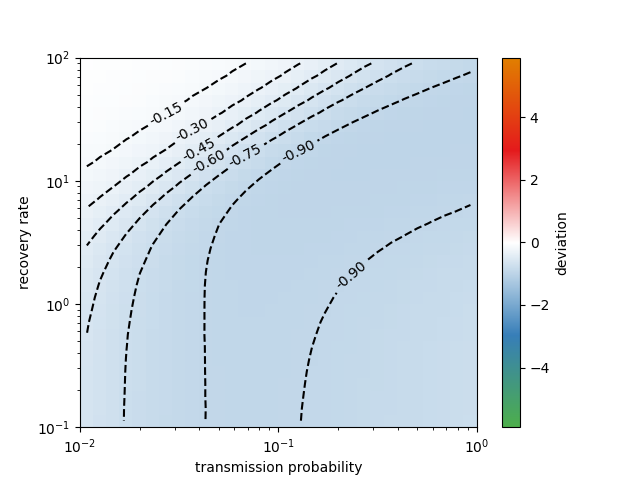}
        \includegraphics[width=.3\linewidth]{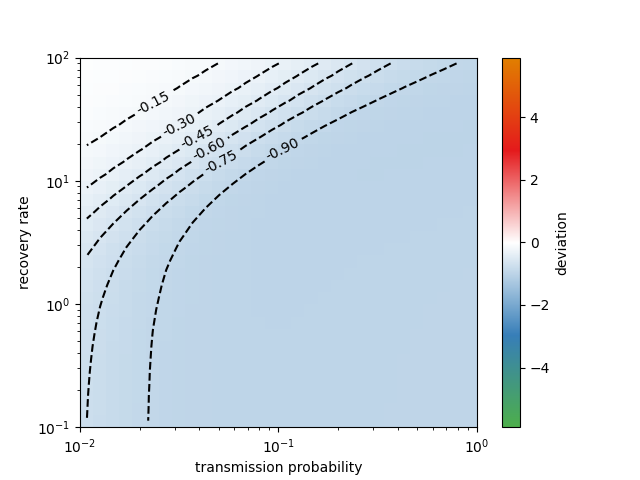}
              \includegraphics[width=.3\linewidth]{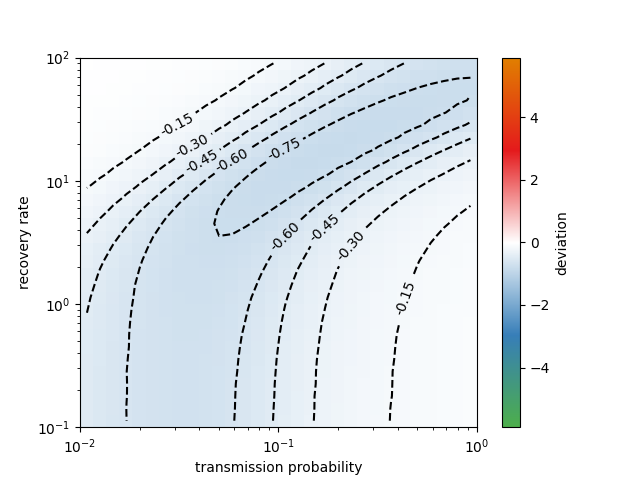}
        \includegraphics[width=.3\linewidth]{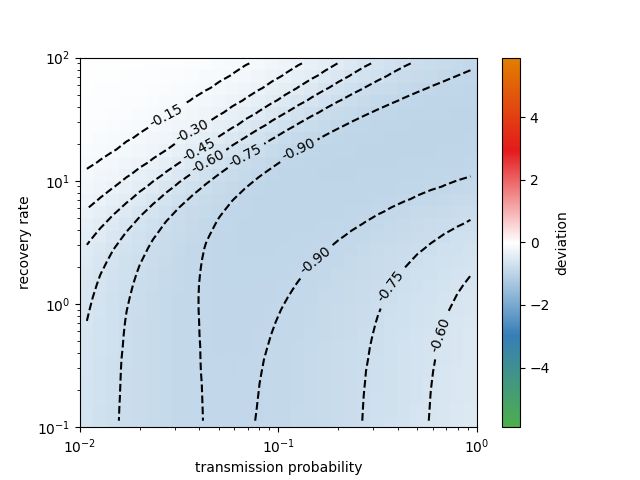}
        \includegraphics[width=.3\linewidth]{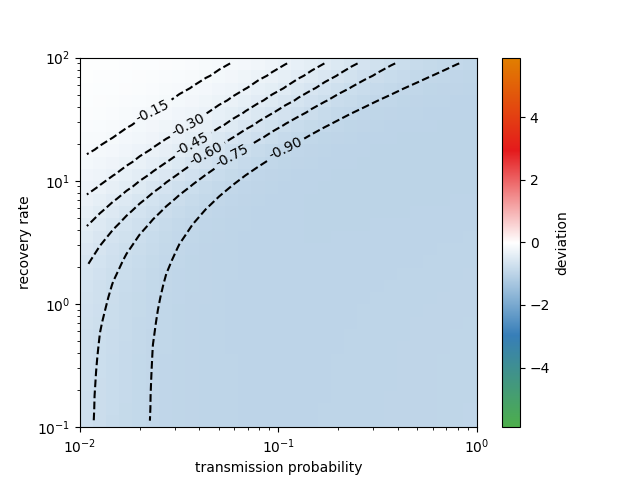} 
                \includegraphics[width=.3\linewidth]{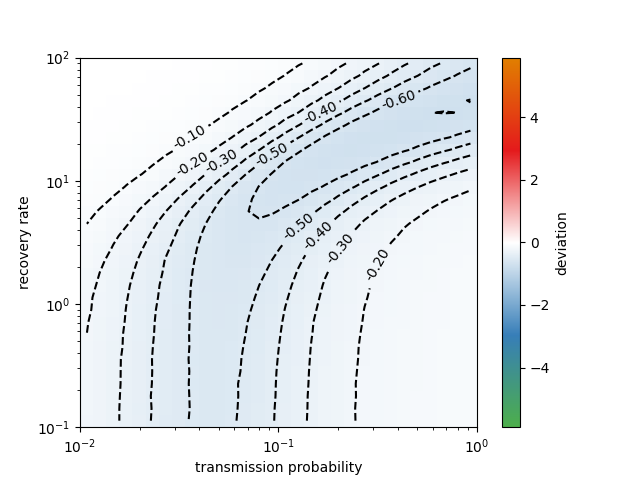}
        \includegraphics[width=.3\linewidth]{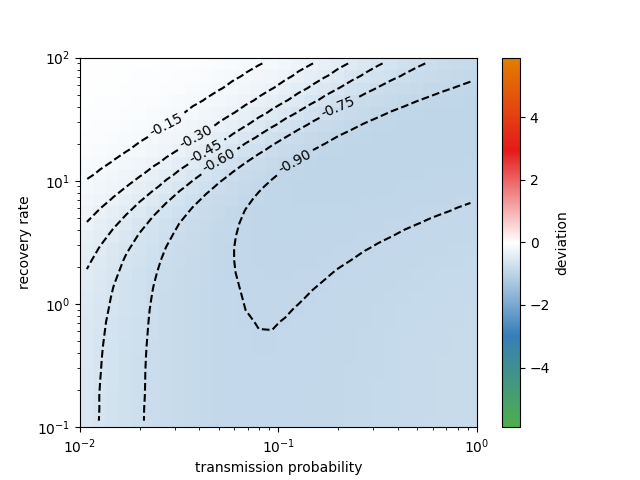}
        \includegraphics[width=.3\linewidth]{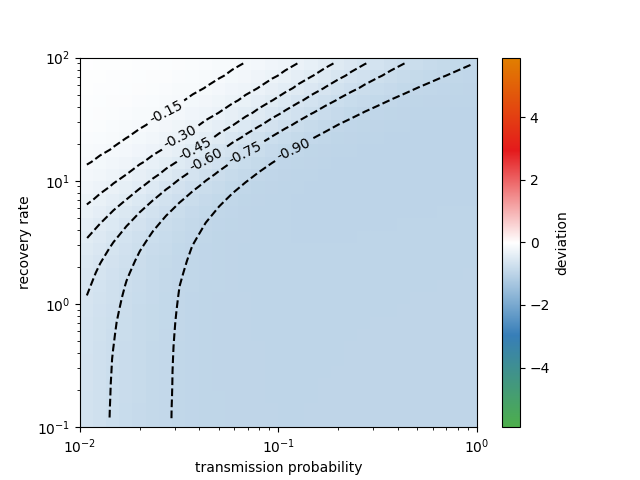}  
    \caption{{\bf Original vs. backbone.} Relative difference in $\Omega$ values obtained from the simulations on the original data and on the backbones, for the Thiers13 data set. First row: ST backbone; 2nd row: GST backbone; 3rd row: TB. Left: $f= 40\%$. Middle : $f=10\%$.  right: $f=5\%$.      
\label{fig:orig_back_omega_reldiff}}
\end{figure}

\clearpage
\newpage
\section{Using vs. not using the group structure in the backbones and surrogates}

\begin{figure}[thb] \centering
\includegraphics[width=.3\linewidth]{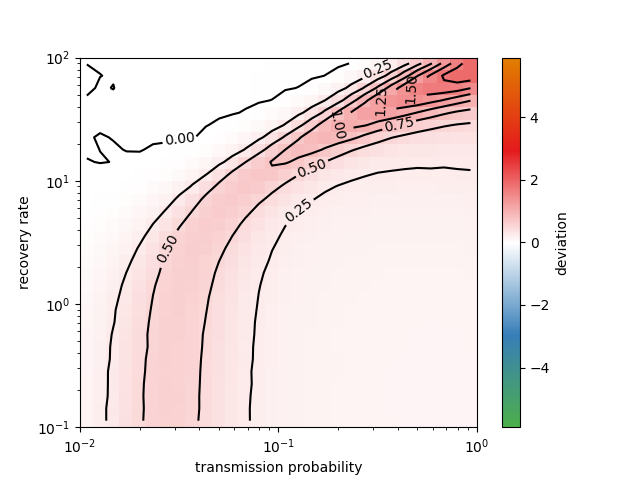}
\includegraphics[width=.3\linewidth]{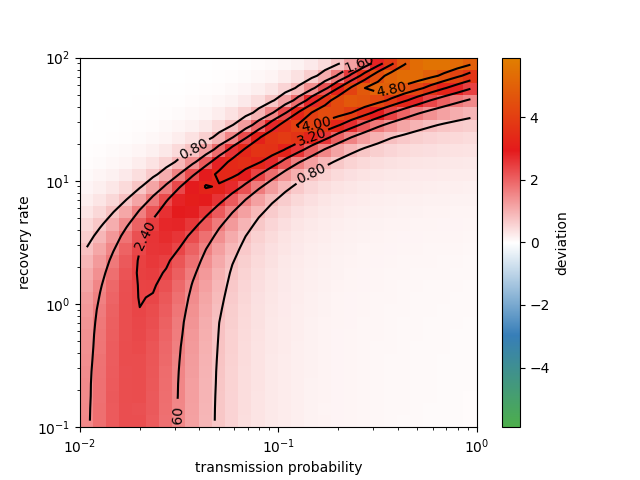}
\includegraphics[width=.3\linewidth]{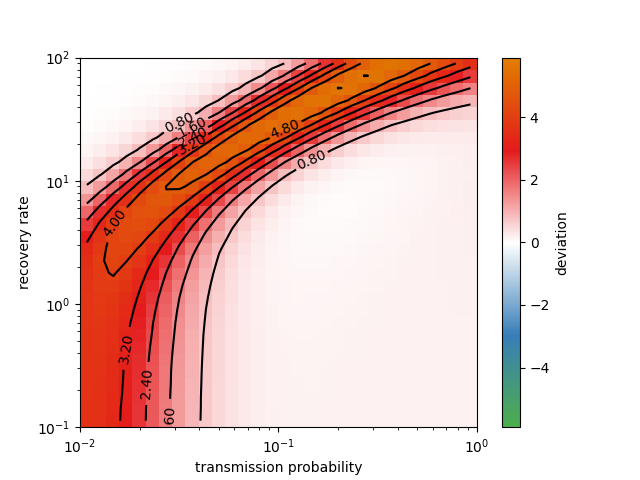}
\includegraphics[width=.3\linewidth]{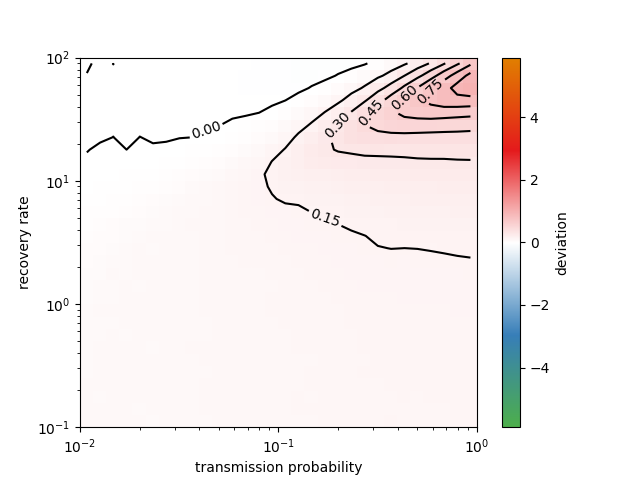}
\includegraphics[width=.3\linewidth]{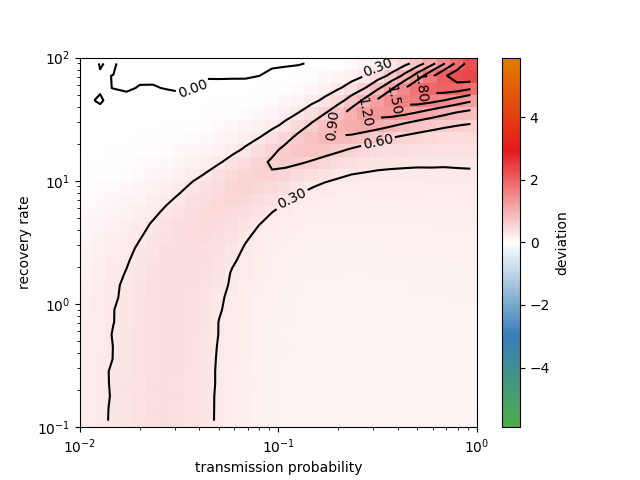}
\includegraphics[width=.3\linewidth]{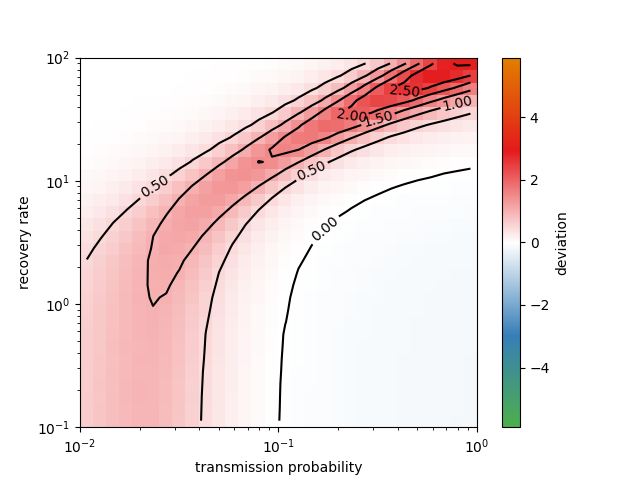}
 \caption{{\bf Surrogate, effect of taking into account group structure for the Thiers13 data set.} Relative difference in $\Omega$ obtained from the simulations on surrogate data with respect to simulations on the original data.
 Top: ST-RA. Bottom: GST-RA. Left $f=40\%$, middle $f=10\%$; right $f=5\%$. In each case the backbone timelines are kept, and timelines respecting the statistics of contact and inter-contact durations are built for the surrogate ties (BTL-Stats method).
 \label{fig:group_reldiff_omega}}
\end{figure}

\begin{figure}[thb]
    \centering
    \includegraphics[width=.3\linewidth]{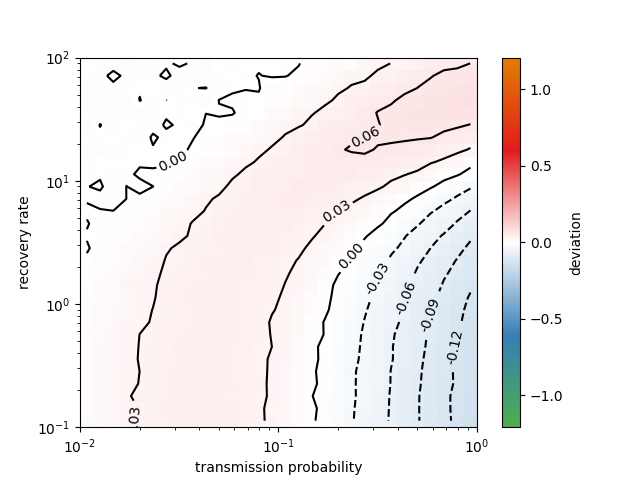}
\includegraphics[width=.3\linewidth]{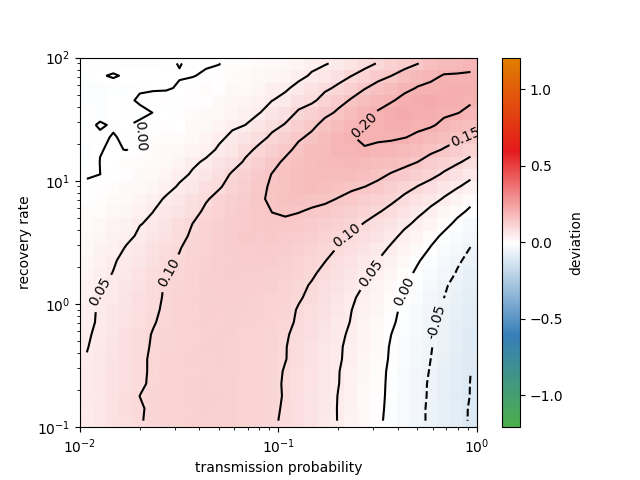}
\includegraphics[width=.3\linewidth]{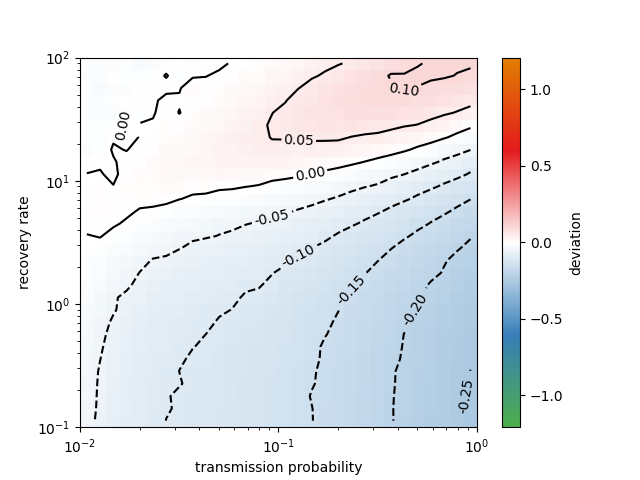}
    \includegraphics[width=.3\linewidth]{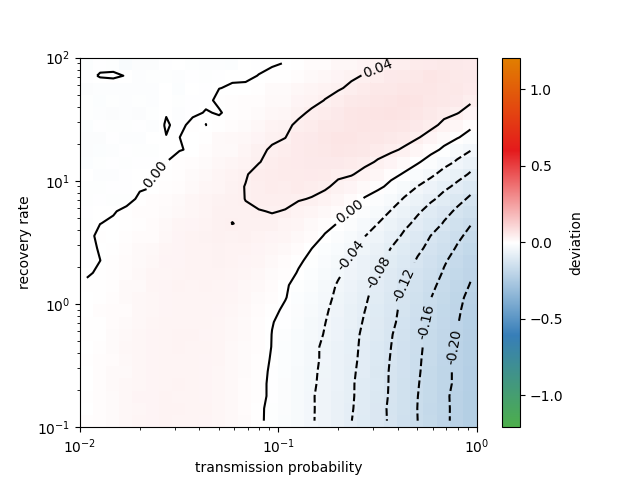}
    \includegraphics[width=.3\linewidth]{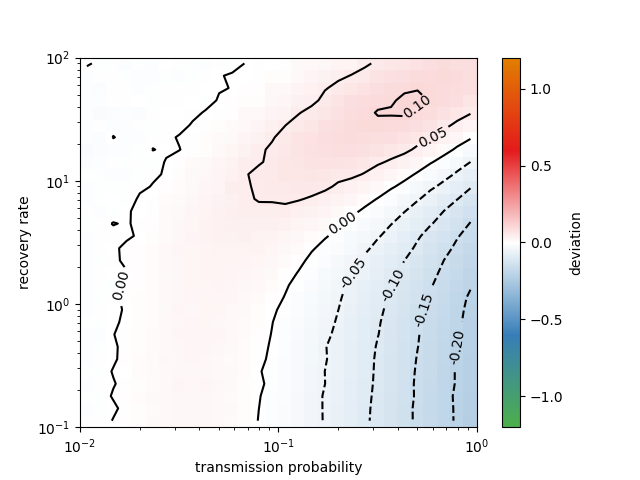}
        \includegraphics[width=.3\linewidth]{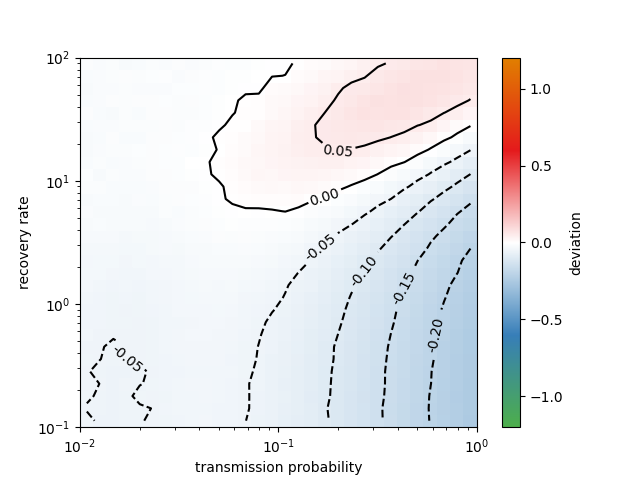}
 \caption{{\bf Surrogate, effect of taking into account group structure for the LyonSchool data set.}   Relative difference in $R_0$ obtained from the simulations on surrogate data
 with respect to simulations on the original data.
 Top: ST-RA. Bottom: GST-RA. Left $f=40\%$, middle $f=10\%$; right $f=5\%$. }
\end{figure}
    
    \begin{figure}[thb]
    \centering
    \includegraphics[width=.3\linewidth]{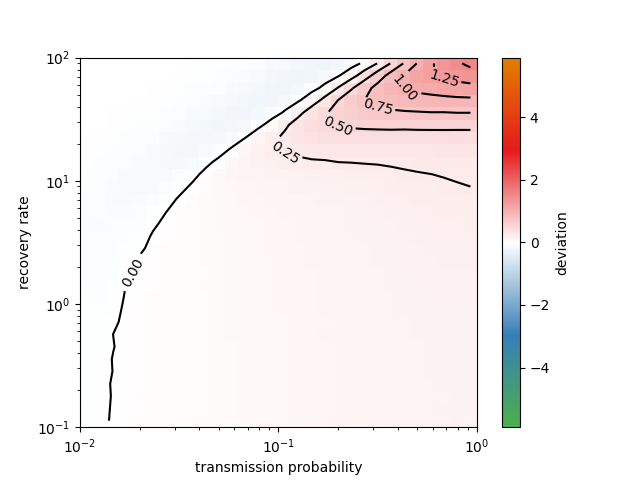}
\includegraphics[width=.3\linewidth]{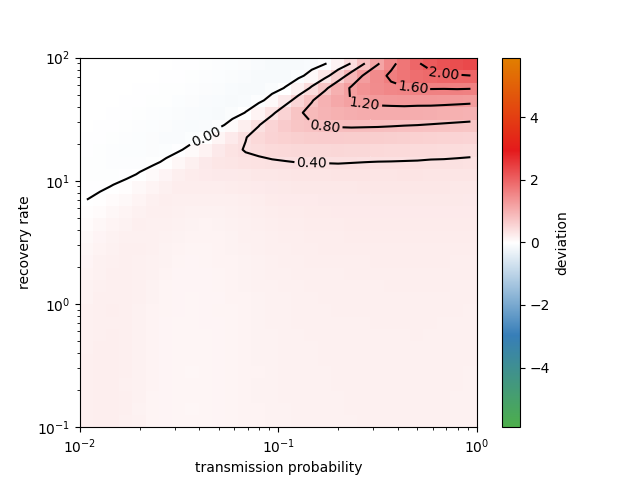}
\includegraphics[width=.3\linewidth]{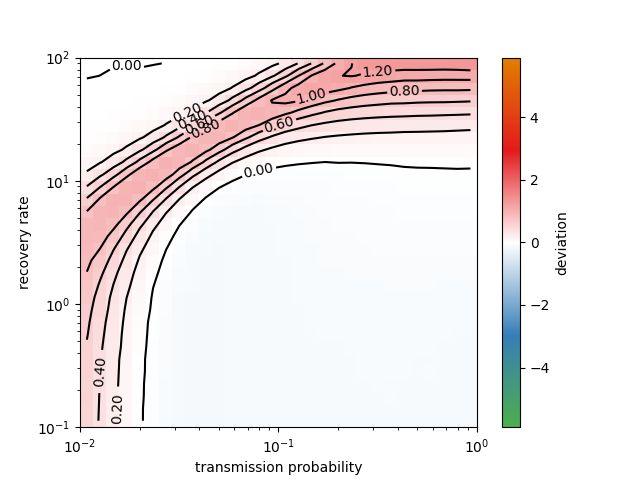}
    \includegraphics[width=.3\linewidth]{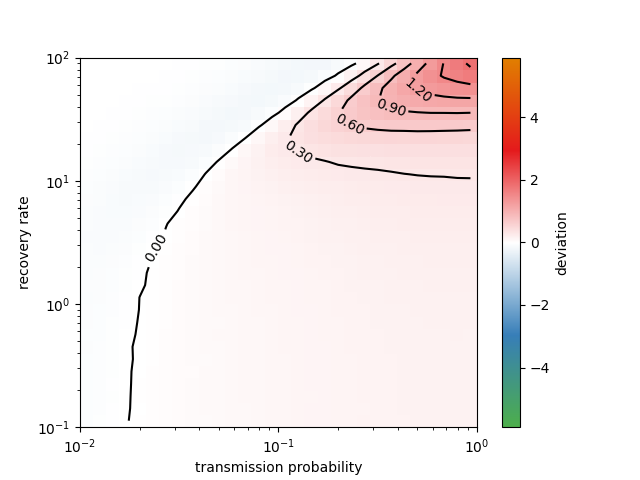}
    \includegraphics[width=.3\linewidth]{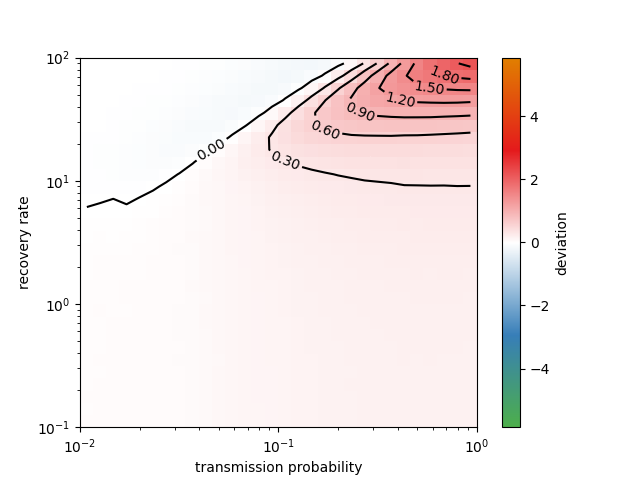}
        \includegraphics[width=.3\linewidth]{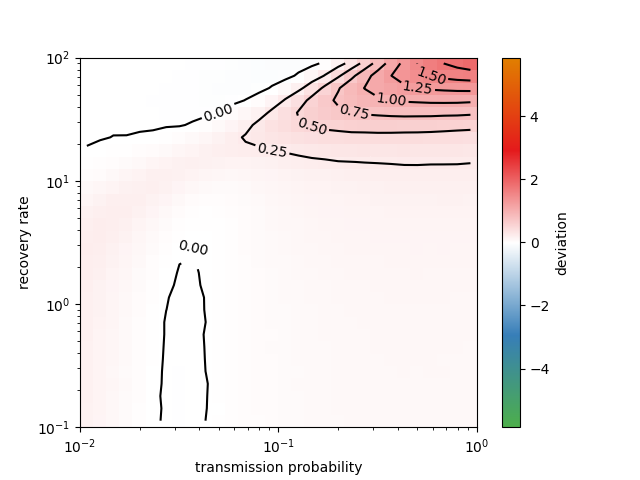}
 \caption{{\bf Surrogate, effect of taking into account group structure for the LyonSchool data set.}  Relative difference in $\Omega$ obtained from the simulations on surrogate data
 and on the original data.
 Top: ST-RA. Bottom: GST-RA. Left $f=40\%$, middle $f=10\%$; right $f=5\%$. }
\end{figure}

\begin{figure}[thb]
    \centering
    \includegraphics[width=.3\linewidth]{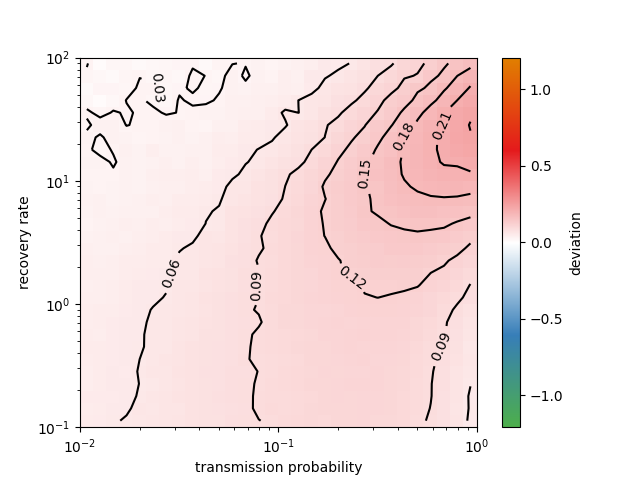}
\includegraphics[width=.3\linewidth]{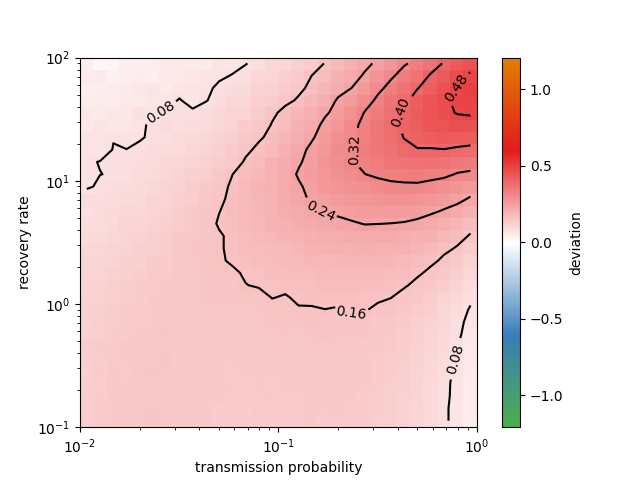}
\includegraphics[width=.3\linewidth]{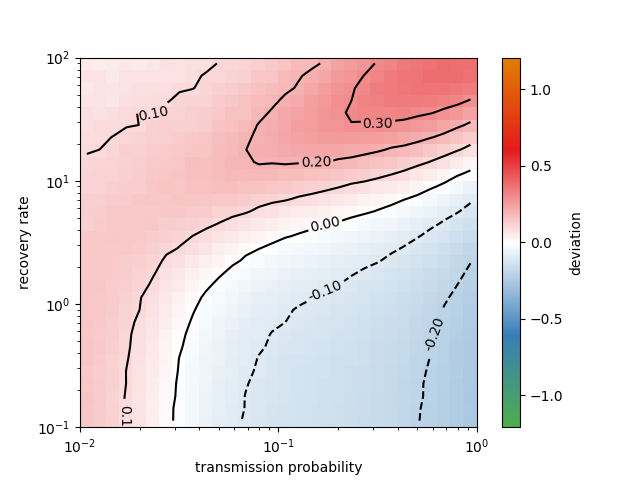}
    \includegraphics[width=.3\linewidth]{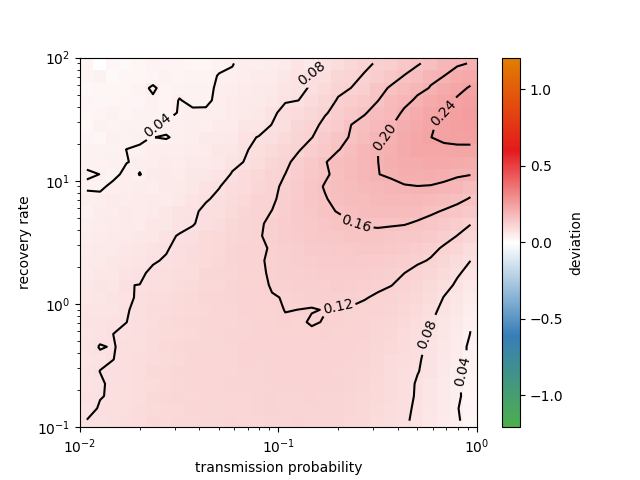}
    \includegraphics[width=.3\linewidth]{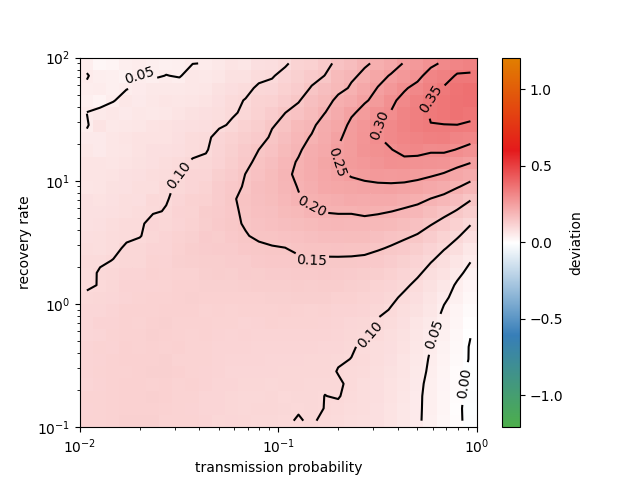}
        \includegraphics[width=.3\linewidth]{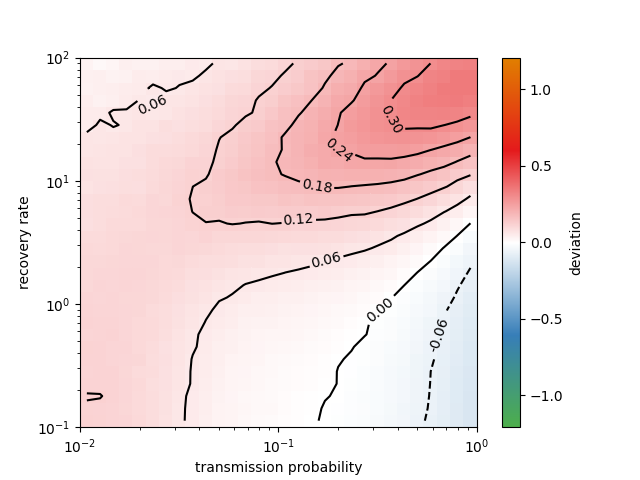}
 \caption{{\bf Surrogate, effect of taking into account group structure for the InVS15 data set}.  Relative difference in $R_0$ obtained from the simulations on surrogate data  and on the original data.
 Top: ST-RA. Bottom: GST-RA. Left $f=40\%$, middle $f=10\%$; right $f=5\%$.}
\end{figure}
    
    \begin{figure}[thb]
    \centering
    \includegraphics[width=.3\linewidth]{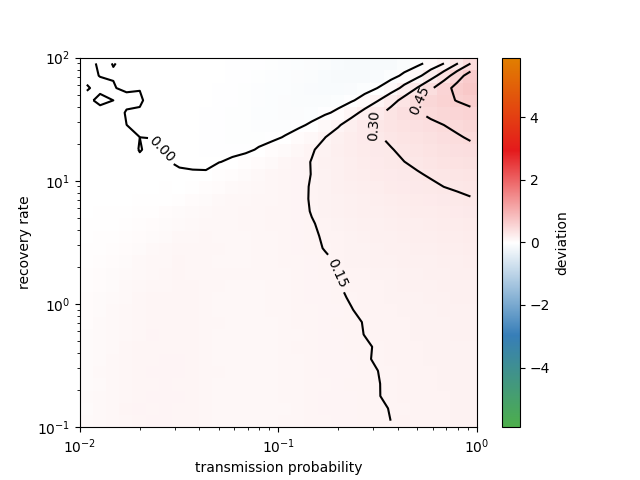}
\includegraphics[width=.3\linewidth]{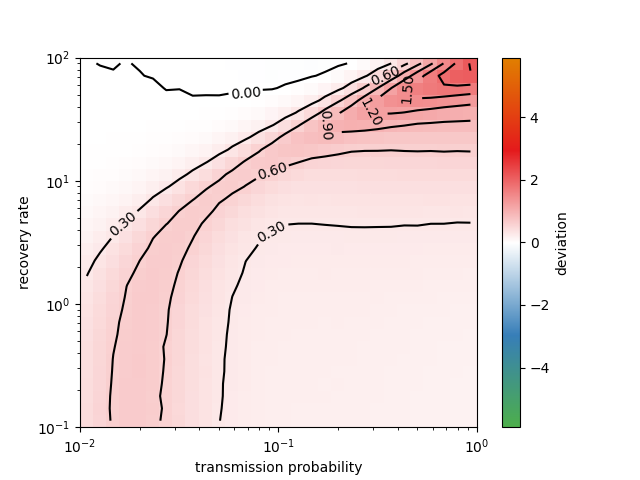}
\includegraphics[width=.3\linewidth]{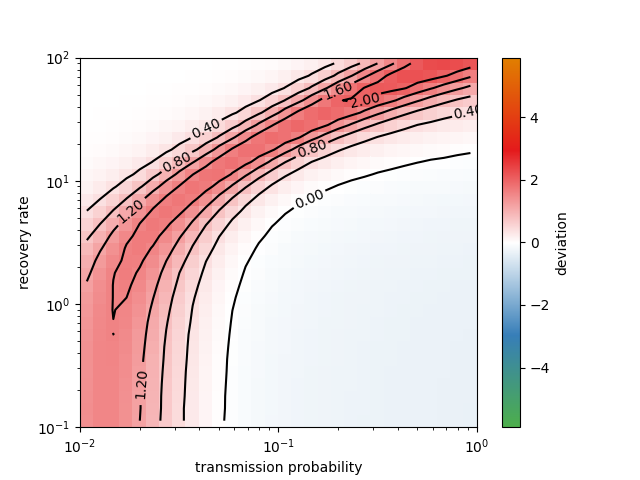}
    \includegraphics[width=.3\linewidth]{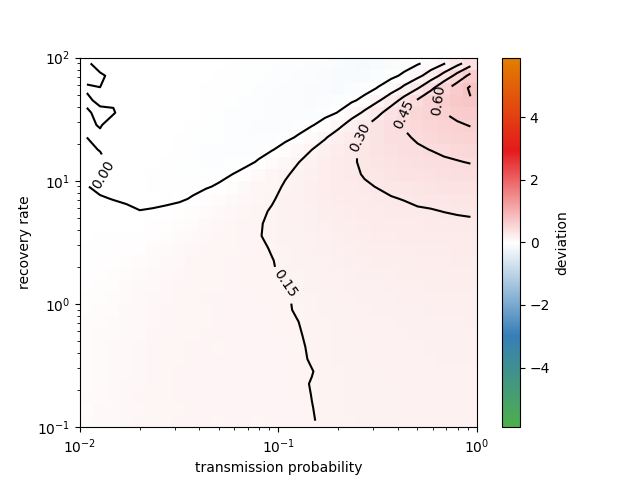}
    \includegraphics[width=.3\linewidth]{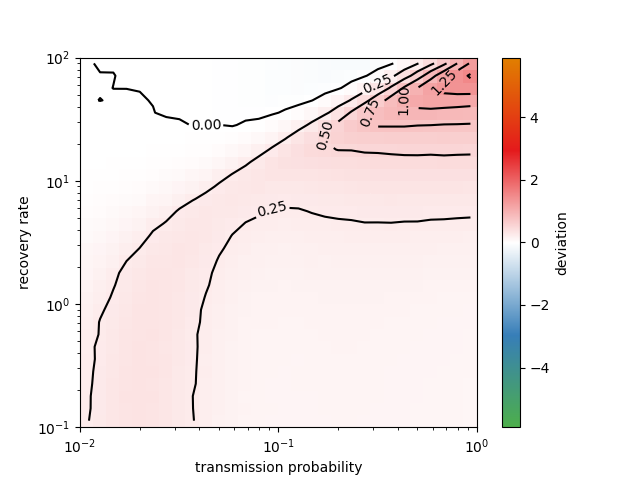}
        \includegraphics[width=.3\linewidth]{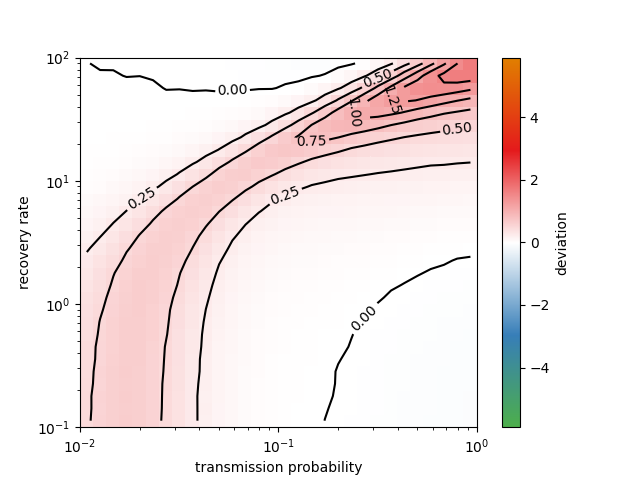}
 \caption{{\bf Surrogate, effect of taking into account group structure for the InVS15 data set.}  Relative difference in $\Omega$ obtained from the simulations on surrogate data
  and on the original data.
 Top: ST-RA. Bottom: GST-RA. Left $f=40\%$, middle $f=10\%$; right $f=5\%$. }
\end{figure}

\clearpage
\newpage
\section{Effect of the timeline reconstruction}

\begin{figure}[thb]
    \centering
        \includegraphics[width=.3\linewidth]{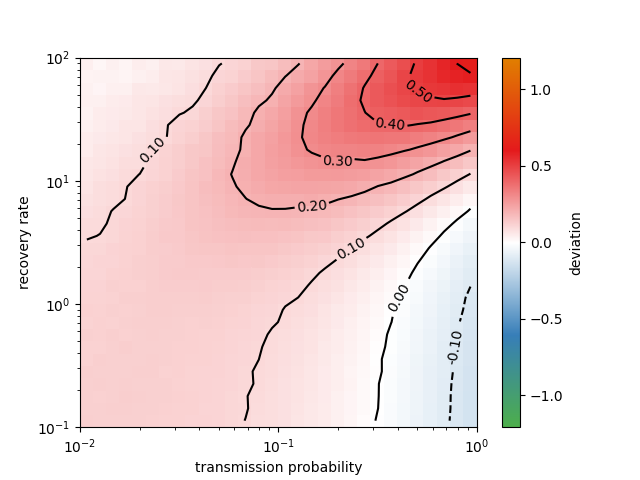}
        \includegraphics[width=.3\linewidth]{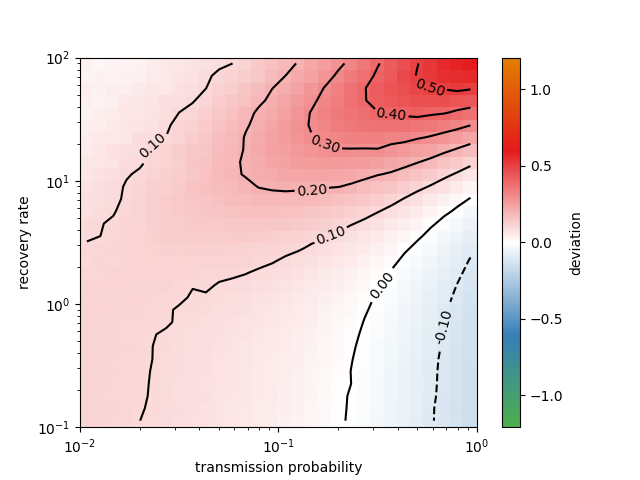}
        \includegraphics[width=.3\linewidth]{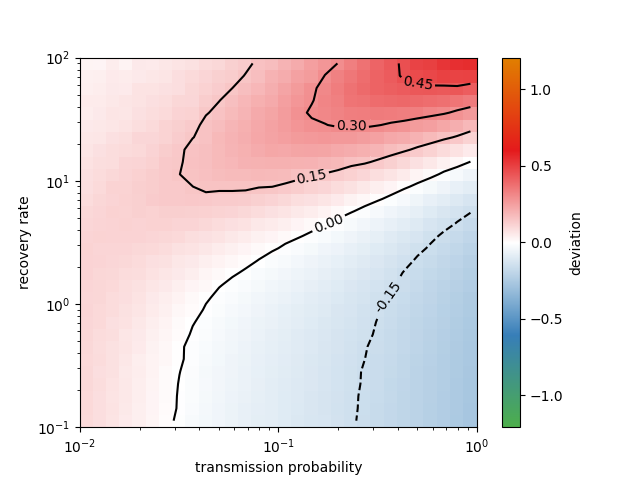}
        \includegraphics[width=.3\linewidth]{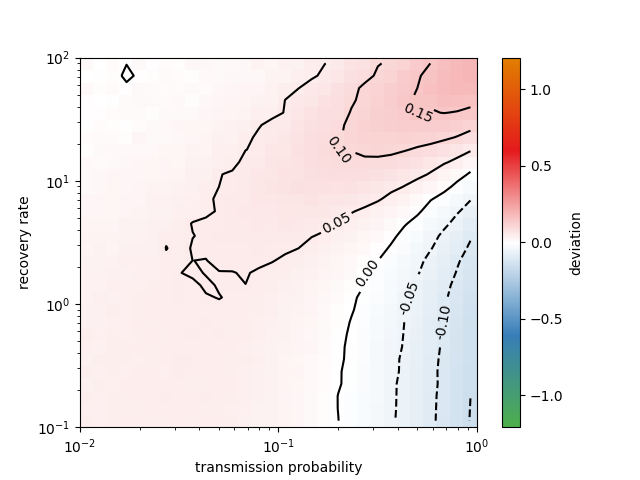}
        \includegraphics[width=.3\linewidth]{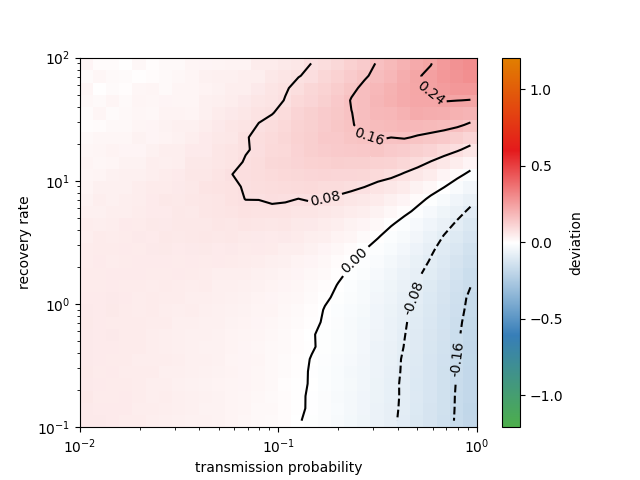}
        \includegraphics[width=.3\linewidth]{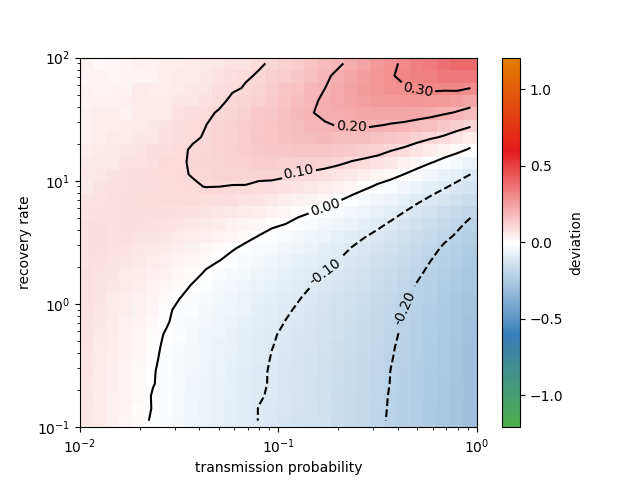}
        \includegraphics[width=.3\linewidth]{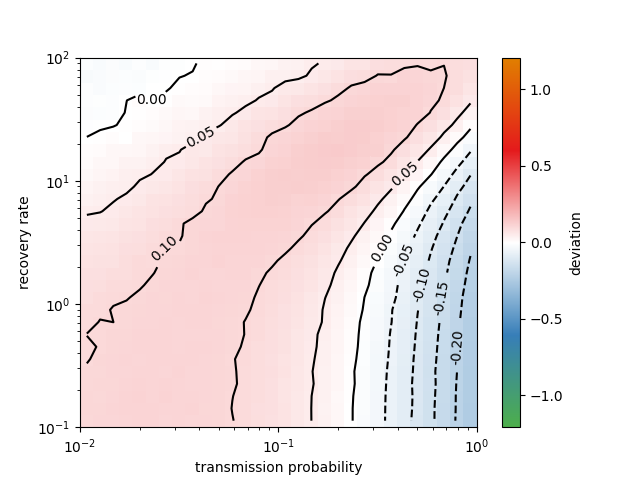}
        \includegraphics[width=.3\linewidth]{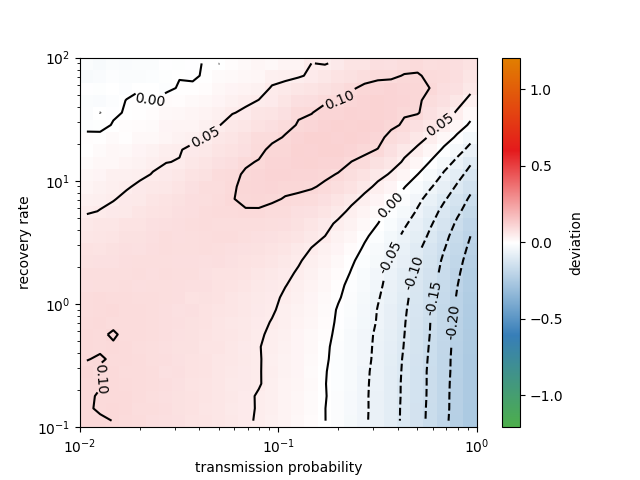}
        \includegraphics[width=.3\linewidth]{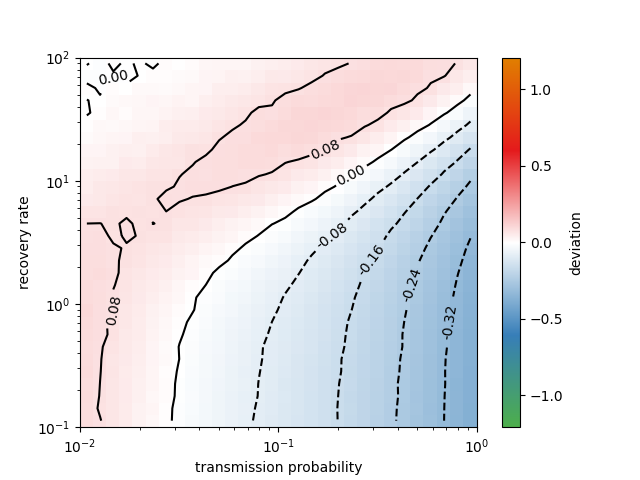}
        \includegraphics[width=.3\linewidth]{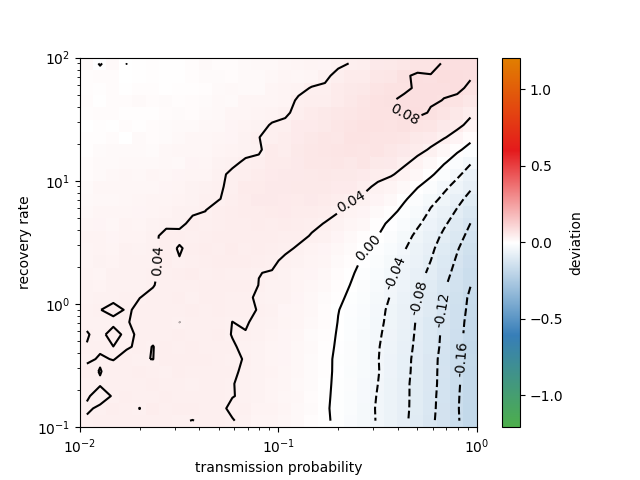}
        \includegraphics[width=.3\linewidth]{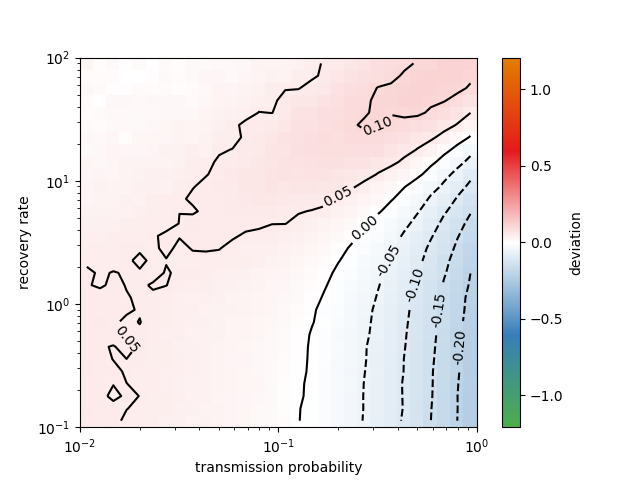}
        \includegraphics[width=.3\linewidth]{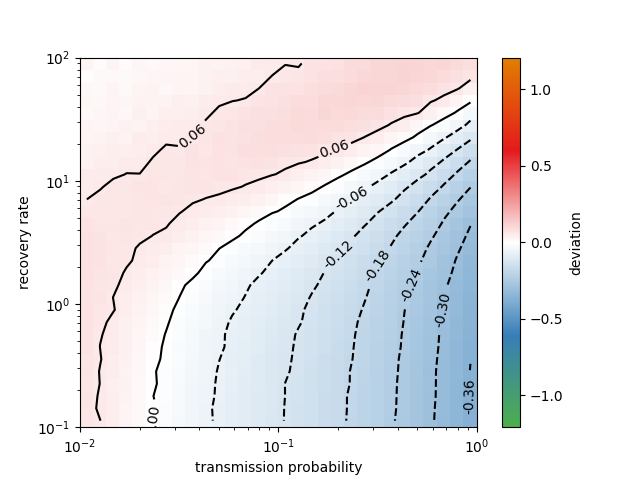}
    \caption{
    {\bf Effect of the surrogate timelines, for the Thiers13 data set.} 
     Relative difference in $R_0$ obtained from the simulations on surrogate data (GST-RA method) and on the original. First row: Poisson timelines.  Second row: BTL-Poisson  (backbone timelines kept, Poisson timelines for surrogate ties).
Third row: Stats (synthetic timelines respecting the data's statistics). Fourth row: BTL-Stats (same, with backbone timelines kept).
    Left column: $f= 40\%$.  Middle column: $f= 10\%$.  Right column: $f= 5\%$. 
    \label{fig:timelines_R0}}
\end{figure}

\begin{figure}[thb]
    \centering
        \includegraphics[width=.31\linewidth]{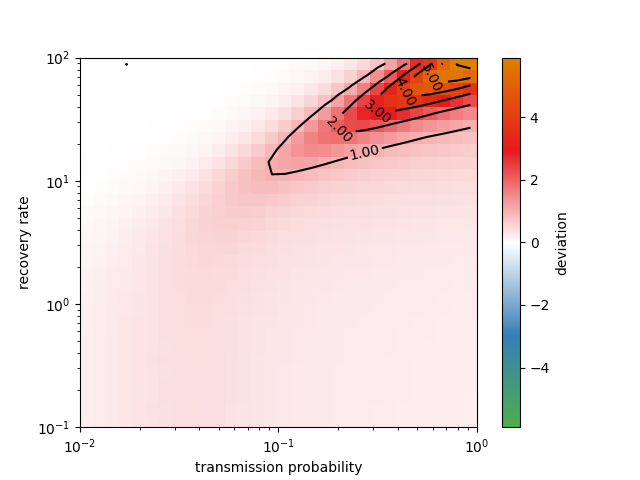}
        \includegraphics[width=.31\linewidth]{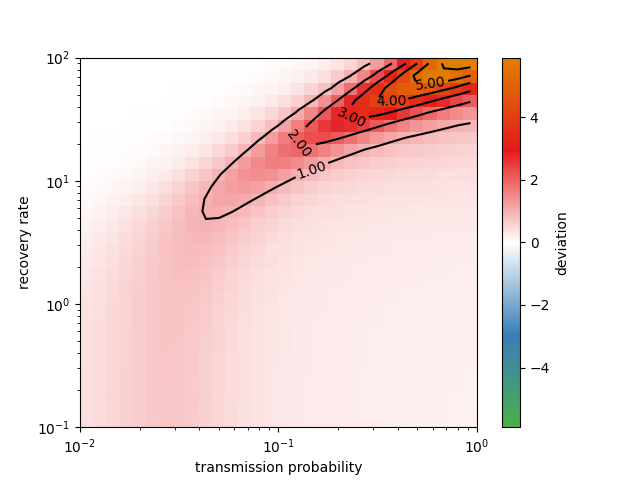}
        \includegraphics[width=.31\linewidth]{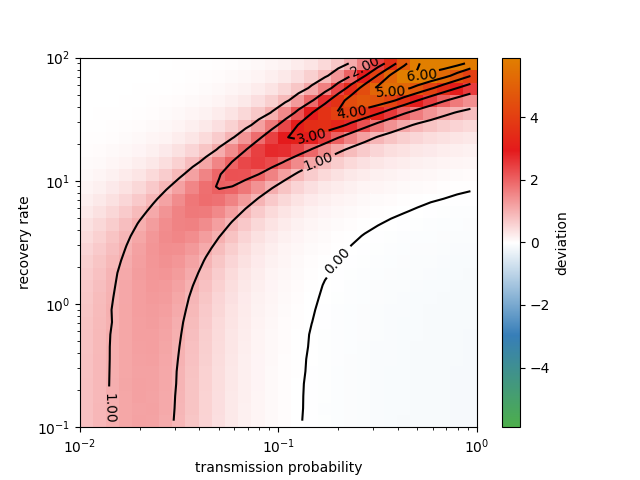}
        \includegraphics[width=.31\linewidth]{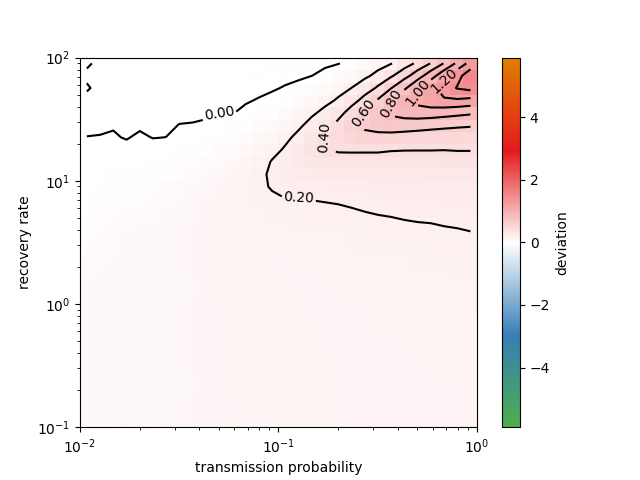}
        \includegraphics[width=.31\linewidth]{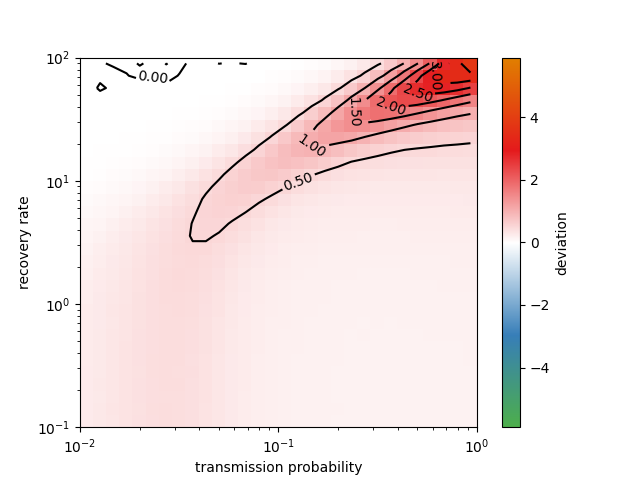}
        \includegraphics[width=.31\linewidth]{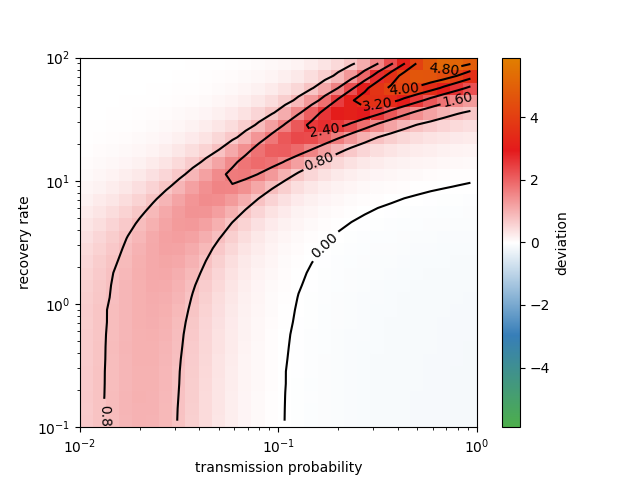}
        \includegraphics[width=.31\linewidth]{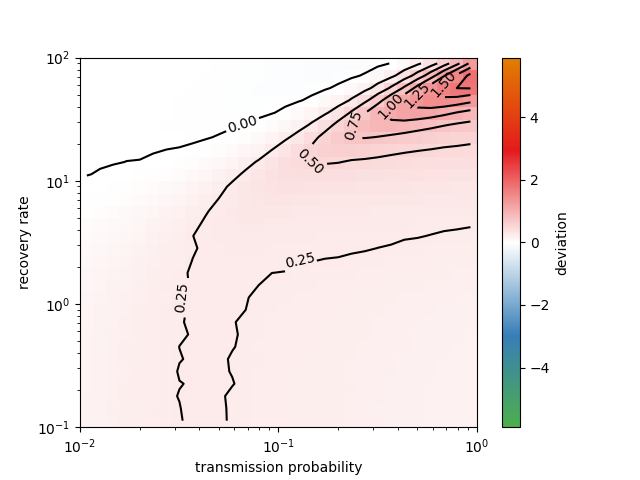}
        \includegraphics[width=.31\linewidth]{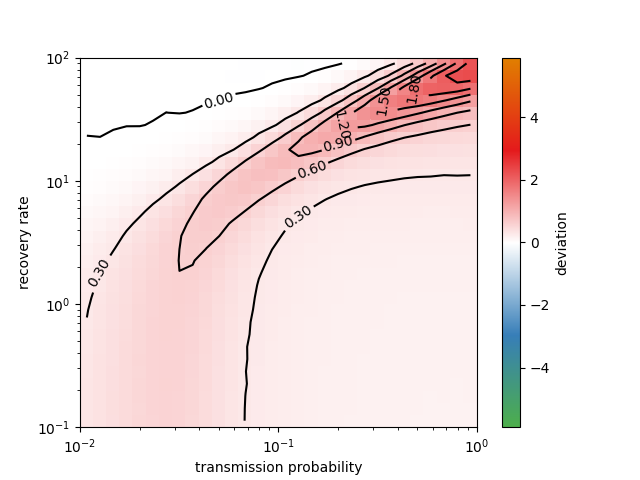}
        \includegraphics[width=.31\linewidth]{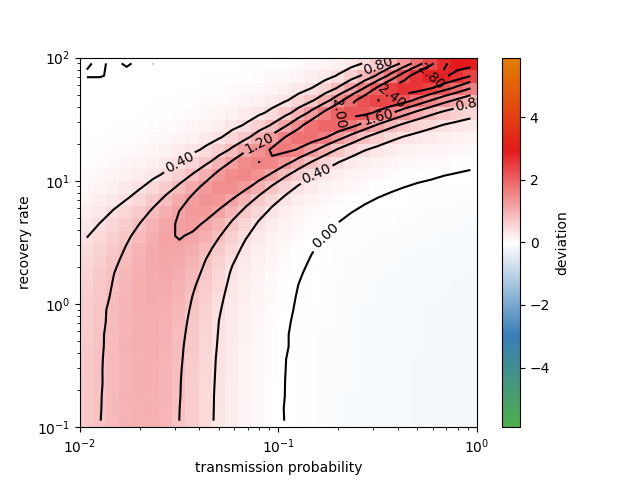}
        \includegraphics[width=.31\linewidth]{Figures/omega/reldiff/SURROGATES_3min/Thiers13_temporal_surrogate_groups_bb_and_stats_size_04.png}
        \includegraphics[width=.31\linewidth]{Figures/omega/reldiff/SURROGATES_3min/Thiers13_temporal_surrogate_groups_bb_and_stats_size_01.png}
        \includegraphics[width=.31\linewidth]{Figures/omega/reldiff/SURROGATES_3min/Thiers13_temporal_surrogate_groups_bb_and_stats_size_005.png}
    \caption{
    {\bf Effect of the surrogate timelines, for the Thiers13 data set.} 
    Each panel shows the relative difference in $\Omega$ obtained from the simulations on surrogate data and on the original. Here we use the
    GST-RA method. First row: Poisson timelines for all links.  Second row: BTL-Poisson method (backbone timelines kept, Poisson timelines for surrogate ties).
Third row: Stats method (synthetic timelines respecting the data's statistics for all ties). Fourth row: BTL-Stats method (backbone timelines kept, synthetic timelines respecting the data's statistics for surrogate ties).
    Left column: $f= 40\%$.  Middle column: $f= 10\%$.  Right column: $f= 5\%$. 
     \label{fig:timelines_omega}}
\end{figure}

\clearpage
\newpage
\section{Various surrogates, other data sets}

 \begin{figure}[thb]
    \centering
        \includegraphics[width=.3\linewidth]{Figures/omega/reldiff/SURROGATES_3min/Thiers13_temporal_surrogate_groups_bb_and_stats_size_04.png}
        \includegraphics[width=.3\linewidth]{Figures/omega/reldiff/SURROGATES_3min/Thiers13_temporal_surrogate_groups_bb_and_stats_size_01.png}
        \includegraphics[width=.3\linewidth]{Figures/omega/reldiff/SURROGATES_3min/Thiers13_temporal_surrogate_groups_bb_and_stats_size_005.png}
        \includegraphics[width=.3\linewidth]{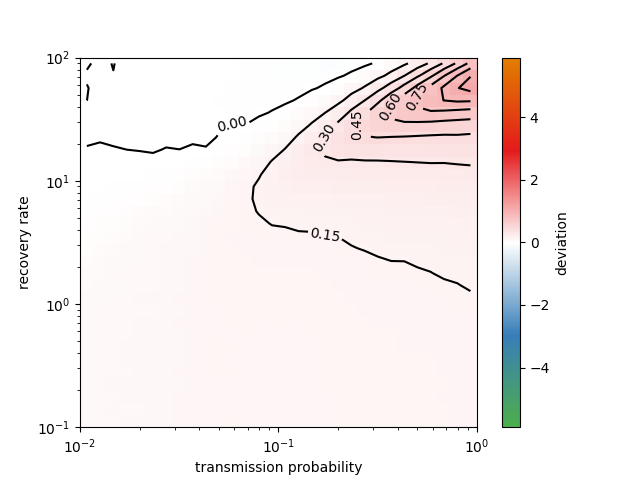}
        \includegraphics[width=.3\linewidth]{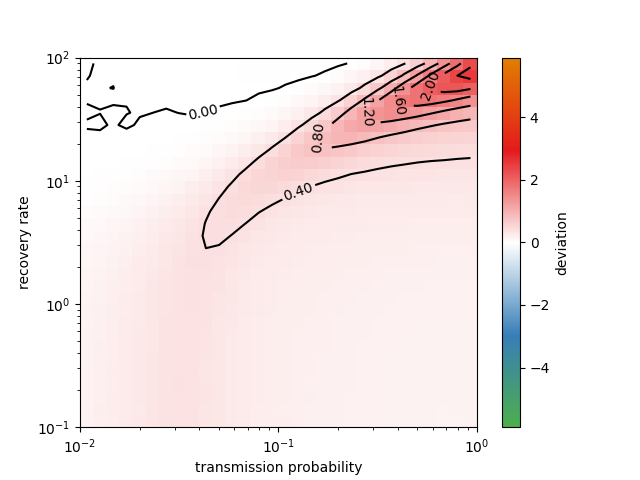}
        \includegraphics[width=.3\linewidth]{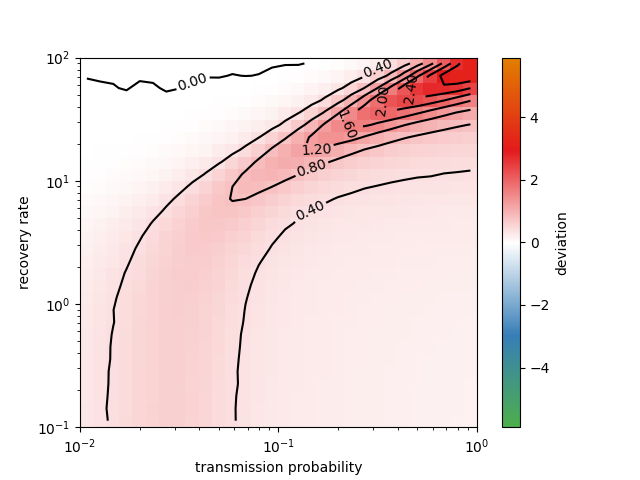}
        \includegraphics[width=.3\linewidth]{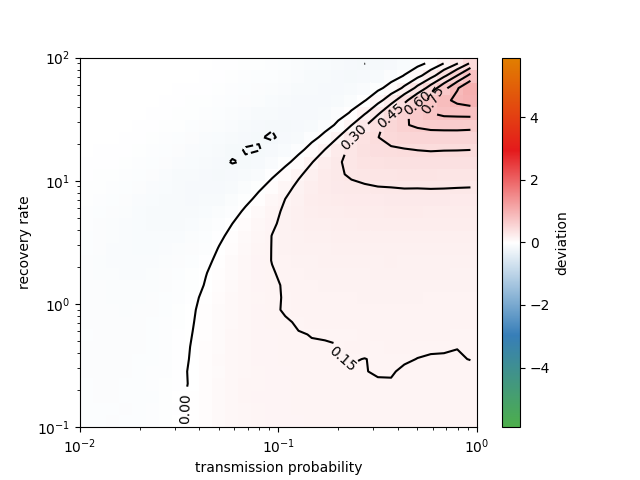}
        \includegraphics[width=.3\linewidth]{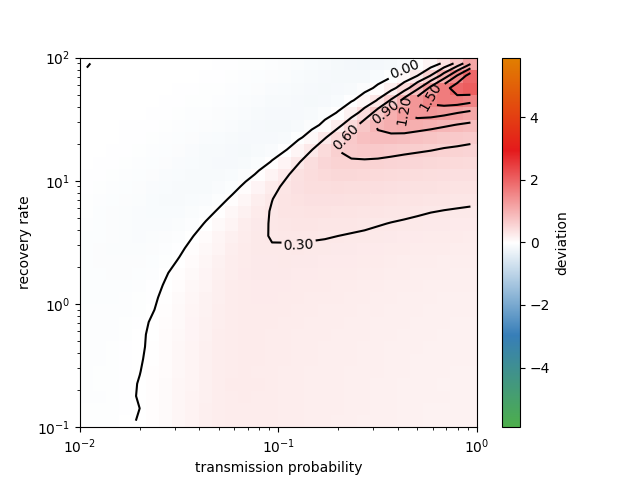}
        \includegraphics[width=.3\linewidth]{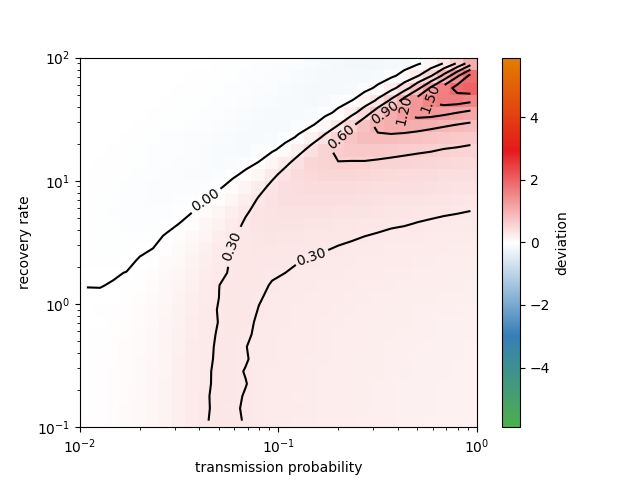}
        \includegraphics[width=.3\linewidth]{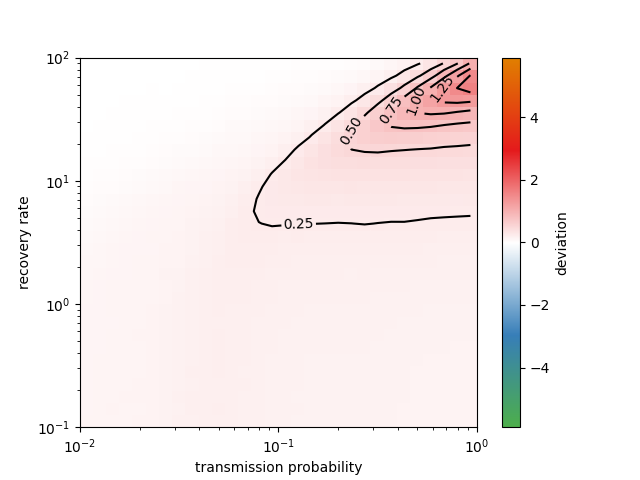}
        \includegraphics[width=.3\linewidth]{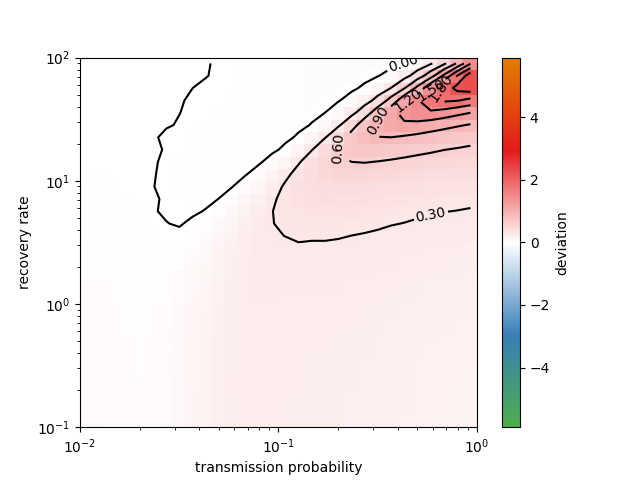}
        \includegraphics[width=.3\linewidth]{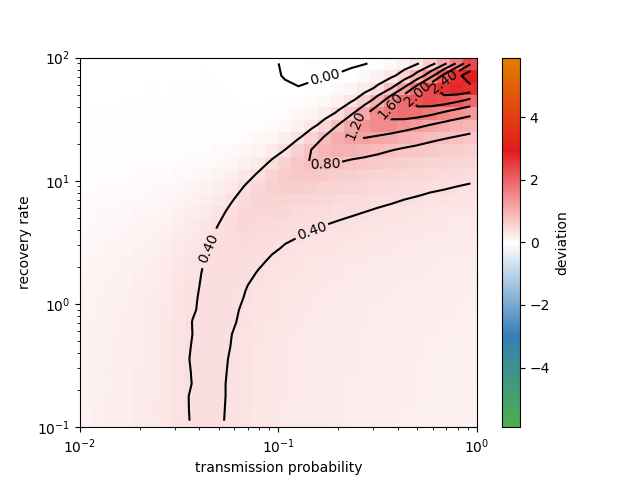}
    \caption{
    {\bf Outcome of SIR processes on surrogate data
obtained by various reconstruction methods, for the Thiers13 data set.} Relative difference in $\Omega$ measured in simulations on the surrogate and original data. 
First row: GST-RA. Second row: GST-OA. Third row: GST-RT. Fourth row: GTB-RT. The backbone timelines are kept, and timelines respecting the statistics of contact and inter-contact durations are built for the surrogate ties (BTL-Stats method).
Left column: $f= 40\%$.  Middle column: $f= 10\%$.  Right column: $f= 5\%$. }
    \end{figure}

\begin{figure}[thb]
    \centering
        \includegraphics[width=.31\linewidth]{Figures/r0/reldiff/SURROGATES_3min/LyonSchool_temporal_surrogate_groups_bb_and_stats_size_04.png}
        \includegraphics[width=.31\linewidth]{Figures/r0/reldiff/SURROGATES_3min/LyonSchool_temporal_surrogate_groups_bb_and_stats_size_01.png}
        \includegraphics[width=.31\linewidth]{Figures/r0/reldiff/SURROGATES_3min/LyonSchool_temporal_surrogate_groups_bb_and_stats_size_005.png}
        \includegraphics[width=.31\linewidth]{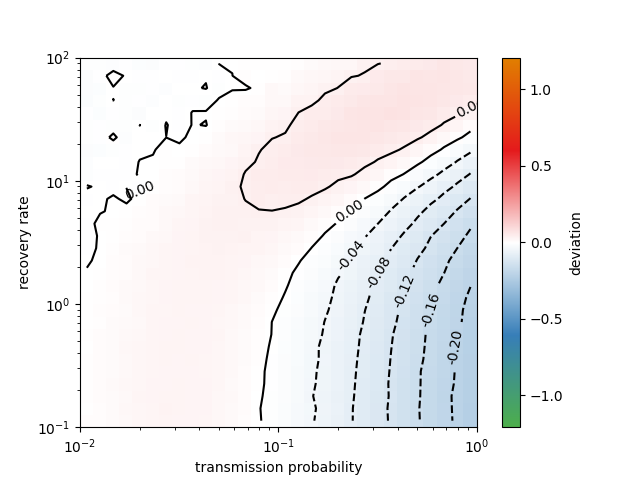}
        \includegraphics[width=.31\linewidth]{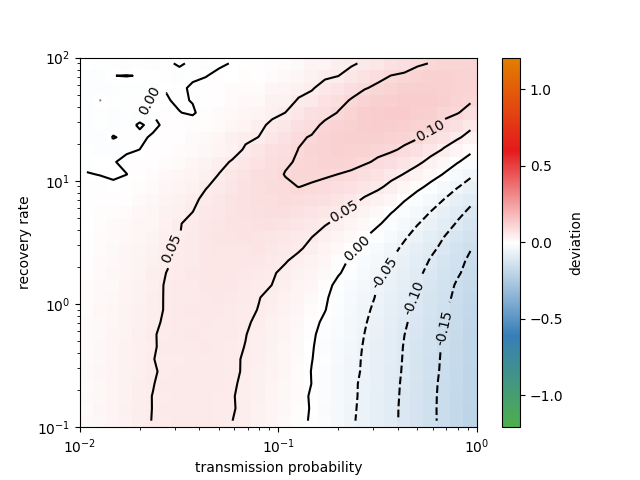}
        \includegraphics[width=.31\linewidth]{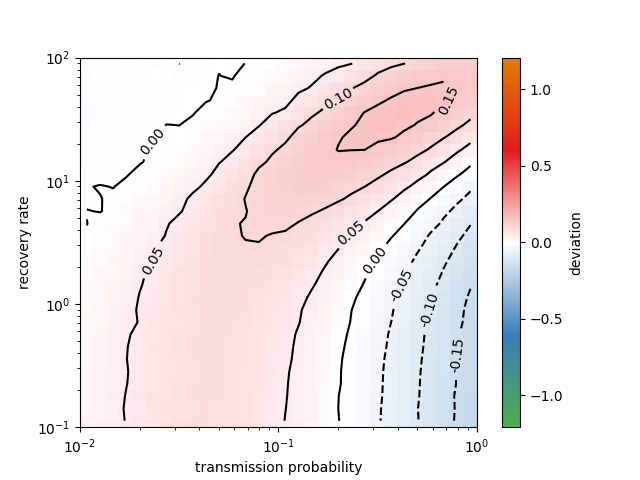}
        \includegraphics[width=.31\linewidth]{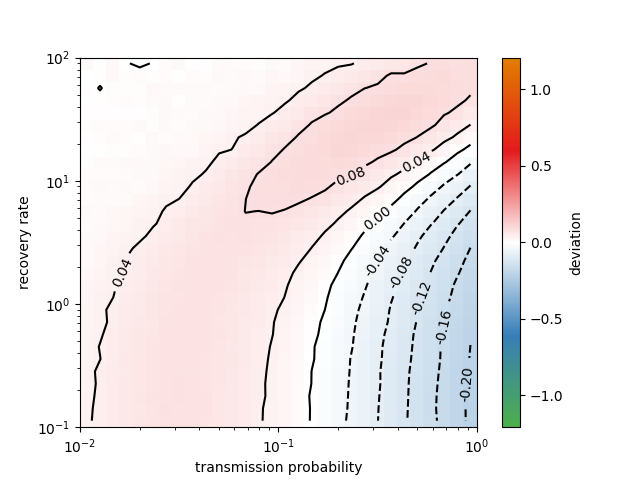}
        \includegraphics[width=.31\linewidth]{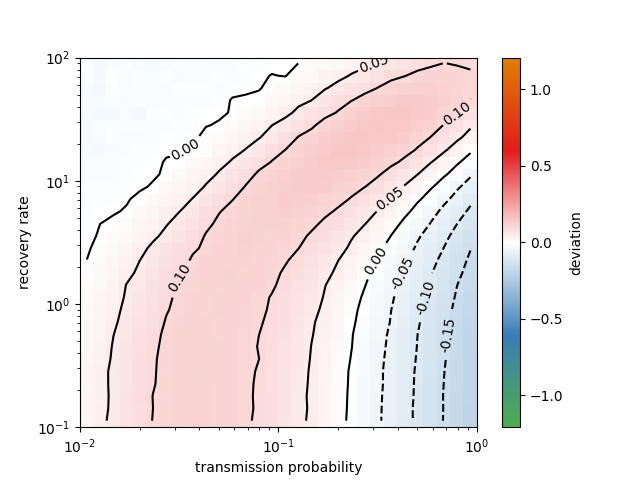}
        \includegraphics[width=.31\linewidth]{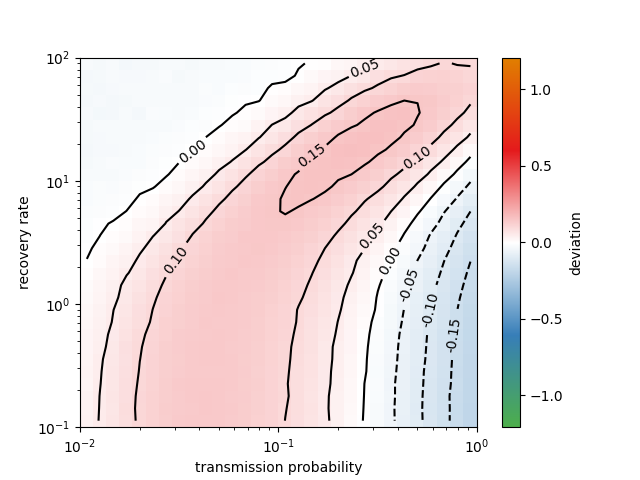}
        \includegraphics[width=.31\linewidth]{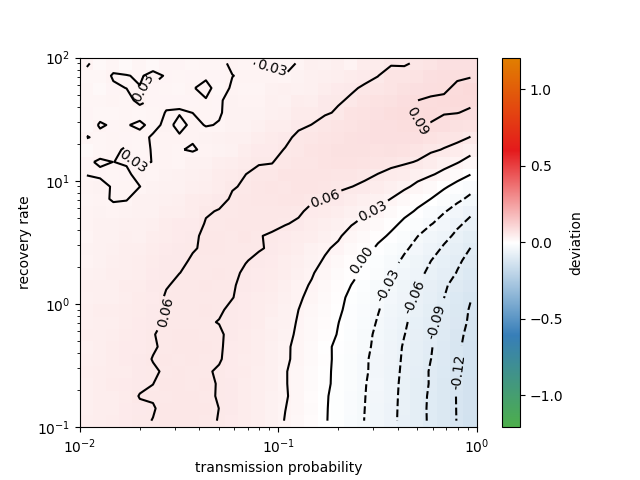}
        \includegraphics[width=.31\linewidth]{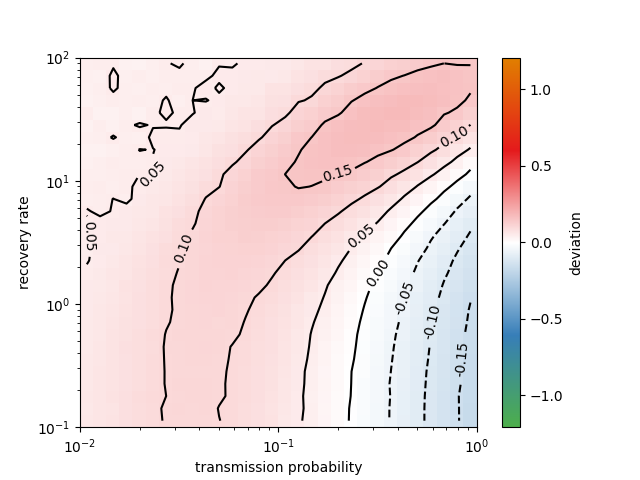}
        \includegraphics[width=.31\linewidth]{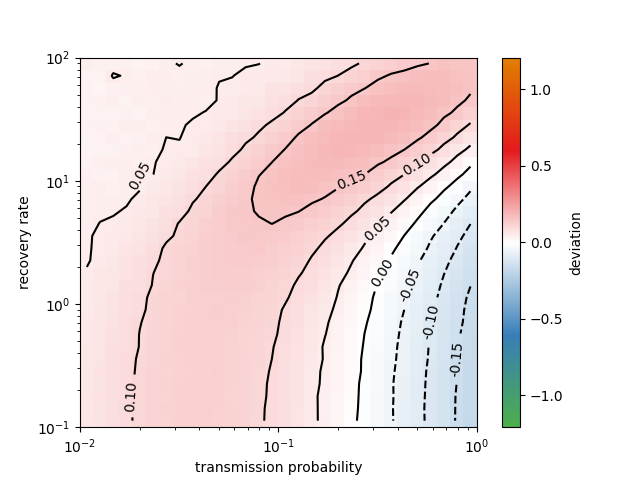}
    \caption{
    {\bf Outcome of SIR processes on surrogate data
obtained by various reconstruction methods, for the LyonSchool data set.} Relative difference in $R_0$ measured in simulations on the surrogate and on the original data. 
First row: GST-RA. Second row: GST-OA. Third row: GST-RT. Fourth row: GTB-RT. In each case the backbone timelines are kept, and timelines respecting the statistics of contact and inter-contact durations are built for the surrogate ties (BTL-Stats method).
Left column: $f= 40\%$.  Middle column: $f= 10\%$.  Right column: $f= 5\%$. }
\end{figure}

\begin{figure}[thb]
    \centering
        \includegraphics[width=.31\linewidth]{Figures/omega/reldiff/SURROGATES_3min/LyonSchool_temporal_surrogate_groups_bb_and_stats_size_04.png}
        \includegraphics[width=.31\linewidth]{Figures/omega/reldiff/SURROGATES_3min/LyonSchool_temporal_surrogate_groups_bb_and_stats_size_01.png}
        \includegraphics[width=.31\linewidth]{Figures/omega/reldiff/SURROGATES_3min/LyonSchool_temporal_surrogate_groups_bb_and_stats_size_005.png}
        \includegraphics[width=.31\linewidth]{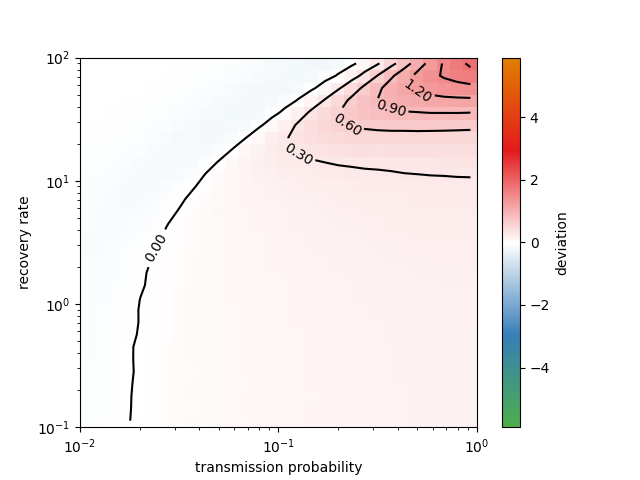}
        \includegraphics[width=.31\linewidth]{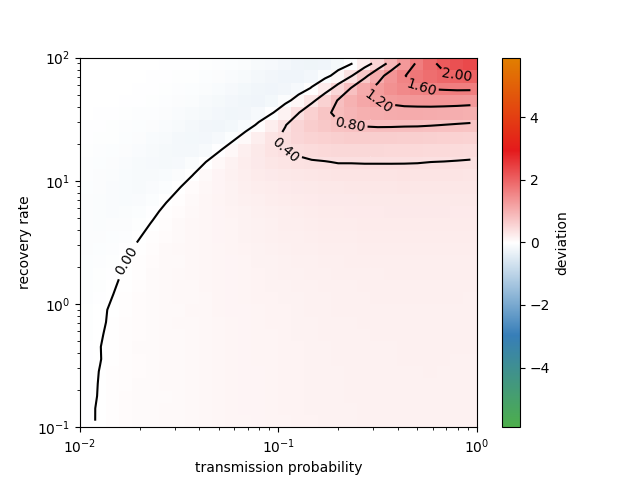}
        \includegraphics[width=.31\linewidth]{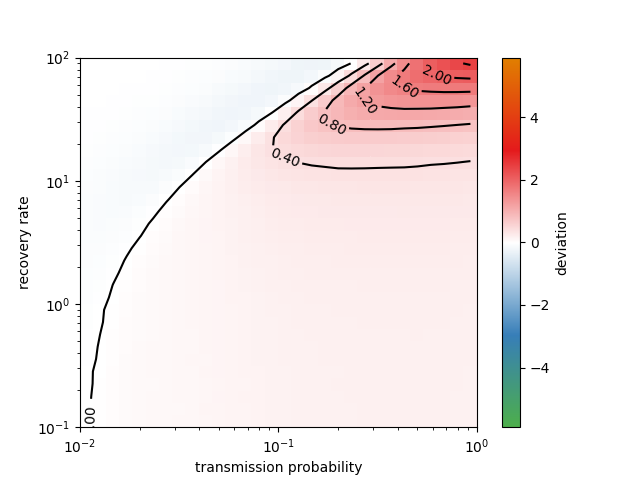}
        \includegraphics[width=.31\linewidth]{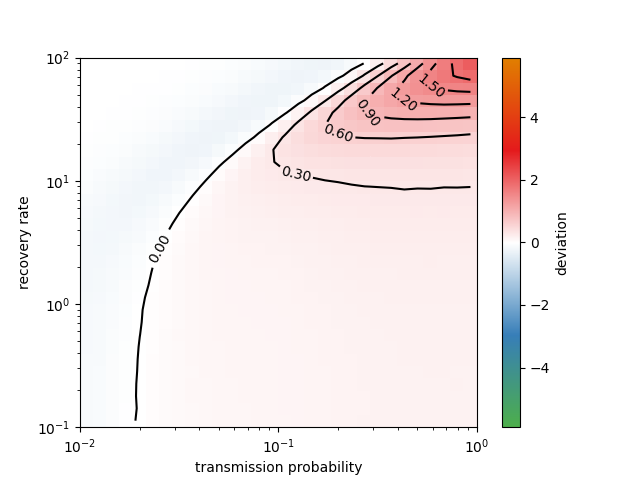}
        \includegraphics[width=.31\linewidth]{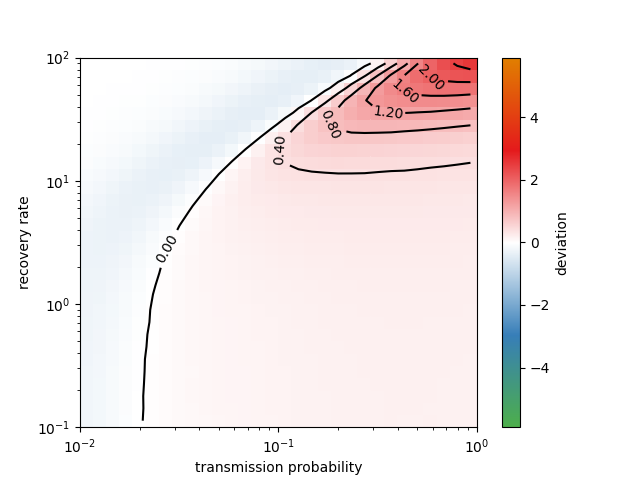}
        \includegraphics[width=.31\linewidth]{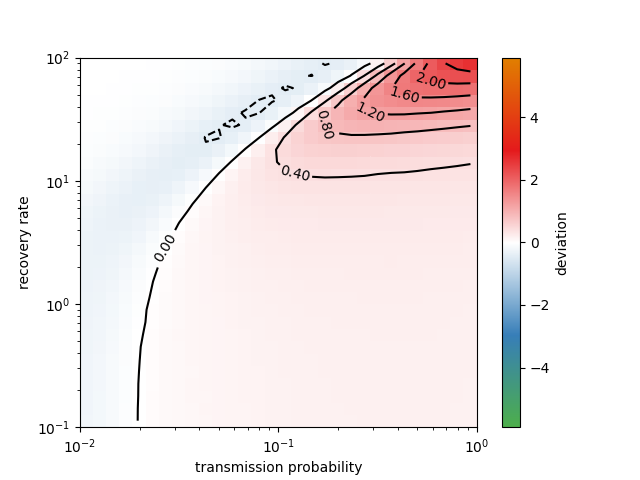}
        \includegraphics[width=.31\linewidth]{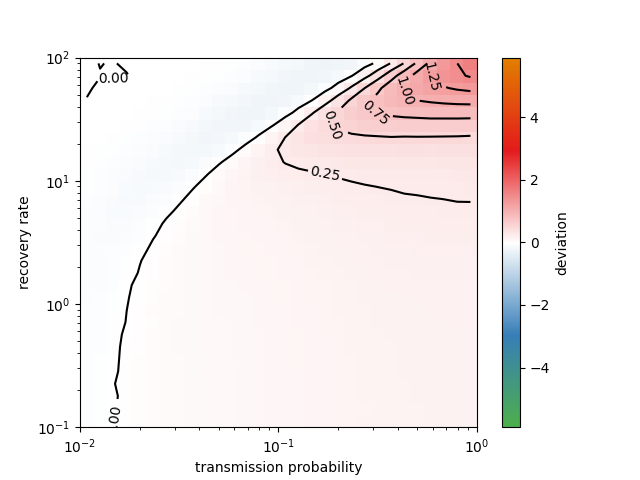}
        \includegraphics[width=.31\linewidth]{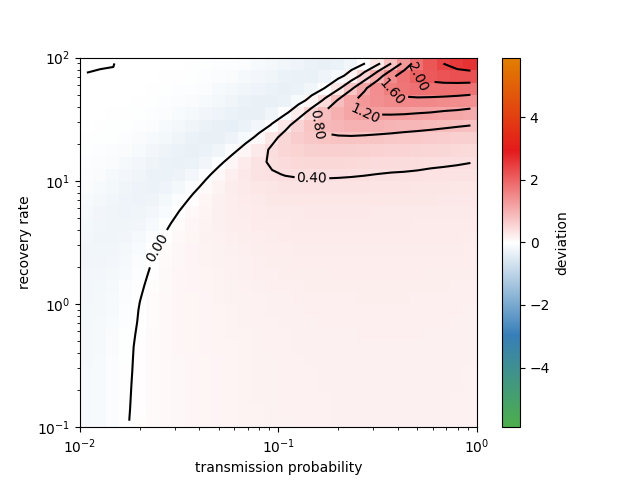}
        \includegraphics[width=.31\linewidth]{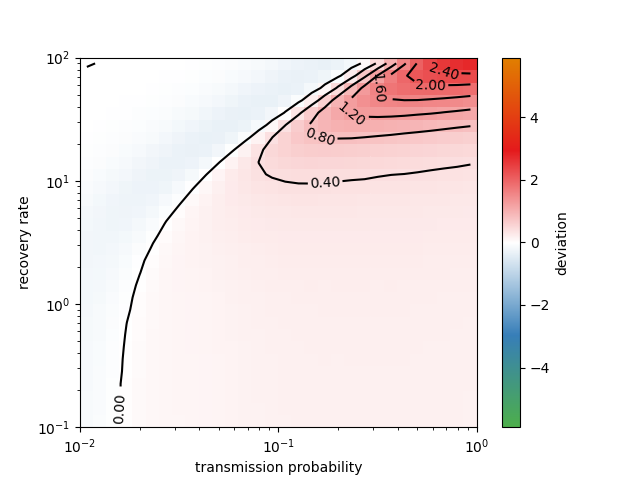}
    \caption{
        {\bf Outcome of SIR processes on surrogate data
obtained by various reconstruction methods, for the LyonSchool data set.} Relative difference in the values of $\Omega$ measured in simulations on the surrogate and on the original data. 
First row: GST-RA. Second row: GST-OA. Third row: GST-RT. Fourth row: GTB-RT. In each case the backbone timelines are kept, and timelines respecting the statistics of contact and inter-contact durations are built for the surrogate ties (BTL-Stats method).
Left column: $f= 40\%$.  Middle column: $f=10\%$.  Right column: $f= 5\%$. }
\end{figure}

\begin{figure}[thb]
    \centering
        \includegraphics[width=.31\linewidth]{Figures/r0/reldiff/SURROGATES_3min/InVS15_temporal_surrogate_groups_bb_and_stats_size_04.png}
        \includegraphics[width=.31\linewidth]{Figures/r0/reldiff/SURROGATES_3min/InVS15_temporal_surrogate_groups_bb_and_stats_size_01.png}
        \includegraphics[width=.31\linewidth]{Figures/r0/reldiff/SURROGATES_3min/InVS15_temporal_surrogate_groups_bb_and_stats_size_005.png}
        \includegraphics[width=.31\linewidth]{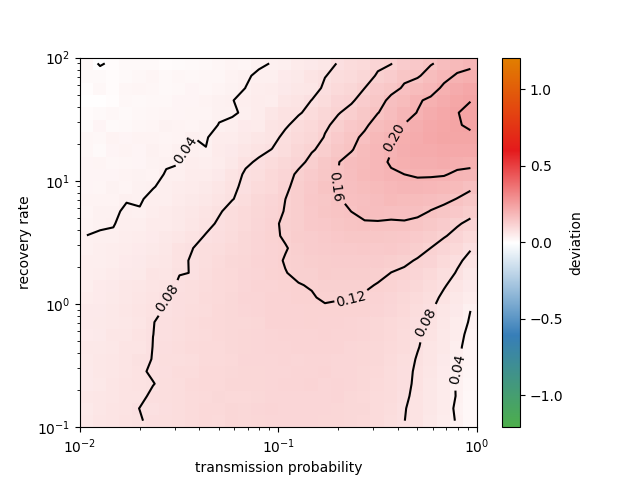}
        \includegraphics[width=.31\linewidth]{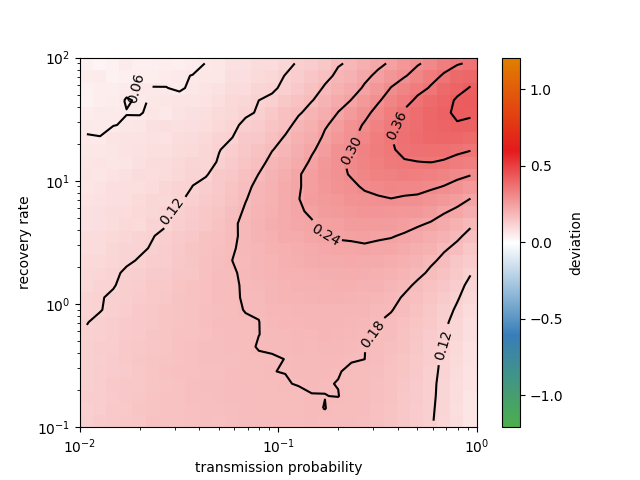}
        \includegraphics[width=.31\linewidth]{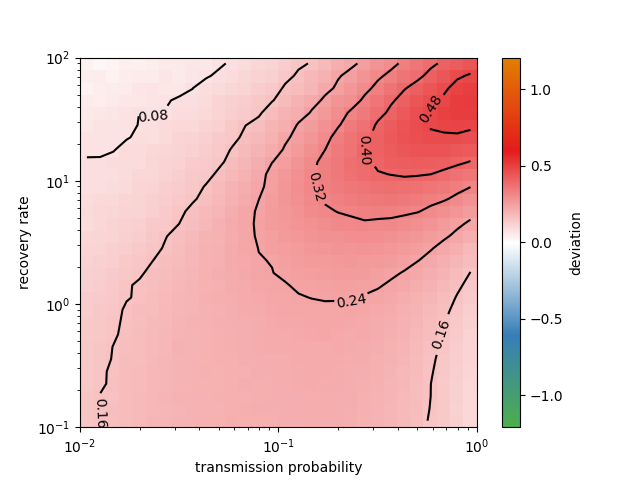}
        \includegraphics[width=.31\linewidth]{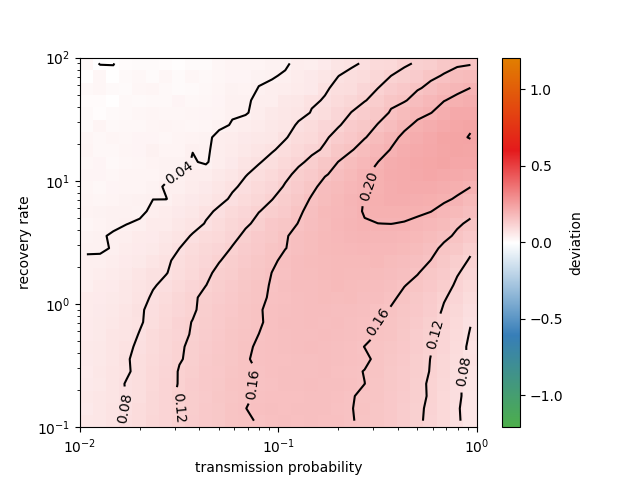}
        \includegraphics[width=.31\linewidth]{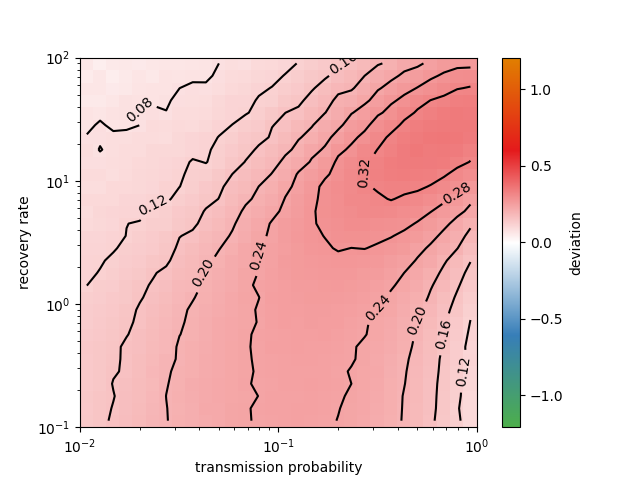}
        \includegraphics[width=.31\linewidth]{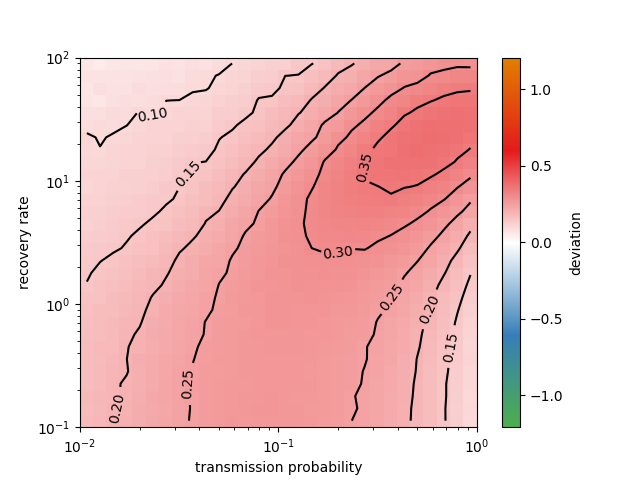}
        \includegraphics[width=.31\linewidth]{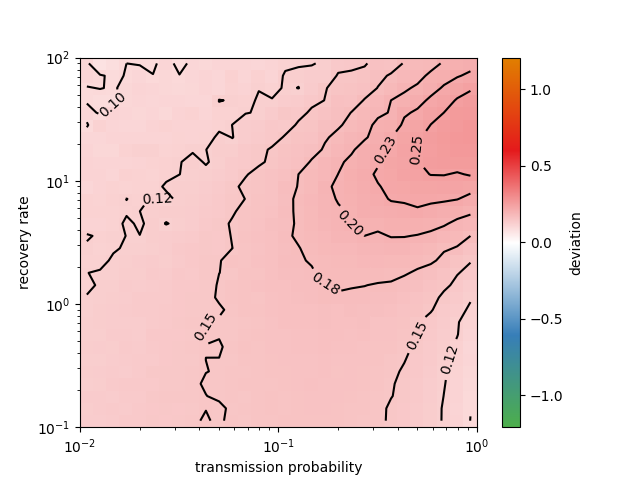}
        \includegraphics[width=.31\linewidth]{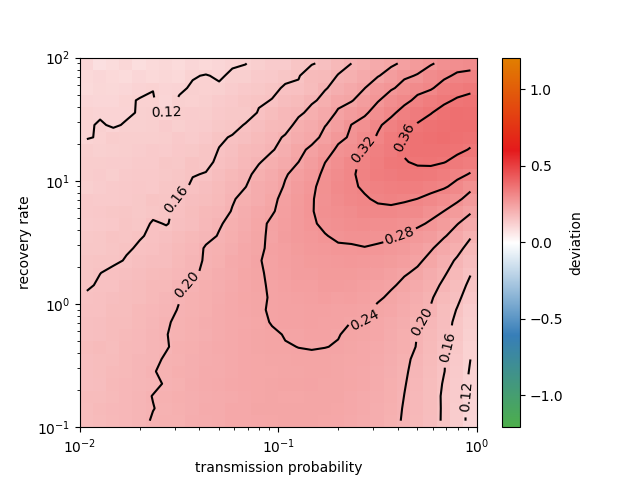}
        \includegraphics[width=.31\linewidth]{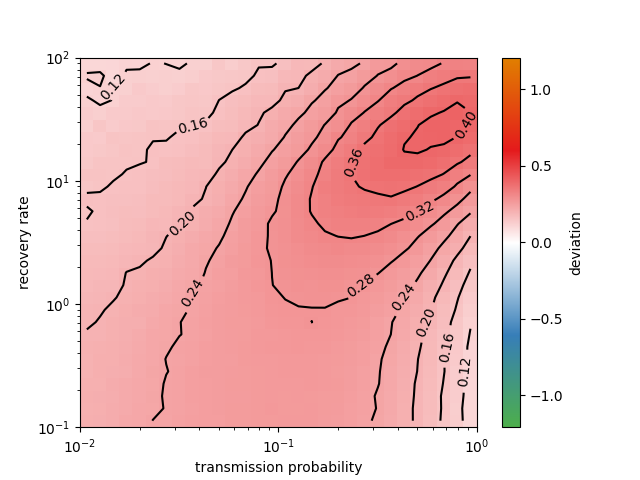}
    \caption{   {\bf Outcome of SIR processes on surrogate data
obtained by various reconstruction methods, for the InVS15 data set.} Relative difference in the values of $R_0$ measured in simulations on the surrogate and on the original data. 
First row: GST-RA. Second row: GST-OA. Third row: GST-RT. Fourth row: GTB-RT. In each case the backbone timelines are kept, and timelines respecting the statistics of contact and inter-contact durations are built for the surrogate ties (BTL-Stats method).
Left column: $f= 40\%$.  Middle column: $f= 10\%$.  Right column: $f= 5\%$. }
\end{figure}

\begin{figure}[thb]
    \centering
        \includegraphics[width=.31\linewidth]{Figures/omega/reldiff/SURROGATES_3min/InVS15_temporal_surrogate_groups_bb_and_stats_size_04.png}
        \includegraphics[width=.31\linewidth]{Figures/omega/reldiff/SURROGATES_3min/InVS15_temporal_surrogate_groups_bb_and_stats_size_01.png}
        \includegraphics[width=.31\linewidth]{Figures/omega/reldiff/SURROGATES_3min/InVS15_temporal_surrogate_groups_bb_and_stats_size_005.png}
        \includegraphics[width=.31\linewidth]{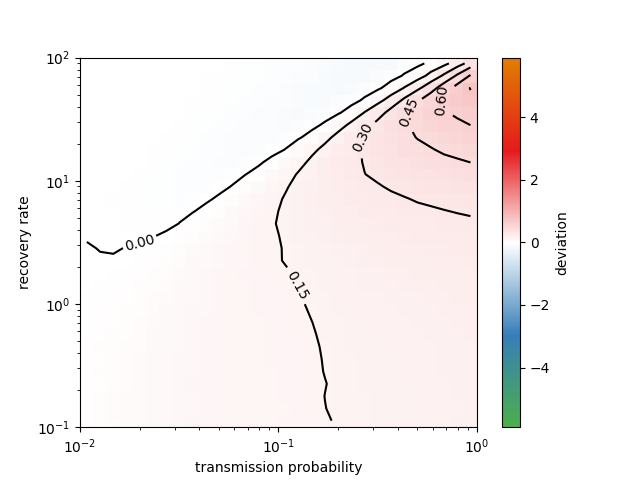}
        \includegraphics[width=.31\linewidth]{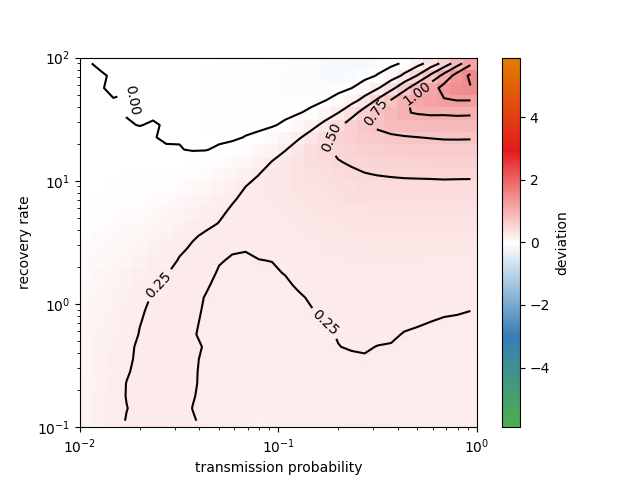}
        \includegraphics[width=.31\linewidth]{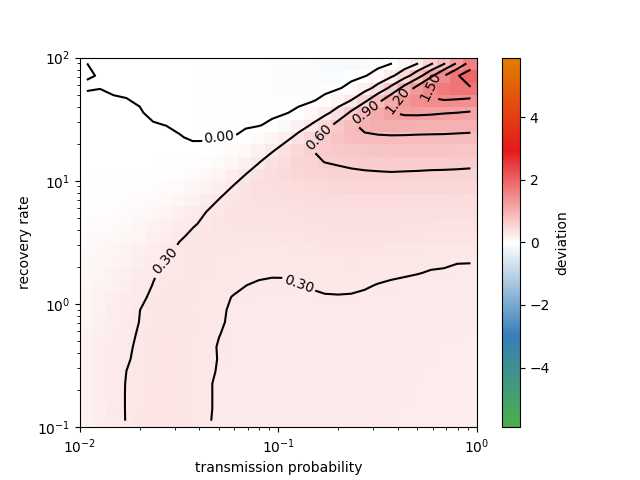}
        \includegraphics[width=.31\linewidth]{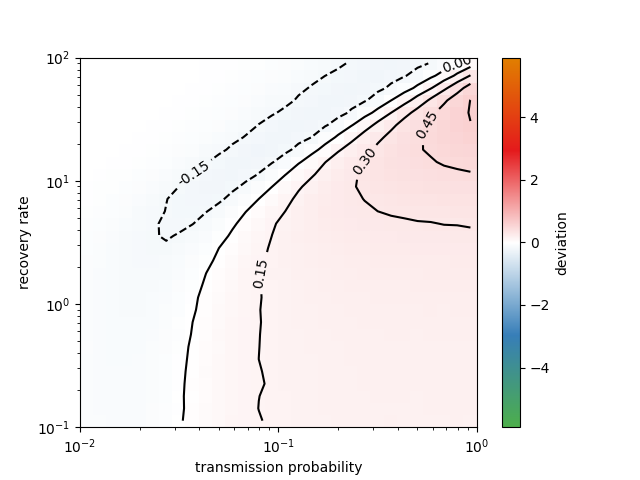}
        \includegraphics[width=.31\linewidth]{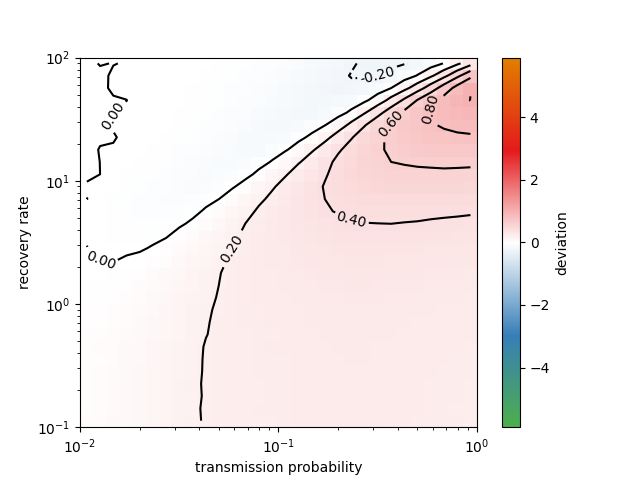}
        \includegraphics[width=.31\linewidth]{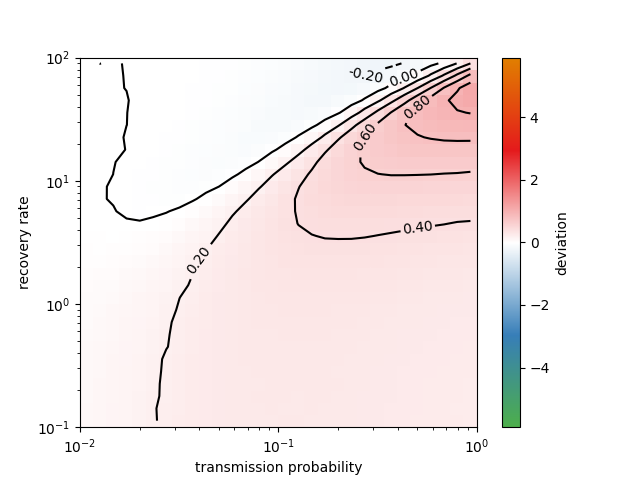}
        \includegraphics[width=.31\linewidth]{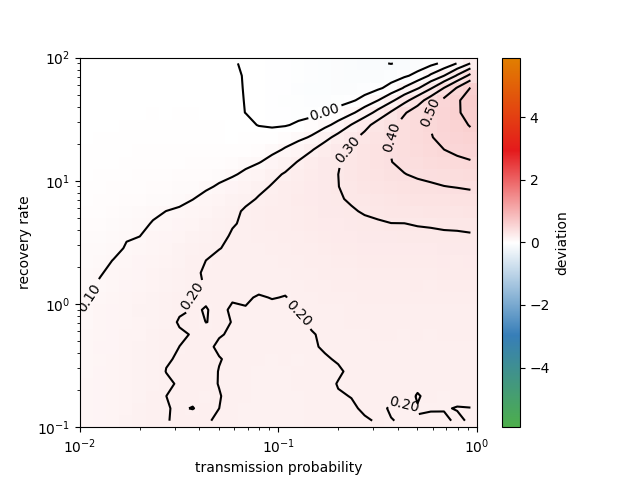}
        \includegraphics[width=.31\linewidth]{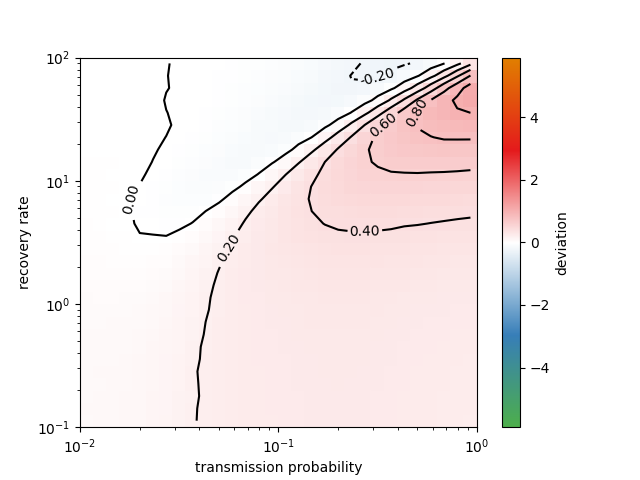}
        \includegraphics[width=.31\linewidth]{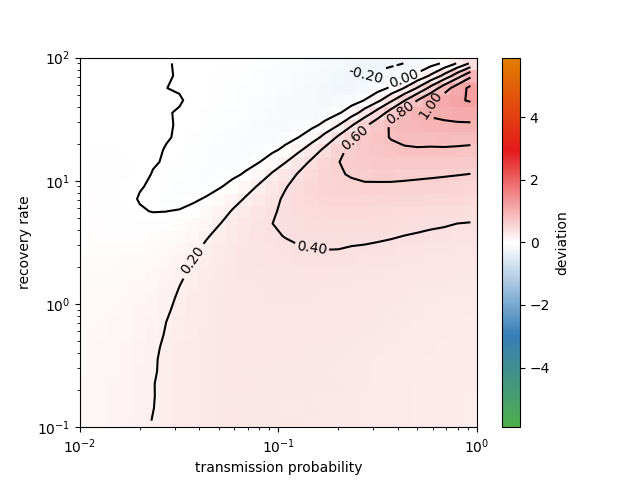}
    \caption{{\bf Outcome of SIR processes on surrogate data
obtained by various reconstruction methods, for the InVS15 data set.} Relative difference in the values of $\Omega$ measured in simulations on the surrogate and on the original data. 
First row: GST-RA. Second row: GST-OA. Third row: GST-RT. Fourth row: GTB-RT. In each case the backbone timelines are kept, and timelines respecting the statistics of contact and inter-contact durations are built for the surrogate ties (BTL-Stats method).
Left column: $f= 40\%$.  Middle column: $f= 10\%$.  Right column: $f= 5\%$. }
\end{figure}

\begin{figure}[thb]
    \centering
     \includegraphics[width=.31\linewidth]{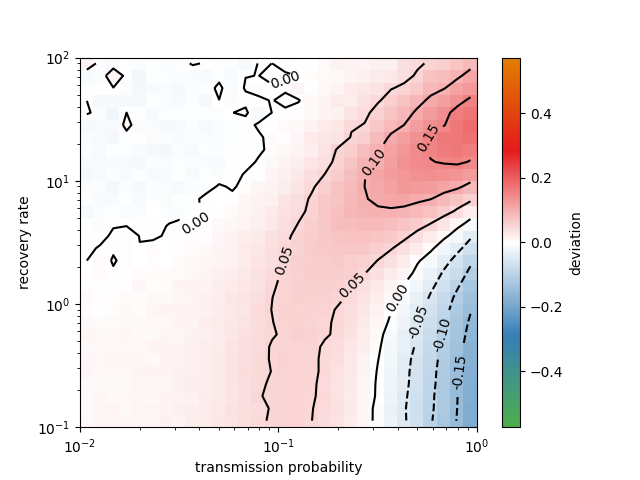}
        \includegraphics[width=.31\linewidth]{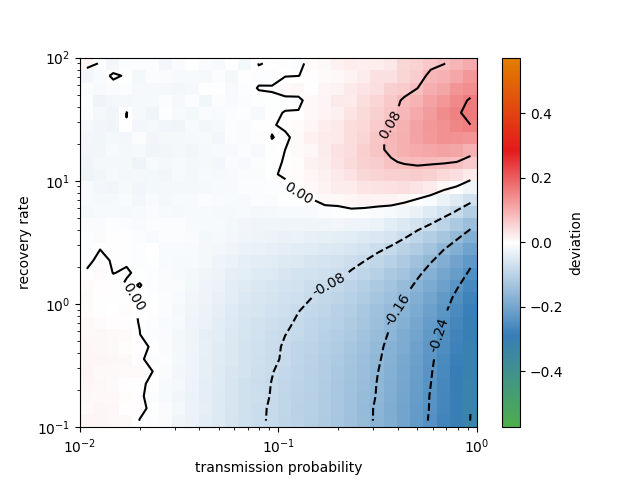}
        \includegraphics[width=.31\linewidth]{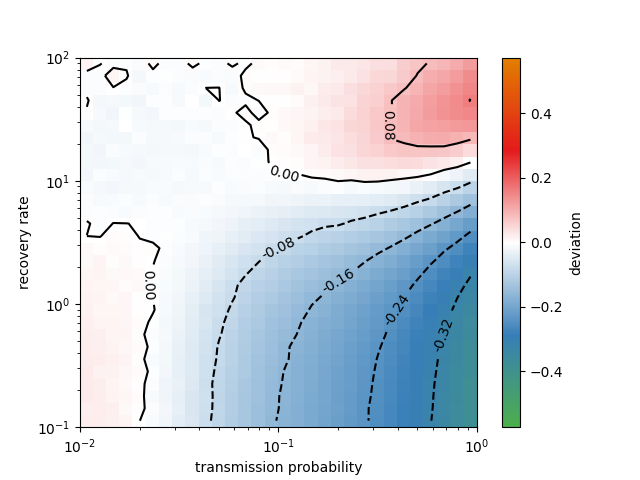}
          \includegraphics[width=.31\linewidth]{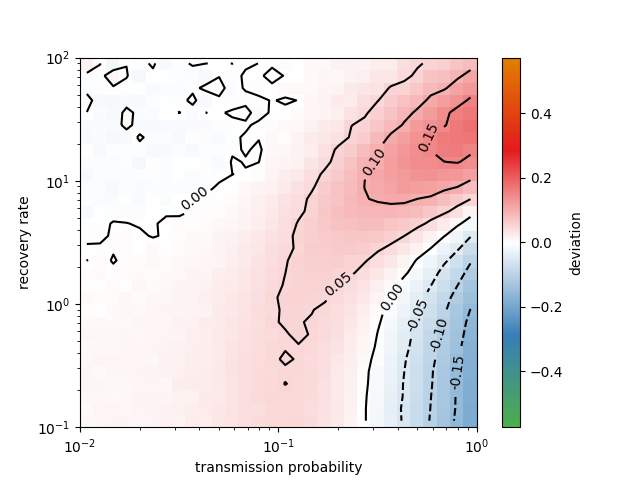}
        \includegraphics[width=.31\linewidth]{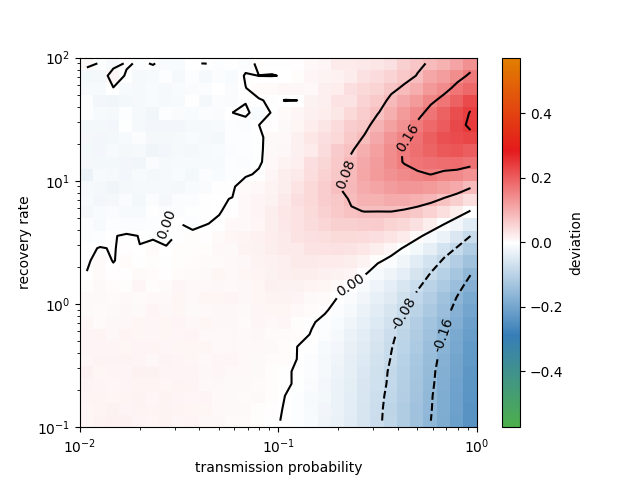}
        \includegraphics[width=.31\linewidth]{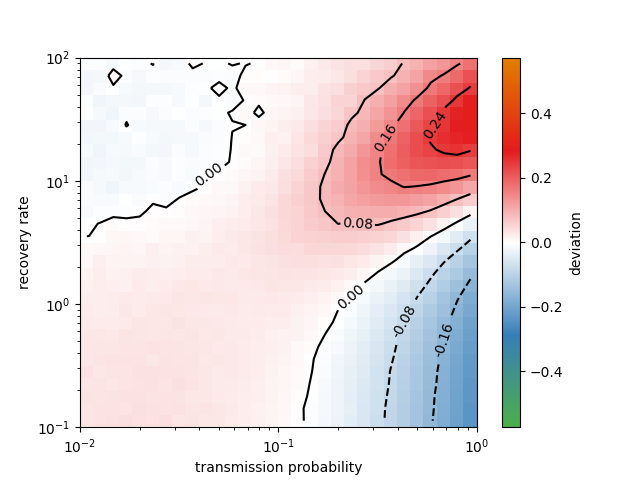}
        \includegraphics[width=.31\linewidth]{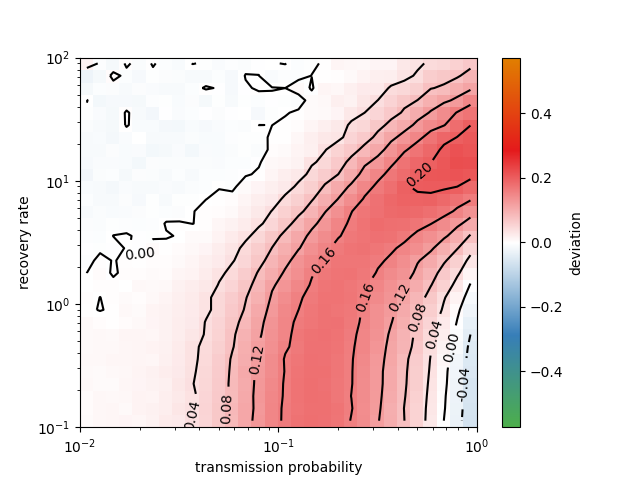}
        \includegraphics[width=.31\linewidth]{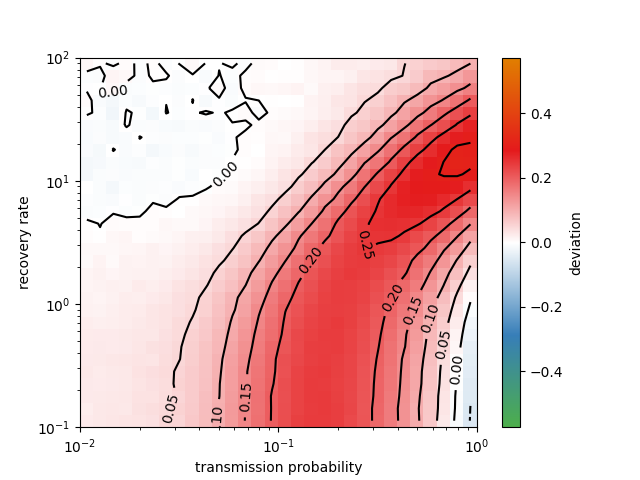}
        \includegraphics[width=.31\linewidth]{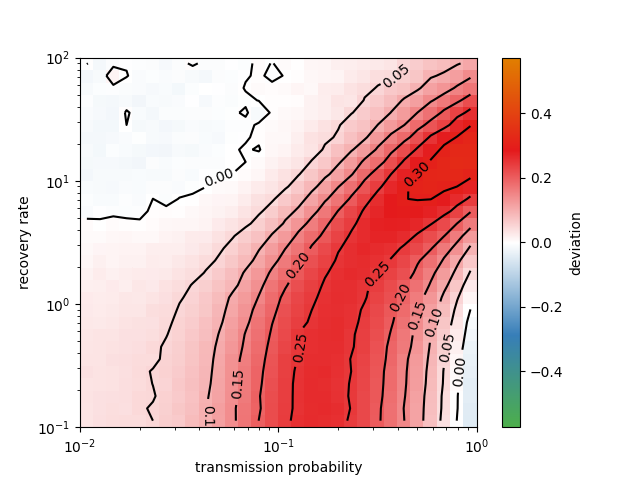}
        \includegraphics[width=.31\linewidth]{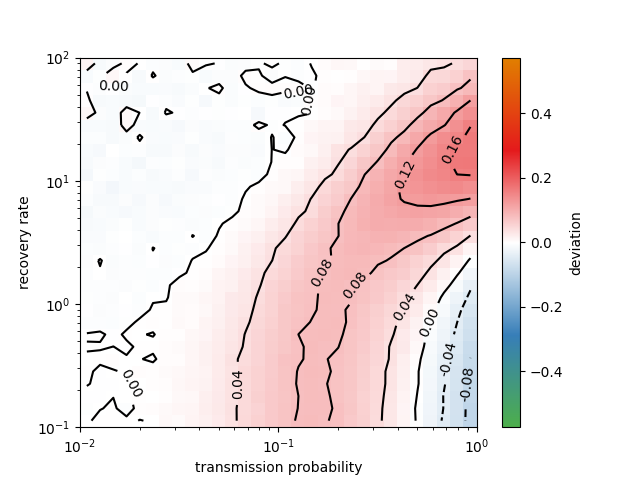}
        \includegraphics[width=.31\linewidth]{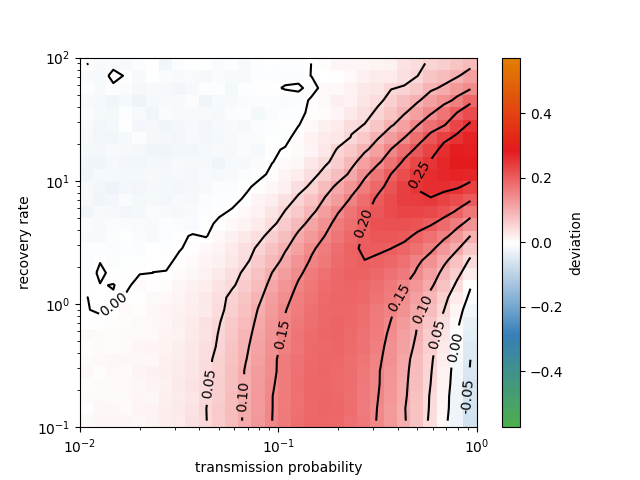}
        \includegraphics[width=.31\linewidth]{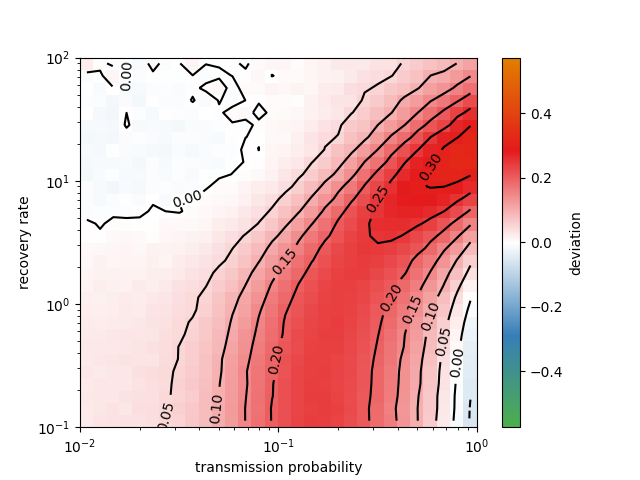}
    \caption{{\bf Outcome of SIR processes on surrogate data
obtained by various reconstruction methods, for the SFHH data set.} Relative difference in the values of $R_0$ measured in simulations on the surrogate and on the original data. 
First row: ST-RA. Second row: ST-OA. Third row: ST-RT.
Fourth row: TB-RT. In each case the backbone timelines are kept, and timelines respecting the statistics of contact and inter-contact durations are built for the surrogate ties (BTL-Stats method).
Left column: $f= 40\%$.  Middle column: $f= 10\%$.  Right column: $f= 5\%$. }
\end{figure}

\end{document}